\documentclass[useAMS,usenatbib,a4paper]{mn2e}

\usepackage{times}
\usepackage{multirow}
\usepackage[dvips]{graphicx}
\usepackage{setspace,psfrag,subfigure, amsfonts,theorem}
\usepackage{amssymb}
\usepackage{bm}
\usepackage[french,english]{babel}
\usepackage{color}
\usepackage{amsmath}
\usepackage{algorithmic}
\usepackage{algorithm}

\title[Tensor optimisation for optical interferometry]
{Tensor optimisation for optical-interferometric imaging}

\author[Aur{\'i}a et al.]
{Anna Aur{\'i}a$^{1}$ , Rafael Carrillo$^{1}$, Jean-Philippe Thiran$^{1,2}$ and Yves Wiaux$^{1,2,3,4}$\\ %
$^{1}$Institute of Electrical Engineering, Ecole Polytechnique F{\'e}d{\'e}rale de Lausanne (EPFL),
      CH-1015 Lausanne, Switzerland\\ %
$^{2}$Department of Medical Radiology, University Hospital Center (CHUV) and University of Lausanne (UNIL), CH-1011 Lausanne, Switzerland\\ %
$^{3}$Department of Radiology and Medical Informatics, University of Geneva (UniGE), 
      CH-1211 Geneva, Switzerland\\
$^{4}$Institute of Sensors, Signals \& Systems, Heriot-Watt University, Edinburgh EH14 4AS, UK
      }
\begin{document}

\date{\today}

\pagerange{\pageref{firstpage}--\pageref{lastpage}} \pubyear{2013}

\maketitle

\label{firstpage}

\begin{abstract}
Image recovery in optical interferometry is an ill-posed nonlinear inverse problem arising from incomplete power spectrum and bispectrum measurements. We reformulate this nonlinear problem as a linear problem for the supersymmetric rank-1 order-3 tensor formed by the tensor product of the vector representing the image under scrutiny with itself. On one hand, we propose a linear convex approach for tensor recovery with built-in supersymmetry, and regularising the inverse problem through a nuclear norm relaxation of a low-rank constraint. On the other hand, we also study a nonlinear nonconvex approach with built-in rank-1 constraint but where supersymmetry is relaxed, formulating the problem for the tensor product of 3 vectors. In this second approach, only linear convex minimisation subproblems are however solved, alternately and iteratively for the 3 vectors. We  provide a comparative analysis of these two novel approaches through numerical simulations on small-size images.
\end{abstract}


\section{Introduction}
\label{sec:intro}

Interferometry is a unique tool to image the sky at otherwise inaccessible resolutions.
Ideally, an interferometer measures complex visibilities identifying the Fourier coefficients of the intensity image $\bm{x}$ of interest. In this context, the visibility associated with a given telescope pair at one instant of observation gives the Fourier transform of the image of interest at a spatial frequency identified by the baseline components in the image plane. At radio wavelengths, these visibilities are indeed accessible, thereby setting a linear inverse problem in the perspective of image recovery. The standard CLEAN algorithm operates by local iterative removal of the convolution kernel associated with the partial Fourier coverage \citep{thompson01}. Convex optimisation methods regularising the inverse problem through sparsity constraints have recently been proposed in the framework of the recent theory of compressive sampling \citep{wia09a, wia09b, wia10, MW11, MNR:MNR21605, li, wenger}.

At optical wavelengths though, atmospheric turbulence induces a random phase delay that implies a systematic cancellation of the visibility values. Power spectrum information can however be retrieved, together with partial phase information through phase closure or bispectrum measurements \citep{thie2010, doi:10.1117/12.551018, Bald2002}. 
These considerations apply both to aperture masking interferometry on a single telescope \citep{Bald1986, Haniff1987, citeulike:11524465}, as well to optical interferometer arrays such as the Very Large Telescope Interferometer (VLTI)\footnote{www.eso.org/sci/facilities/paranal/telescopes/vlti/}. Providing detailed images of complex astrophysical phenomena is an important challenge for optical interferometry today \citep{Bald2002}. In the perspective of image recovery, prior constraints are also essential to regularise this nonlinear ill-posed inverse problem.

The state-of-the-art MiRA method \citep{Thie2008} takes a maximum a posteriori (MAP) approach where the image is the solution of an optimisation problem with an objective function $f(\bm{x})=f_{\rm data}(\bm{x})+\ell f_{\rm prior}(\bm{x})$, for some arbitrary parameter $\ell$ to be tuned, and with additional positivity and total flux constraints. Sparsity priors have in particular been promoted \citep{Thie2008, Renard:arXiv1106.4508}. The data nonlinearity induces nonconvexity of the objective function. The adopted strategy is to perform only local optimisation, in the context of which the solution depends not only on the data and on the priors but also strongly on the initial image and on the path followed by the local optimisation method. 
The WISARD alternative \citep{Meimon:05} takes a two-step alternate minimisation self-calibration approach. Firstly, the missing Fourier phases are recovered on the basis of a current estimate and phase closure information enabling to build pseudocomplex visibilities. Secondly, the image is recovered from the pseudocomplex visibilities as in radio interferometry. While the second step is convex and leads to a unique image independently of the initialisation, the first step is not. The overall procedure remains nonconvex and the final solution depends on the initial guess.
In summary, state-of-the-art methods are nonconvex due to the intrinsic data nonlinearity \citep{thie2010}, and therefore known to suffer from a strong sensitivity to initialisation.

The approaches proposed here stem from a different perspective. We firstly formulate a linear version of the problem for the real and positive supersymmetric rank-1 order-3 tensor $\mathcal{X}=\bm{x}\circ\bm{x}\circ\bm{x}$ formed by the tensor product of the size-$N$ vector $\bm{x}$ representing the image under scrutiny with itself. This allows us to pose a linear convex problem for recovery of a size-$N^3$ tensor $\mathcal{X}$ with built-in supersymmetry, and regularising the inverse problem through a nuclear norm relaxation of a low-rank constraint, also enforcing reality and positivity constraints. We also study a nonlinear nonconvex approach with built-in rank-1 constraint but where supersymmetry is relaxed, formulating the problem for the tensor product $\bm{u_1}\circ\bm{u_2}\circ\bm{u_3}$ of 3 size-$N$ vectors. In contrast with the state of the art though, only linear convex minimisation subproblems are solved, alternately and iteratively for the vectors, also enforcing reality and positivity\footnote{We also attempted an alternative nonconvex approach consisting in solving the nonlinear problem directly for $\bm{x}$, using the nonconvex projected gradient method proposed by \cite{Att}. First simulations did not produce any meaningful reconstruction and this approach was discarded.}. 
While the former approach is much heavier than the latter in terms of memory requirements and computation complexity due to the drastically increased dimensionality of the unknown, the underlying convexity ensures essential properties of convergence to a global minimum of the objective function and independence to initialisation, justifying a comparative analysis. For numerical experiments, we consider a generic discrete measurement setting where measurements identify with triple products of discrete Fourier coefficients of $\bm{x}$. These triple products are selected randomly according to a variable-density scheme sampling more densely low spatial frequencies, and are affected by simple additive Gaussian noise.

In Section \ref{sec:PSM}, we review convex optimisation and proximal splitting methods. In section \ref{sec:IP}, we introduce our generic discrete data model and describe our new linear tensor formulation of the optical-interferometric imaging problem. In sections \ref{sec:PF} and \ref{sec:NN}, the new AM and NM approaches are discussed. Our simulation setting for comparison of these two methods and corresponding results are presented in section \ref{sec:Simulations}. Finally, we conclude in section \ref{sec:Conclusion}.

\section{Convex optimisation and proximal splitting methods}\label{sec:PSM}
A real valued function $f(\bm{x})$, from $\mathbb{R}^{N}$ to $\mathbb{R}$, is called convex if 
\begin{equation}
f( (1-\beta)\bm{x}_1+\beta\bm{x}_2)\leq (1-\beta)f(\bm{x}_1)+\beta f(\bm{x}_2)
\end{equation}
for any $\bm{x}_1,\bm{x}_2\in \mathbb{R}^N$ and any $\beta\in[0,1]$. Optimisation problems including convex objective functions and convex constraints, called convex optimisation problems, have many attractive properties, in particular the essential property that any local minimum must be a global minimum, which comes directly from the definition of a convex function. Also, convex problems can be efficiently solved, both  in theory (i.e., via algorithms with worst-case  polynomial complexity) and in practice \citep{Boyd2004}. Among the broad range of convex optimisation methods, proximal splitting methods offer great flexibility and are shown to capture and extend several well-known algorithms in a unifying framework. Examples of proximal splitting algorithms include Douglas-Rachford, iterative thresholding, projected Landweber, projected gradient, forward-backward, alternating projections, alternating direction method of multipliers and alternating split Bregman \citep{Com}. They solve optimisation problems of the form
\begin{equation}\label{cvx1}
\min_{\bm{x}\in\mathbb{R}^{N}} f_1(\bm{x})+\ldots +f_K(\bm{x}),
\end{equation}
where $f_1(\bm{x}),\ldots,f_K(\bm{x})$ are convex lower semicontinuous functions from $\mathbb{R}^{N}$ to $\mathbb{R}$. In the case of convex constrained problems, they can be reformulated as unconstrained problems by using the indicator function of the convex constraint set as one of the functions in \eqref{cvx1}, i.e. $f_k(\bm{x})=i_{C}(\bm{x})$ where $C$ represents the convex constraint set. The indicator function, defined as $i_{C}(\bm{x})=0$ if $\bm{x}\in C$ or $i_{C}(\bm{x})=+\infty$ otherwise, belongs to the class of convex lower semicontinuous functions. Note that complex-valued vectors are treated as real-valued vectors with twice the dimension accounting for real and imaginary parts \citep{carrillo13b}. 

Proximal splitting methods proceed by splitting the contribution of the functions $f_1(\bm{x}),\ldots,f_K(\bm{x})$ individually so as to yield an easily implementable algorithm. They are called proximal because each non-smooth function in \eqref{cvx1} is incorporated in the minimisation via its proximity operator \citep{Com}. Let $f$ be a convex lower semicontinuous function from $\mathbb{R}^{N}$ to $\mathbb{R}$, then the proximity operator of $f$ is defined as:
\begin{equation}\label{proxdef}
\mathrm{prox}_{f}(\bm{x}) \triangleq \arg\min_{\bm{z}\in\mathbb{R}^{N}} f(\bm{z})+\frac{1}{2}\| \bm{x}-\bm{z} \|_2^2.
\end{equation}
In the case of indicator functions of convex sets, the proximity operator is the projection operator onto the set. Most proximal splitting algorithms reach a solution to \eqref{cvx1} by alternately applying the proximity operator associated with each function. For example, in the case that all functions in \eqref{cvx1} are indicator functions, the algorithm reduces to the classical projection onto convex sets algorithm \citep{Boyd2004}, which performs alternate projections to reach the solution. An important feature of proximal splitting methods is that they offer a powerful framework for solving convex problems in terms of speed and scalability of the techniques to very high dimensions. See \cite{Com} for a review of proximal splitting methods and their applications in signal and image processing. The reader is also referred to \citet{carrillo13b} for a description of proximal splitting algorithms and their use in radio-interferometric imaging.

\section{Data model and tensor formulation}\label{sec:IP}

For the sake of simplicity, we adopt a discrete setting where the intensity image of interest is represented by the real and positive vector $\bm{x}\in\mathbb{R}_+^N$ with components $x_i$. Its 2D discrete Fourier transform is denoted $\bm{\hat{x}}\in\mathbb{C}^N$ with components $\hat{x}_i$. By abuse of notation, we denote $\hat{x}_{-i}$ the component of $\bm{\hat{x}}$ at the opposite spatial frequency to that associated with $\hat{x}_{i}$. Signal reality implies $\hat{x}_{-i}=\hat{x}_{i}^*$, where $^*$ stands for complex conjugation.

The optical interferometry inverse problem is simplified considering a generic discrete measurement setting where the closure constraint is relaxed and optical-interferometric measurements take the generic form of a triple product of Fourier coefficients of the image: $ \hat{x}_i\hat{x}_{j}\hat{x}_k$. Power spectrum measurements follow with $j=-i$, and $k=0$ ($\hat{x}_0$ stands for the Fourier coefficient at zero frequency), and explicit bispectrum measurements would follow from the constraint that the spatial frequencies associated with $\hat{x}_i$, $\hat{x}_{j}$, and $\hat{x}_k$ sum to zero. In this context,  measurements are performed on the frequencies of a discrete grid in the Fourier plane, the so-called \textit{frequels}. In a real scenario the Fourier transform should be evaluated at (non-equispaced) continuous frequencies \citep{carrillo13b}.
We write the measurement equation in compact form as
\begin{equation}
\bm{y}=\mathcal{V}(\bm{x})+\bm{n},\label{mm1}
\end{equation}
where $\mathcal{V}$ is a nonlinear operator providing an undersampled set of triple products of Fourier coefficients of $\bm{x}$. The measurement vector $\bm{y}\in \mathbb{C}^M$, with components $y_a$ ($1\leq a \leq M$) is assumed to be affected by a simple noise vector $\bm{n}\in \mathbb{C}^M$ with i.i.d.~Gaussian components $n_a$. The number of measurements is typically smaller than the signal dimension: $M<N$. Finally, we assume that the total flux is measured independently and consider a normalised signal such that $\sum_i x_i = \hat{x}_0=1$.  This flux normalization is approximately enforced by adding the data point $\hat{x}_0^3=1$.

In what follows, we show how to bring the linearity of the measurement scheme by lifting the image model from a vector to a tensor formulation.
We start by reviewing some tensor definitions and notations. Firstly, the order (or number of dimensions, ways or modes) of a tensor $\mathcal{X}\in \mathbb{C}^{N_1 \times ... \times N_d}$ with components $\mathcal{X}_{i_1, ...., i_d}$ is the number $d$ of the indices characterising its components. For the sake of simplicity, we will present the formulation only for tensors of order 3. A 3-way tensor $\mathcal{X} \in \mathbb{C}^{N_1 \times N_2 \times N_3}$ is rank-1 if it can be written as the outer product of 3 vectors, i.e. $\mathcal{X}=\bm{a} \circ \bm{b} \circ \bm{c}$, or component-wise $\mathcal{X}_{ijk}=a_ib_jc_k$. 
Secondly, the rank of a tensor, rank$(\mathcal{X})$, is defined as the smallest number of rank-1 tensors that generate $\mathcal{X}$ as their sum. In other words, if $\mathcal{X}$ can be expressed as $\mathcal{X}=\sum_{r=1}^R \bm{a_r} \circ \bm{b_r} \circ \bm{c_r}$, then rank$(\mathcal{X}) \le R$.
The notion of rank when applied to a tensor is analogue to the matrix rank though most of the common properties of the latter do not hold when dealing with objects of a dimension higher than 2. One of the main differences is that there is no algorithm to compute the rank of a given tensor. In fact the problem is NP-hard \citep{Hastad90}. The well-known method to find a rank-$k$ approximation of a matrix through the largest $k$ values of its singular value decomposition (SVD) \citep{Stewart92onthe} does not apply or have an equivalent for the case of high-dimension tensors. Thirdly, \textit{matricisation} is the process of transforming a tensor into a matrix. The mode-$n$ matricisation of a tensor $\mathcal{X}$ is denoted by $\mathsf{X}_{(n)}$ and results from unfolding all its modes but the mode $n$ into the rows of a matrix.
The $n$-rank of a tensor follows as $n\text{-rank}(\mathcal{X})=(\text{rank}(\mathsf{X}_{(1)}),\text{rank}(\mathsf{X}_{(2)}),\text{rank}(\mathsf{X}_{(3)}))$. In contrast to the rank function, it is easier to handle, since the problem is reduced to calculations with matrices which are already well-known objects with nice properties. The reader can refer to the review from \cite{Kol} for a more detailed explanation on different notions of tensor rank and their associated decomposition methods.
Finally, a tensor is called cubical if every mode has the same size, i.e. $\mathcal{X} \in \mathbb{R}^{N \times N \times N}$. A cubical tensor $\mathcal{X}$ is called supersymmetric if its entries are invariant under permutation of their indices: $\mathcal{X}_{ijk}=\mathcal{X}_{ikj}=\mathcal{X}_{jik}=\mathcal{X}_{jki}=\mathcal{X}_{kij}=\mathcal{X}_{kji}$.

The measurement model (\ref{mm1}) can be recast as the following linear model for the real and positive supersymmetric rank-1 order-3 tensor $\mathcal{X}= \bm{x} \circ \bm{x} \circ \bm{x} \in\mathbb{R}_+ ^{ N \times N \times N}$:
\begin{equation}
\bm{y}=\mathcal{T}(\mathcal{X})+\bm{n},\label{mm2}
\end{equation}
where the linear operator $\mathcal{T}$ consists in performing a 2D discrete Fourier transform along each of the 3 dimensions, identified by an operator $\mathcal{F}$, followed by a selection and vectorisation operator  $\mathcal{M}$ providing variable-density undersampling in this 6D Fourier space: $\mathcal{T}=\mathcal{MF}$. 
The unit flux measurement is also included in the mask as a measurement on the ``triple-zero frequency''. Note that this formulation is a generalisation of the Phase Lift approach for the well-known phase retrieval problem \citep{Can2011b}. In that framework, quadratic measurements of the form $|\left \langle \bm{x},\bm{a_i} \right \rangle|^2$ for given projection vectors $\bm{a_i}$, are seen as linear measurements on the rank-1 matrix $\mathsf{X}= \bm{x}\bm{x}^{\dagger}$ representing the outer product of the signal with itself ($^{\dagger}$ stands for the conjugate-transpose operation).

We note however that the rank-1 and supersymmetry properties are not explicitly built-in in the tensor formulation (\ref{mm2}), which thereby presents a drastically increased dimensionality, $N^3$, of the unknown $\mathcal{X}$ compared to the original $\bm{x}$ of size $N$ in (\ref{mm1}). In the following sections, we discuss our two different regularisation schemes for tensor recovery. We firstly study a nonconvex alternate minimisation (AM) approach where the rank-1 constraint is built-in, and subsequently move to a convex nuclear minimisation (NM) scheme with built-in supersymmetry.

\vspace{-0.5cm}

\section{Rank-1 alternate minimisation (AM)}\label{sec:PF}

\subsection{Algorithm formulation}

We consider the following explicit rank-1 formulation of (\ref{mm2}), where supersymmetry is relaxed:
\begin{equation}
\bm{y}=\mathcal{T}(\bm{u_1} \circ \bm{u_2} \circ \bm{u_3})+\bm{n}.\label{mm3}
\end{equation}
The measurements can now be understood as an undersampled set of products of Fourier coefficients of $\bm{u_1}$, $\bm{u_2}$, and $\bm{u_3}$, thus bringing back nonlinearity. We consider the following nonconvex minimisation problem for tensor recovery:
\begin{equation}
\min_{\bm{u_1},\bm{u_2},\bm{u_3} \in \mathbb{R}_+ ^{ N}} \|\mathcal{T}(\bm{u_1} \circ \bm{u_2} \circ \bm{u_3}) - \bm{y}\|_2^2. \label{mm3'}
\end{equation}
A priori this problem seems as nonlinear and nonconvex as the initial problem (\ref{mm1}). Thanks to the nonsupersymmetric relaxation though, an alternate minimisation algorithm can be designed, solving sequentially for each variable ($\bm{u_1}$, $\bm{u_2}$ or $\bm{u_3}$) while keeping the other two fixed, and iterating until convergence. At each iteration, the 3 linear and convex subproblems
\begin{equation}
\min_{\bm{u_p} \in \mathbb{R}_+ ^{ N}} \|\mathsf{T}_{(\bm{u_qu_s})}\bm{u_p} - \bm{y}\|_2^2 \label{mm3''},
\end{equation}
are therefore solved sequentially for $1\leq p\neq q \neq s \leq 3$, where the linear operators $\mathsf{T}_{(\bm{u_qu_s})}$ are defined by $\mathsf{T}_{(\bm{u_qu_s})}\bm{u_p} \equiv \mathcal{T}(\bm{u_p} \circ \bm{u_q} \circ \bm{u_s})$. In each subproblem the linear operator is computed using the values of the fixed variables at the current step. The final AM algorithm is depicted in Algorithm \ref{alg1}. The algorithm is initialised with the same random vector for each of the 3 subproblems. The algorithm is stopped when the relative variation between the objective function in \eqref{mm3'} evaluated at successive solutions is smaller than some predefined bound or after the maximum number of iterations allowed is reached. At convergence, the tensor solution takes the form of 3 vectors $\bm{u}_{1}$, $\bm{u}_{2}$, and $\bm{u}_{3}$. We have no guarantee that the 3 solution vectors are identical and heuristically choose the final solution to be their mean as shown in step 8 of Algorithm \ref{alg1}\footnote{Note that \cite{Attouch08} prove that this alternate minimisation approach converges to a critical point of the objective function (\ref{mm3'}), provided that terms of the form $\gamma\|\bm{u_p}-\bm{\bar{u}_p}\|_2^2$ controlling the distance between the current unknown $\bm{u_p}$ with respect to its value at the previous iteration $\bm{\bar{u}_p}$ are added to the objective function in (\ref{mm3''}), for any $\gamma>0$. Simulations in the context of the setting described in Section \ref{sec:Simulations} show that the algorithm converges to the same solution for $\gamma\neq0$ and $\gamma=0$. Other simulations also show that starting the minimisation of the three variables with the same random initial point leads to very similar solutions for the 3 vectors, or for their mean, both in terms of signal-to-noise ratio and visual quality.}.

\begin{algorithm}[h!]
\caption{AM algorithm}\label{alg1}
\begin{algorithmic}[1]
\STATE Initialize $k=1$, $\bm{u_1}^{(0)},\bm{u_2}^{(0)},\bm{u_3}^{(0)}\in\mathbb{R}^N$.
\WHILE{not converged}
\STATE $\bm{u_1}^{(k)}=\arg\min_{\bm{u_1}} \|\mathsf{T}_{(\bm{u_2}^{(k-1)}\bm{u_3}^{(k-1)})}\bm{u_1} - \bm{y}\|_2^2$.
\STATE $\bm{u_2}^{(k)}=\arg\min_{\bm{u_2}} \|\mathsf{T}_{(\bm{u_1}^{(k)}\bm{u_3}^{(k-1)})}\bm{u_2} - \bm{y}\|_2^2$.
\STATE $\bm{u_3}^{(k)}=\arg\min_{\bm{u_3}} \|\mathsf{T}_{(\bm{u_1}^{(k)}\bm{u_2}^{(k)})}\bm{u_3} - \bm{y}\|_2^2$.
\STATE $k\leftarrow k+1$
\ENDWHILE
\STATE $\bm{x}_{\rm AM}=\frac{1}{3}(\bm{u_1}^{(k)}+\bm{u_2}^{(k)}+\bm{u_3}^{(k)})$
\RETURN $\bm{x}_{\rm AM}$
\end{algorithmic}
\end{algorithm}

\subsection{Optimisation details}
To solve each of the subproblems in Algorithm \ref{alg1} (steps 3--5) we resort to a forward-backward (projected gradient) algorithm \citep{Com}. The forward-backward algorithm solves \eqref{mm3''} using a two step procedure: a gradient descent step (forward step) to minimize the quadratic function in \eqref{mm3''}, and a projection step (backward step) to bring back the current update to the constraint set. The algorithm uses the following recursion:
\begin{equation}
\bm{u_p}^{(t+1)}=\mathrm{prox}_{i_{C}} \left( \bm{u_p}^{(t)}  + \mu_p^{(t)} \mathsf{T}_{(\bm{u_qu_s})}^{\dagger} \left(\bm{y}-\mathsf{T}_{(\bm{u_qu_s})}\bm{u_p}^{(t)} \right)\right),
\end{equation}
where $t$ denotes the iteration variable, $C=\mathbb{R}_+ ^{N}$ and $\mu^{(t)}$ is a variable step size that controls the gradient descent update. The step size is adapted using a backtracking line-search procedure \citep{beck09}. The proximity operator $\mathrm{prox}_{i_{C}}$ is nothing but the projector onto the positive orthant $\mathbb{R}_+ ^{N}$, i.e. setting the imaginary part and the negative values of the real part to zero \citep{Boyd2004}.

The memory requirement to solve this minimisation problem is dominated by the storage of the 3 vectors, which is of size $\mathcal{O}(N)$. In terms of computation time, the algorithm is dominated at it each iteration by the application of the operator $\mathcal{T}$ which computes 3 2D FFTs of size $N$, with an asymptotic complexity of order $\mathcal{O}(N\log N)$. This approach is computationally efficient. In contrast with the state of the art approaches such as MiRA and WISARD, it brings convexity to the subproblems. But the global problem remains nonconvex and the solution may still depend on the initialisation. One can easily identify convergence to a local minimum through large residual values of the objective function. With the aim to mitigate the dependence to initialisation, and as suggested by \cite{Hal}, we propose to run the algorithm $n_{\rm ri}$ times with random initialisations, choosing \textit{a posteriori} the solution with minimum objective function value. 

\section{Supersymmetric nuclear minimisation (NM)}\label{sec:NN}

\subsection{Algorithm formulation}

Tensor supersymmetry can be embedded in various ways. One approach is to formulate the inverse problem (\ref{mm2}) only for the subset of variables $\mathcal{X}_{ijk}$ with $i \leq j \leq k$. The collection of these values define the ``subtensor'' $\mathcal{X}_{\rm s}$, which can be related to $\mathcal{X}$ by an operator $\mathcal{R}$ replicating tensor components over all permutations for each triplet $(i,j,k)$: $\mathcal{X}=\mathcal{R}(\mathcal{X}_{\rm s})$. The inverse problem would thus read $\bm{y}=[\mathcal{TR}](\mathcal{X}_{\rm s})+\bm{n}$. We adopt an alternative and equivalent approach consisting in substituting the original measurement vector $\bm{y}$ by its replicated version $\mathcal{R}(\bm{y})$, and using a symmetrised version $\mathcal{M}_{\rm s}$ of the selection mask, ensuring that all permutations of a triplet $(i,j,k)$ are assumed to be measured. We will see below why a symmetrised data vector together with a symmetrised measurement operator represent a sufficient condition to impose the tensor symmetry at each step of the algorithm in our approach, and in particular supersymmetry of the solution. The modified inverse problem thus reads as:
\begin{equation}
\bm{y}_{\rm s}=\mathcal{T}_{\rm s}(\mathcal{X})+\bm{n}_{\rm s},\label{mm4}
\end{equation}
with $\bm{y}_{\rm s}=\mathcal{R}(\bm{y})$, $\bm{n}_{\rm s}=\mathcal{R}(\bm{n})$ and $\mathcal{T}_{\rm s}=\mathcal{M}_{\rm s}\mathcal{F}$ denoting the symmetrised versions of the measurement vector, noise vector and measurement operator, respectively. Without loss of generality, we assume that the initial selection operator $\mathcal{M}$ contains no redundant measurements, i.e.~ $i \leq j \leq k$. This ensures that  $\mathcal{R}$ is well-defined. Also note that the noise statistics remains unaltered and only concerns the entries before replication.

Low-rankness, reality and positivity will be imposed as regularisation priors in the convex minimisation problem to be defined. 
As pointed out, the rank of a tensor is difficult to handle since the problem of finding rank($\mathcal{X}$) is NP-hard. Computing the rank of different matricisations of the tensor is an easier task. The unfoldings of a rank-1 tensor are actually rank-1 matrices, so that a low $n$-rank constraint can be used as a proxy for low-rankness. The rank of a matrix is however a nonconvex function. The nuclear norm, defined as the $\ell_1-$norm of its singular values, is a well-known convex relaxation of the rank function that was recently promoted in matrix recovery theory \citep{candes2009}.
Building on those results, \cite{Gan} tackle the low-$n$-rank tensor recovery problem through the minimisation of the sum of the nuclear norms of the mode-$n$ matricisations $\mathsf{X}_{(n)}$ for all $n$. In the supersymmetric case, the mode-$n$ matricisations are all identical and denoted $\mathsf{X}_{(n)}=\mathcal{U}(\mathcal{X})\in\mathbb{C}^{N\times N^2}$, where $\mathcal{U}$ stands for the unfolding operator. We propose here to exploit the symmetry of the tensor under scrutiny, together with the signal normalisation, to promote a computationally more efficient low-rank prior.
Relying on these properties, we note that summations over one index of a tensor of the form $\bm{x} \circ \bm{x} \circ \bm{x} $ with $\sum_i x_i = 1$ leads to the order-2 tensor $\bm{x} \circ \bm{x}$, which is real, positive, symmetric, as well as rank-1 and positive-semidefinite. We define $\mathcal{C}$ as the operator performing the summation over one dimension. Once more supersymmetry ensures that the resulting matrix is independent of the choice of the dimension along which components are summed up: $\mathcal{C}(\mathcal{X})\in\mathbb{C}^{N\times N}$ with $[\mathcal{C}(\mathcal{X})]_{ij}= \sum_k \mathcal{X}_{ijk}$. A low-rank constraint on $\mathcal{C}(\mathcal{X})$ will be promoted, through a nuclear norm minimisation, as a convex proxy for the low-rankness of $\mathcal{X}$. Positive-semidefiniteness of $\mathcal{C}(\mathcal{X})$, i.e.~positivity of the eigenvalues, which are then identical to the singular values, may also be explicitly added as a convex prior, denoted $\mathcal{C} (\mathcal{X}) \succeq 0$, together with the convex reality and positivity constraints of $\mathcal{X}$: $\mathcal{X} \in \mathbb{R}_+^{N\times N \times N}$. This summation approach is \textit{a priori} computationally significantly more efficient given the reduced matrix size of $\mathcal{C}(\mathcal{X})$ compared to that of the unfolded matrix $\mathcal{U}(\mathcal{X})$.

The resulting convex nuclear norm minimisation problem (NM) for $\mathcal{X}$ thus reads as:
\begin{equation}
\min_{\mathcal{X}\in S} \|\mathcal{C}(\mathcal{X})\|_* \quad  
 \text{subject to}  \quad  \|\bm{y}_{\rm s}-\mathcal{T}_{\rm s}(\mathcal{X})\|_2 \leq \epsilon, \label{eq:minNN}
\end{equation}
where $S= S_1 \cap S_2$, with $S_1=\mathbb{R}_+^{N\times N \times N}$ and $S_2=\{\mathcal{X} \, \vert \, \mathcal{C} (\mathcal{X}) \succeq 0 \}$. Recalling that the measurements $\bm{y}$ are assumed to be corrupted with simple i.i.d. complex Gaussian noise with variance $\sigma_n^2/2$ on real and imaginary parts, the residual estimator $\|\bm{y}-\mathcal{T}(\mathcal{X})\|_2^2$ follows a $\chi^2$ distribution with $2M$ degrees of freedom, with expectation $2M$ and standard deviation is $(4M )^{1/2}$. For a large number of degrees of freedom the distribution is extremely peaked around its expectation value. This fact is related to the well-known phenomenon of the concentration of measure \citep{MNR:MNR21605}. The value $\epsilon_0^2=(2M + 4\sqrt{M})\sigma_n^2/2$, i.e.~2 standard deviations above the expectation, represents a high percentile of the distribution (in practice extremely close to $2M$), and consequently a likely bound for $\|\bm{n}\|_2^2$. An equivalent bound for the symmetrised residual noise term $\|\bm{y}_{\rm s}-\mathcal{T}_{\rm s}(\mathcal{X})\|_2^2$ may simply be inferred as $\epsilon^2\simeq\alpha\epsilon_0^2$, where $\alpha$ is simply the ratio of number of components in $\bm{y}_{\rm s}$ and $\bm{y}$. We take the value $\alpha=6$ as the relative number of $(i,j,k)$ triplets with repeated indices in the mask is very small. Note that this last consideration only arises from the discrete setting adopted.

Once the tensor solution $\mathcal{X}_{\rm NM}$ is recovered, the problem of extracting the sought signal $\bm{x}_{\rm NM}$ remains. 
If the tensor solution was actually a real positive rank-1 supersymmetric tensor whose elements sum up to unity, the retrieval of $\bm{x}_{\rm NM}$ could be done in different ways, such as directly extracting the first eigenvector of matrix $\mathcal{C}(\mathcal{X}_{\rm NM})$ or simply performing a sum over two dimensions $\sum_{jk}[\mathcal{X}_{\rm NM}]_{ijk}$. The nuclear norm minimisation approach however does not guarantee that the final solution is indeed rank-1. We therefore resort to the generic algorithm proposed by \citet{Kofidis02onthe} to find the best rank-1 supersymmetric approximation $\mathcal{P}_1(\mathcal{X}_{\rm NM})$ of a supersymmetric tensor $\mathcal{X}_{\rm NM}$ in the least square sense. This algorithm is a generalisation for the tensor case of the power method applied to find the dominant eigenvector of matrices \citep{Golub89}. It boils down to determining a unitary vector $\bm{x}$ and a scalar $\lambda$, such that $\|\mathcal{X} - \lambda \bm{x} \circ \bm{x} \circ \bm{x} \| $ is minimised, where $\|\cdot\|$ indicates simply the sum of the square of the components of the tensor. We denote the resulting solution as
\begin{equation}
\bm{x}_{\rm NM}= [\mathcal{EP}_1](\mathcal{X}_{\rm NM}),
\end{equation}
where $\mathcal{E}$ formally represents the operator retrieving from a supersymmetric rank-1 order-3 tensor its underpinning vector. Note that this vector extraction problem is not convex \footnote{Note that \cite{Kofidis02onthe} provide a proof of convergence of their algorithm for even-order tensors only. Simulations in the context of the setting described in Section \ref{sec:Simulations} show that the this procedure systematically converges for our order-3 tensors, and provides significantly better results than a heuristic procedure based on extracting the first eigenvector of $\mathcal{C}(\mathcal{X}_{\rm NM})$ or performing a sum over two dimensions $\sum_{jk}[\mathcal{X}_{\rm NM}]_{ijk}$.}. 

The final NM algorithm is shown in Algorithm \ref{alg2}. To solve the complex optimisation problem in \eqref{eq:minNN} we use the Douglas-Rachford splitting algorithm, which is tailored to solve problems of the form in \eqref{cvx1} with $K=2$. The problem in \eqref{eq:minNN} can be reformulated as in \eqref{cvx1} by setting $f_1(\mathcal{X})=\|\mathcal{C}(\mathcal{X})\|_*+i_{S}(\mathcal{X})$ and $f_2(\bm{x})=i_{C_{\epsilon}}(\mathcal{X})$, where $C_{\epsilon}=\{ \mathcal{X} \in \mathbb{C}^{N\times N \times N} : \|\bm{y}_{\rm s}-\mathcal{T}_{\rm s}(\mathcal{X})\|_2 \leq \epsilon \}$. The main recursion of the Douglas-Rachford algorithm is detailed in steps 3-4 of Algorithm \ref{alg2}, where $\nu>0$ and $\tau_k\in(0,2)$ are convergence parameters. The sequence $\{\mathcal{X}^{(k)} \}$ generated by the recursion in Algorithm \ref{alg2} converges to a solution of the problem \eqref{eq:minNN}~\citep{Com}. The algorithm is stopped when the relative variation between successive solutions, $\left \|\mathcal{X}^{(k)}-\mathcal{X}^{(k-1)} \right\|/\left \| \mathcal{X}^{(k-1)} \right \|$, is smaller than some bound $\xi\in(0,1)$, or after the maximum number of iterations allowed, $T_{\rm{max}}$, is reached. In our implementation we use the values $\tau_k = 1$, $\forall t$, $\xi=10^{-3}$ and $\nu = 10^{-1}$. In the following subsection we detail the computation of the proximity operators for $f_1$ and $f_2$.

\begin{algorithm}[h!]
\caption{NM algorithm}\label{alg2}
\begin{algorithmic}[1]
\STATE Initialize $k=1$, $\mathcal{X}^{(1)}\in\mathbb{R}^{N \times N \times N}$, $\tau_k\in(0,2)$ and $\nu>0$.
\WHILE{not converged}
\STATE $\mathcal{Z}^{(k)}=\mathrm{prox}_{\nu f_2}\left( \mathcal{X}^{(k)}\right)$.
\STATE $\mathcal{X}^{(k+1)}=\mathcal{X}^{(k)}+\tau_k\left(\mathrm{prox}_{\nu f_1}\left(2\mathcal{Z}^{(k)}-\mathcal{X}^{(k)}\right)-\mathcal{Z}^{(k)}\right)$.
\STATE $k\leftarrow k+1$
\ENDWHILE
\STATE $\bm{x}_{\rm NM}= [\mathcal{EP}_1](\mathcal{X}^{(k)})$.
\RETURN $\bm{x}_{\rm NM}$ 
\end{algorithmic}
\end{algorithm}

\vspace{0.5cm}
\subsection{Optimisation details}
The computation of the proximal operator of $f_1$, which includes the nuclear norm prior, as well as the positive-semidefiniteness, reality and positivity constraints, is itself a complicated optimisation problem. Therefore the dual foward-backward algorithm \citep{Com} is used at each iteration of the Douglas-Rachford recursion to compute the proximal operator of $f_1$. We can decompose $f_1$ as $f_1(\mathcal{X})=g_1(\mathcal{X})+g_2(\mathcal{X})$, where $g_1(\mathcal{X})=\|\mathcal{C}(\mathcal{X})\|_*+i_{S_1}(\mathcal{X})$ and $g_2(\mathcal{X})=i_{S_2}(\mathcal{X})$. Let $\mathsf{Q}^{(0)}\in \mathbb{C}^{N \times N}$ and $\mathcal{S}^{(0)}\in\mathbb{C}^{N\times N\times N}$ be the all zero matrix and the all zero tensor respectively. The dual forward-backward algorithm uses the following recursion to compute $\mathrm{prox}_{\nu f_1}(\mathcal{X})$:
\begin{align}\label{proxf2}
\mathsf{Q}^{(t+1)}&=\left ( \mathsf{I}-\mathrm{prox}_{\nu g_1}\right ) \left ( \mathsf{Q}^{(t)}+\gamma_t\mathcal{C}(\mathcal{S}^{(t)})\right )\\ \nonumber
\mathcal{S}^{(t+1)}&=\mathrm{prox}_{ \nu g_2} \left(\mathcal{X}-\mathcal{C}^{\dagger}(\mathsf{Q}^{(t+1)}) \right),
\end{align}
where $\mathsf{I}\in \mathbb{R}^{N \times N}$ is the identity operator and $\gamma_t \in (0,2)$ is a step size. The sequence $\{\mathcal{S}^{(t)} \}$ converges linearly to $\mathrm{prox}_{\nu f_1}(\mathcal{X})$. 

The computation of $\mathrm{prox}_{\nu g_1}$ and $\mathrm{prox}_{\nu g_2}$ are very simple operations. We start by computing $\mathrm{prox}_{\nu g_1}$. Let $\mathsf{Q}\in \mathbb{C}^{N \times N}$ be a symmetric matrix and suppose it has an eigenvalue decomposition $\mathsf{U} \mathsf{\Lambda} \mathsf{U}^{\dagger}$, where $\mathsf{U}$ is the orthogonal matrix of eigenvectors and $\mathsf{\Lambda}=\mathrm{diag}(\lambda_1,\ldots, \lambda_N)$ is the diagonal matrix with the eigenvalues. Then, the proximity operator of $\nu g_1$ is computed as:
\begin{equation}
\mathrm{prox}_{\nu g_1}(\mathsf{Q})=\mathsf{U} \bar{\mathsf{\Lambda}}_{\nu} \mathsf{U}^{\dagger},
\end{equation} 
where $\bar{\mathsf{\Lambda}}_{\nu}=\mathrm{diag}((\lambda_1-\nu)^+,\ldots, (\lambda_N-\nu)^+)$ and $(a)^+=\max(0,a)$ denotes the positive part of $a$. The operator $\bar{\mathsf{\Lambda}}_{\nu}$ performs a soft thresholding on the eigenvalues of $\mathsf{Q}$, to minimise the nuclear norm, and also preserves only the positive eigenvalues, to project onto the set of positive-semidefinite matrices \citep{Cai2010,Vanden96}. The proximal operator of $\nu g_2$ is the projector onto the set of positive tensors in $\mathbb{R}^{N \times N \times N}$ which is computed by setting the imaginary part and the negative values of the real part of the input tensor to zero, i.e.
\begin{equation}
\mathrm{prox}_{\nu g_2}(\mathcal{S})=\{ (\mathbf{Re}(\mathcal{S}_{i,j,k}))^+\}_{1\leq i,j,k \leq N},
\end{equation}
where $\mathbf{Re}(\cdot)$ denotes the real part of a complex number \citep{Boyd2004}.

The proximal operator of $f_2$ is the projector operator onto the set $C_{\epsilon}$, which is computed as:
\begin{equation}
\mathrm{prox}_{\nu f_2}(\mathcal{X})=\mathcal{X}+\mathcal{T}_{\rm s}^{\dagger}\left( \mathcal{P}_{\epsilon}\left( \mathcal{T}_{\rm s}(\mathcal{X})-\bm{y}_{\rm s}\right)-\mathcal{T}_{\rm s}(\mathcal{X})+\bm{y}_{\rm s}\right),
\end{equation}
where $\mathcal{P}_{\epsilon}(\bm{r})=\min(1,\epsilon/\|\bm{r}\|_2)\bm{r}$. 

All the operations done in the computation of the proximal operators of $f_1$ and $f_2$ preserve tensor symmetry, provided that the symmetrised version $\mathcal{T}_{\rm s}$ of the measurement operator and a symmetrised data vector are used. These two are sufficient conditions to impose supersymmetry at each iteration of Algorithm \ref{alg2}, and consequently for the final tensor solution.

The memory requirement to solve this NM problem is dominated by the storage of the tensor, which is of size $\mathcal{O}(N^3)$. In terms of computation time, the algorithm is dominated at it each iteration by the application of the operator $\mathcal{T}_{\rm s}$ which computes $N^2$ 2D FFTs of size $N$ along each of the three dimensions, with an asymptotic complexity of $\mathcal{O}(N^3\log N)$. These orders of magnitude obviously stand in starck contrast with those for the AM approach.

While the NM approach is much heavier than the AM approach in terms of memory requirements and computation complexity due to the drastically increased dimensionality of the unknown, the underlying convexity at the tensor level ensures essential properties of convergence to a global minimum of the objective function and independence to initialisation, justifying a comparative analysis.

\section{Simulations and results}
\label{sec:Simulations}

In this section we evaluate the performance of the NM and AM algorithms through numerical simulations. Our optimisation code\footnote{Code and test data are available at https://github.com/basp-group/co-oi.} was implemented in MATLAB and run on a standard 2.4 GHz Intel Xeon processor. Given the expected large memory requirements and long reconstructions time for the NM formulation, we consider small-size images with $N=16^2=256$ for which the image vector occupies the order of 4 KB in double precision, while the size-$N^3$ tensor variable already takes the order of 100 MB. The memory requirement for the simple tensor variable would already rise to the order of 8 GB for a $32^2=1024$ image size.

\begin{figure}
\centering
   
   \includegraphics[
    clip, keepaspectratio, width = 6.5cm]{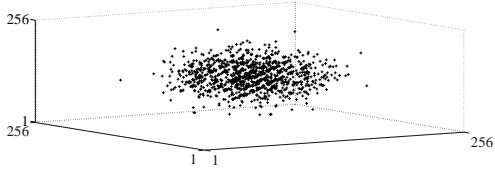}
  
\caption{Example of variable-density sampling pattern in the discrete 6D Fourier space of $\mathcal{X}$ of dimension $N^3$, for a $N=16^2$ image size and an undersampling regime of $M/N=0.75$.}
\label{fig:sp}
\end{figure}

For what the measurement setting is concerned, we assume random variable-density sampling in the 6D Fourier space, where low spatial frequencies are more likely to by sampled than high frequencies. In practice the sampling pattern is obtained by sampling \textit{frequels} independently along each of the 3 tensor dimensions from a bidimensional random Gaussian profile in the corresponding fourier plane,  associating the originally continuous random points with the nearest discrete frequency. The sampling is carried out progressively, noting that if a product is sampled twice the result is discarded and repeating this procedure until $M$ samples are obtained. Again this consideration only arises from the discrete setting adopted. Figure \ref{fig:sp} presents a typical sampling pattern.

In all experiments we define the input signal-to-noise ratio as ${\rm ISNR}= - 10 \log (\sigma_n^2/e_y^2)$ where $e_y^2=(1/M)\sum_a^M | y_a|^2$. The signal-to-noise ratio of a reconstruction $\bm{\bar{x}}$ is defined as ${\rm SNR}= - 10 \log (\|\bm{\bar{x}}-\bm{x} \|^2/\|\bm{x} \|^2)$. With this definition, the higher the SNR, the closer the recovered signal $\bm{\bar{x}}$ is from the original $\bm{x}$.

\begin{figure}
\centering
   
   \includegraphics[
    clip, keepaspectratio, width = 6.5cm]{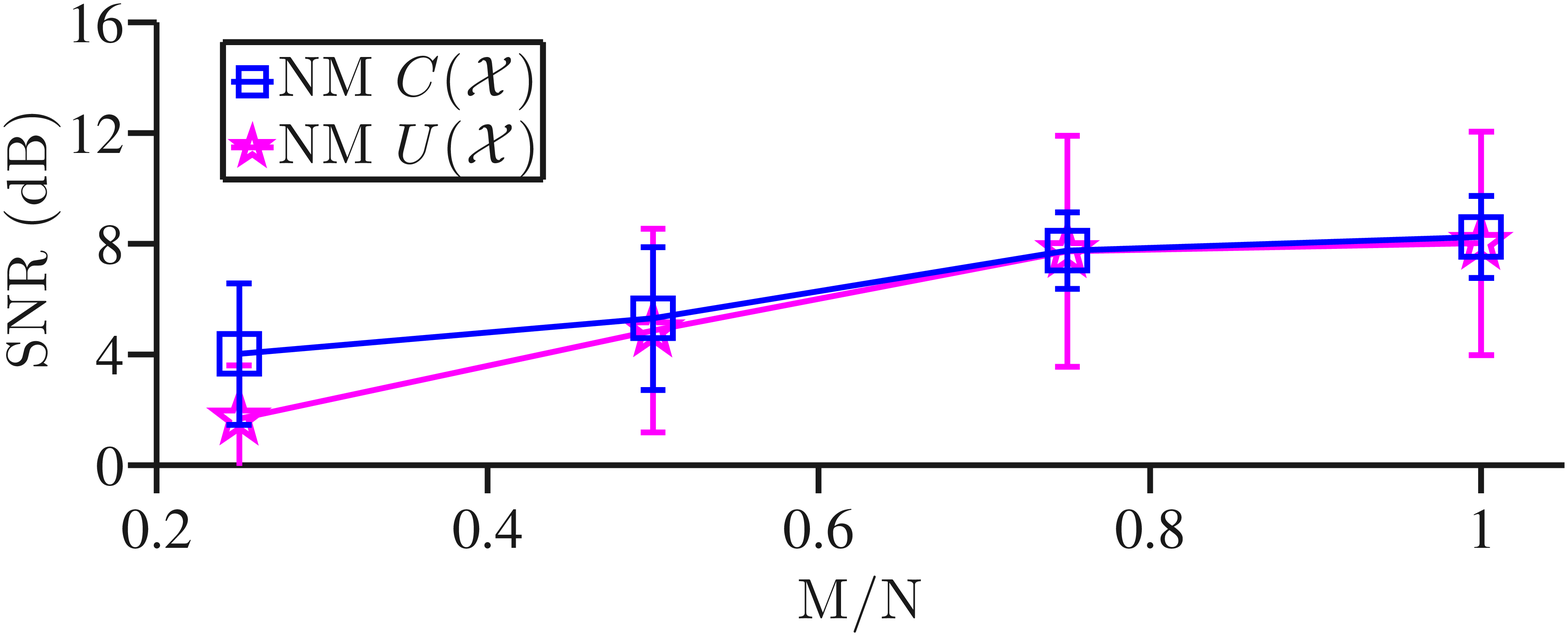}
    \includegraphics[
     clip, keepaspectratio, width = 6.5cm]{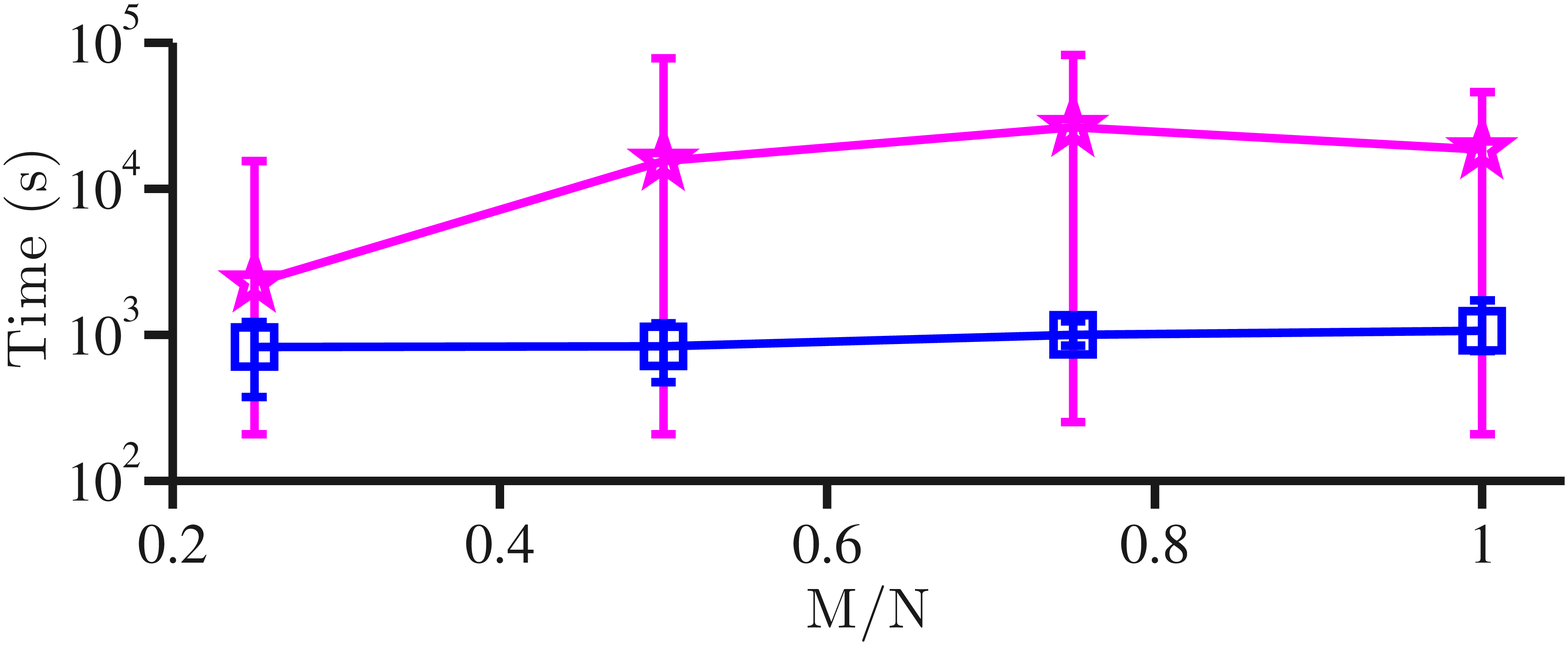}
  
\caption{Reconstruction quality and timing comparison between the NM approach defined in (\ref{eq:minNN}), with the equivalent minimisation problem where the summation operator $\mathcal{C}$ is replaced by the unfolding operator $\mathcal{U}$. Tests done on $N=16^2$ images with $32$ randomly located spikes and ISNR $=$ 30dB, for undersampling ratios $M/N$ in the range $[0.25, 1]$. The SNR curves (top panel) represent average values over $10$ simulations and corresponding 1-standard-deviation error bars. The timing curves (bottom panel) represent average values over $10$ simulations and min-max error bars.}
\label{fig:1}
\end{figure}

As a preliminary experiment, we provide a comparison of the performance of the NM approach defined in (\ref{eq:minNN}), with the equivalent minimisation problem where the summation operator $\mathcal{C}$ is replaced by the unfolding operator $\mathcal{U}$ in the nuclear norm and where the positive-semidefiniteness constraint is discarded as it does not apply for non-square matrices. Both algorithms were tested on images constructed from 32 random spikes, with ISNR $=$ 30dB. The positive spike values are taken uniformly at random and normalised to get unit flux, while positions are drawn at random from a Gaussian profile centred on the image. The graphs in Figure \ref{fig:1} represent the SNR and timing curves as a function of undersampling in the range $[0.25, 1]$. A total of $10$ simulations per point are performed, varying the signal, as well as the sampling and noise realisations. Both approaches provide similar reconstruction qualities, with a smaller variability of the component summation approach, which is also slightly superior at low undersamplings. The component summation approach, running in the order of $10^3$ seconds, is as expected significantly faster than the unfolding approach, running on average more than $10$ times more slowly in the range $[0.5, 1]$. We therefore discard further consideration of the latter.

\begin{figure}
\centering
   
    \includegraphics[
     clip, keepaspectratio, width = 6.5cm]{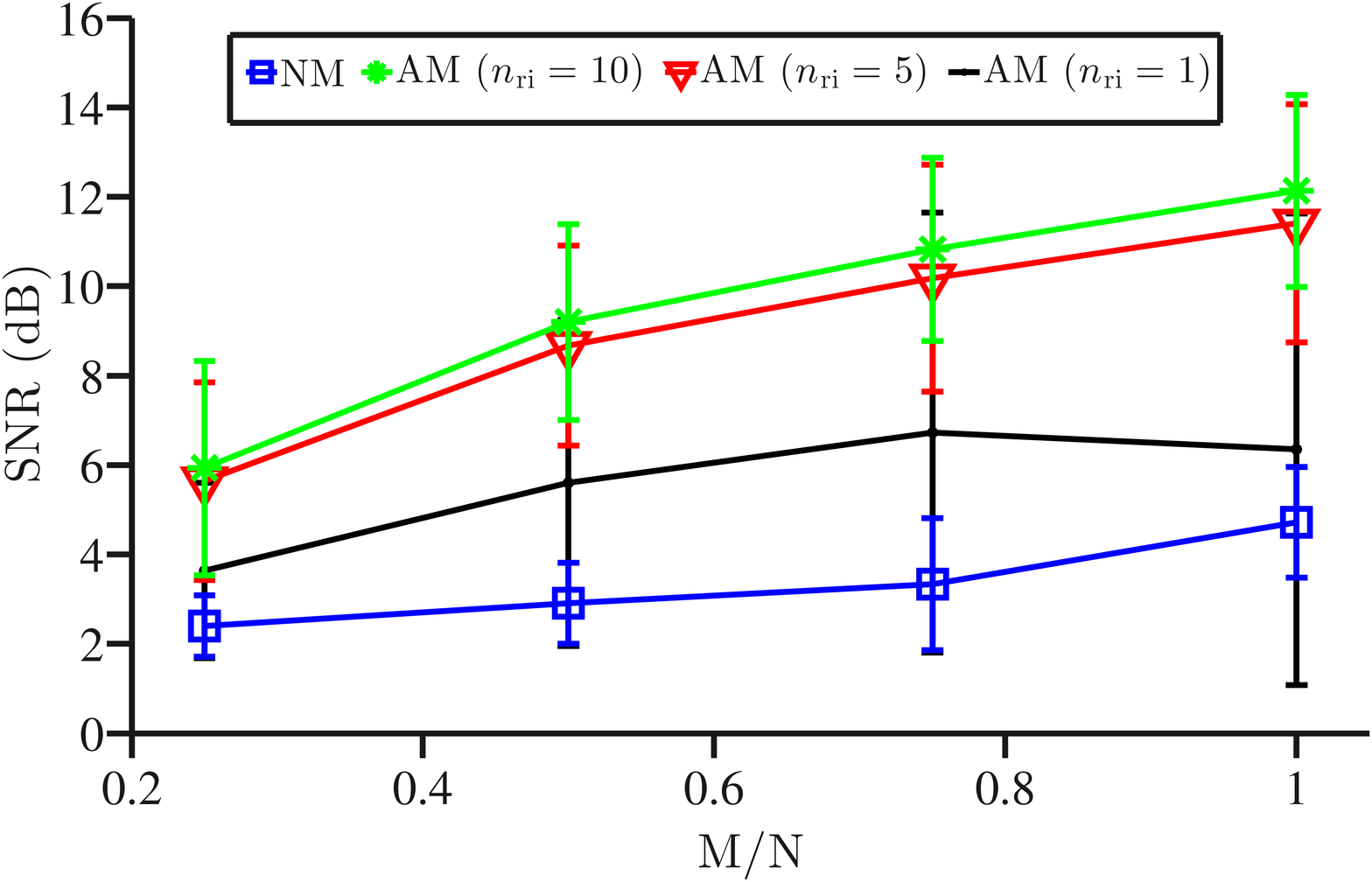}
    \includegraphics[
    clip, keepaspectratio, width = 6.5cm]{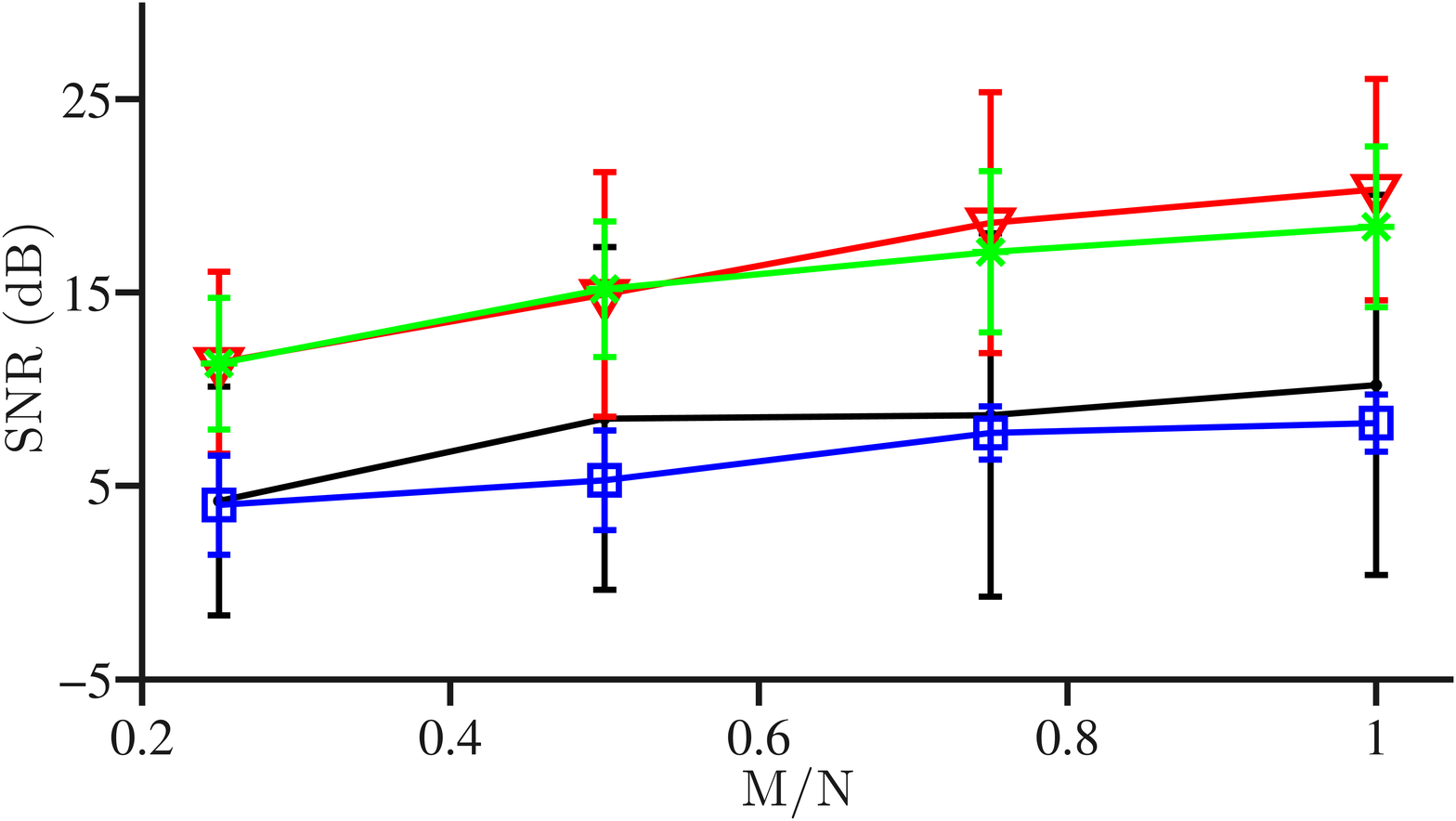}
\caption{Reconstruction quality results for synthetic images of size $N=16^2$ with randomly distributed spikes and ISNR $=$ 30dB for undersampling ratios $M/N$ in the range $[0.25, 1]$. Top panel: $64$ spikes. Bottom panel: $32$ spikes. The curves represent the average SNR values over multiple simulations ($50$ for AM and $10$ for NM) and corresponding 1-standard-deviation error bars.}
\label{fig:nosparse}

\end{figure}

Having validated our NM approach in comparison with alternative state-of-the-art low tensor rank approaches, we compare its performance with that of the AM scheme. Firstly, we evaluate the reconstruction quality on images constructed from $32$ and $64$ randomly located spikes. The AM approach is also evaluated for varying reinitialisation numbers: $n_{\rm ri}\in \{1, 5, 10\}$. The graphs in Figure \ref{fig:nosparse} represent the SNR curves as a function of undersampling in the range $[0.25, 1]$. A total of $50$ and $10$ simulations per point are performed for AM and NM respectively, varying the signal, as well as the sampling and noise realisations. The results show a clear superiority of AM relative to NM in terms of average reconstruction quality. Both approaches exhibit nonnegligible variability. The dependency of the nonconvex AM approach to initialisation is clearly illustrated by the $n_{\rm ri}=1$ and $n_{\rm ri}=5$ curves, confirming the importance of the multiple reinitialisations. We also observe a saturation between $n_{\rm ri}=5$ and $n_{\rm ri}=10$. As expected from asymptotic complexity considerations, AM runs significantly faster than NM, with reconstructions in the order of $10^2$ seconds for $n_{\rm ri}=5$, approximately $10$ times faster than NM.

\begin{figure}

\centering

\raisebox{-.5\height}{\includegraphics[
     clip, keepaspectratio, width = 0.15\textwidth]{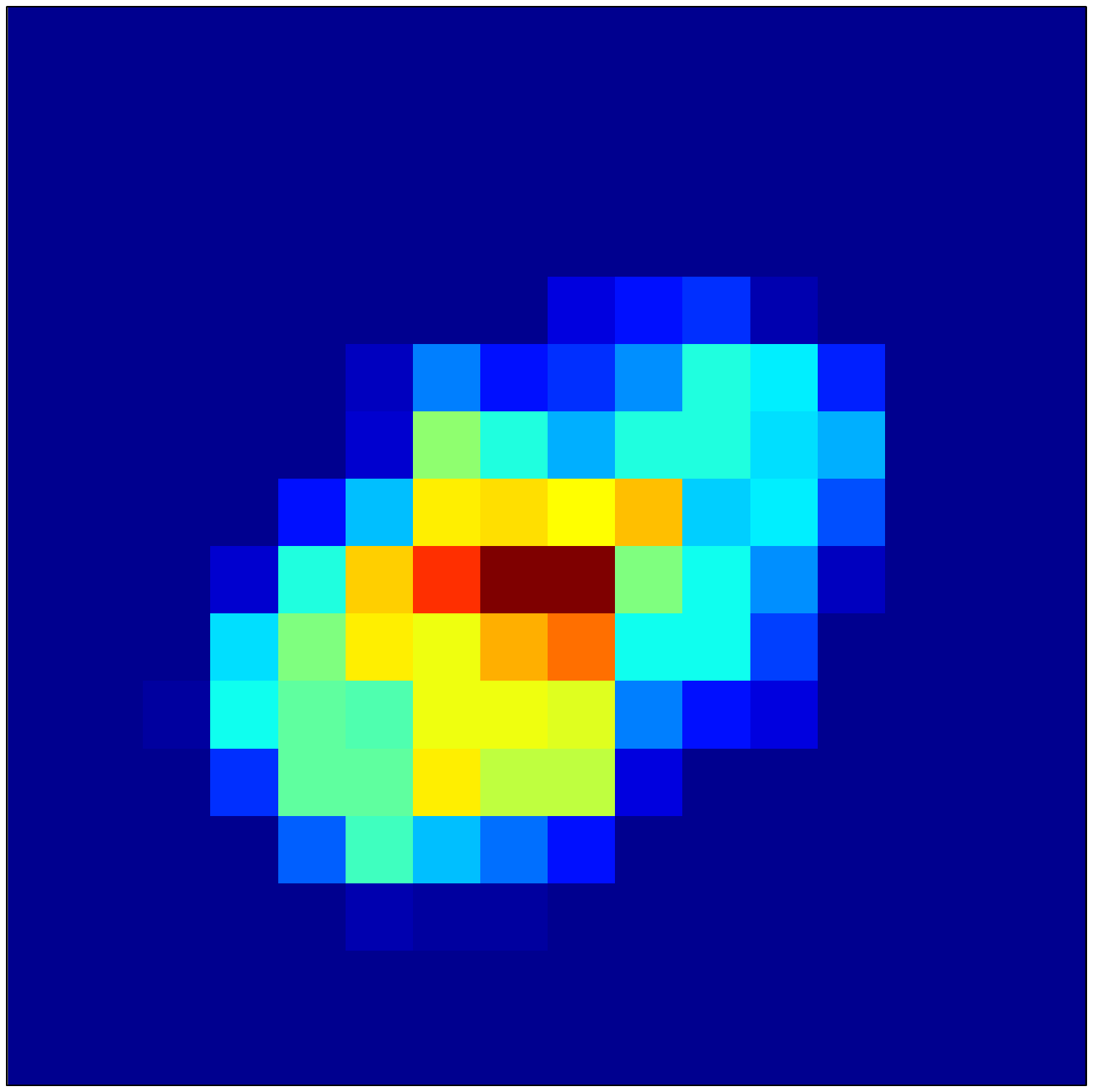}}\label{fig:4a}
\hfill \raisebox{-.5\height}{\includegraphics[
     clip,keepaspectratio, height = 0.16\textwidth]{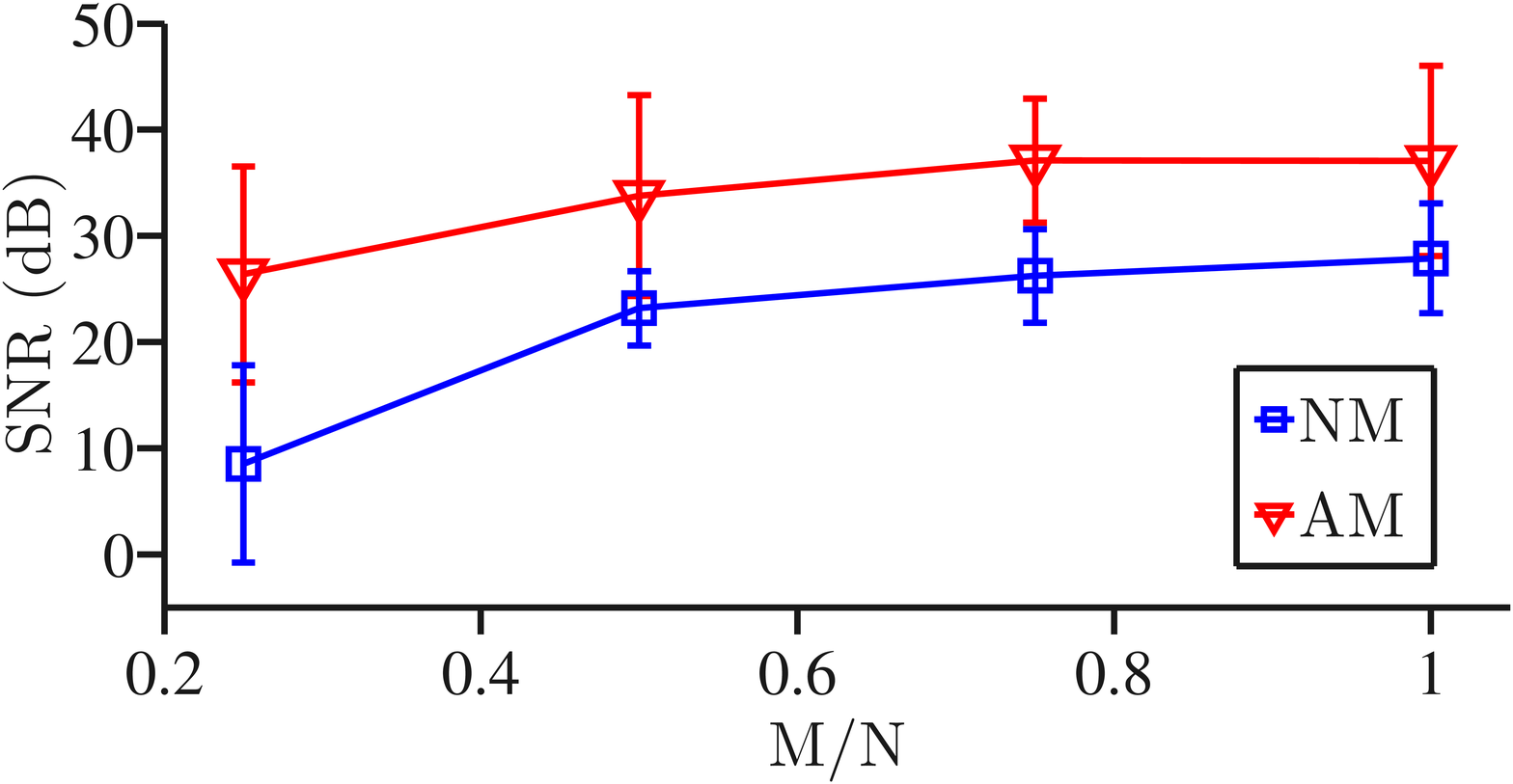}}\label{fig:4b} \\ 
\vspace{2mm}
\includegraphics[
    clip, keepaspectratio, width = 0.15\textwidth]{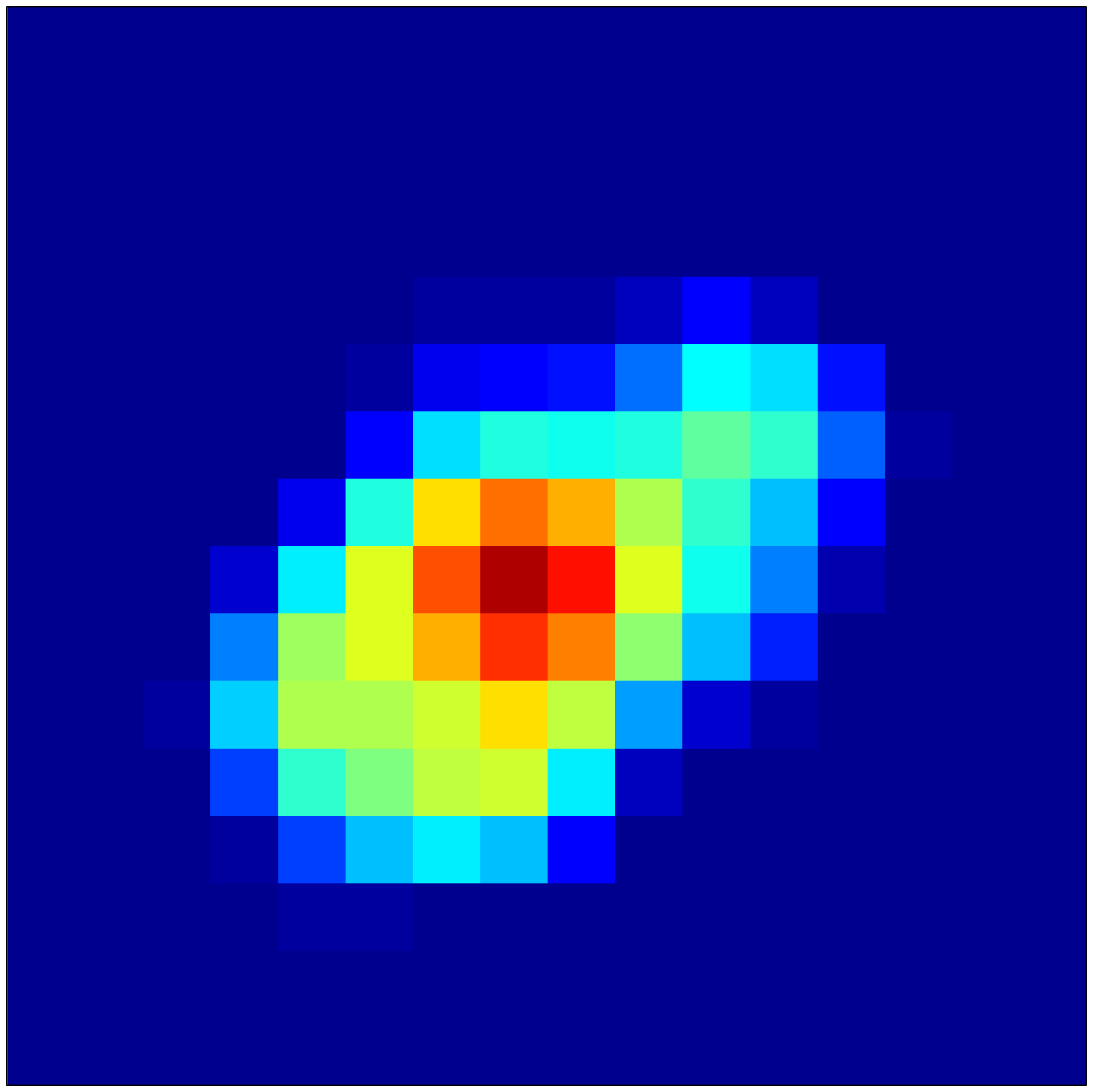}\label{fig:4c} 
\hfill \includegraphics[
     clip, keepaspectratio, width = 0.15\textwidth]{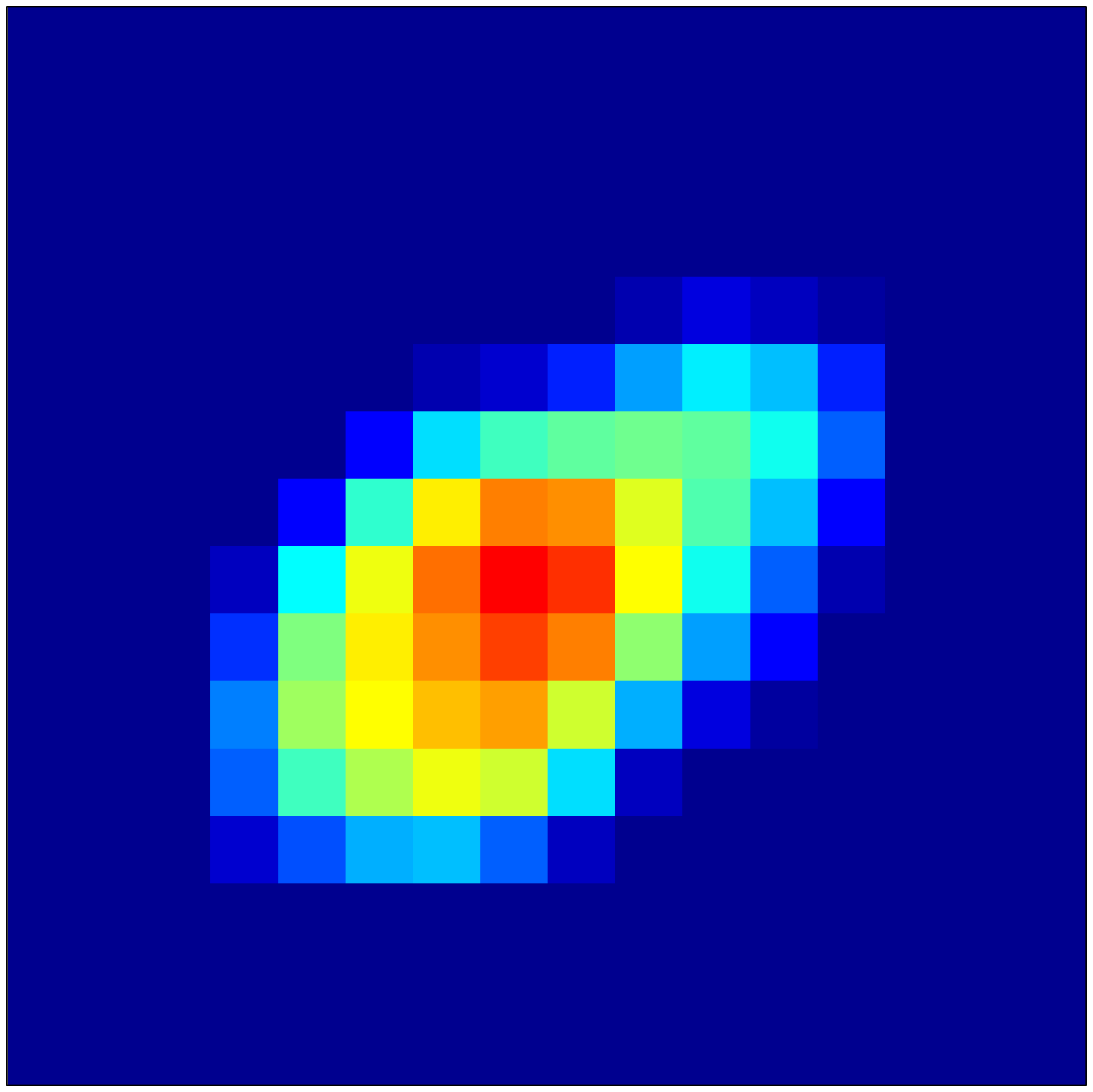}\label{fig:4d}
\hfill \includegraphics[
     clip, keepaspectratio, width = 0.15\textwidth]{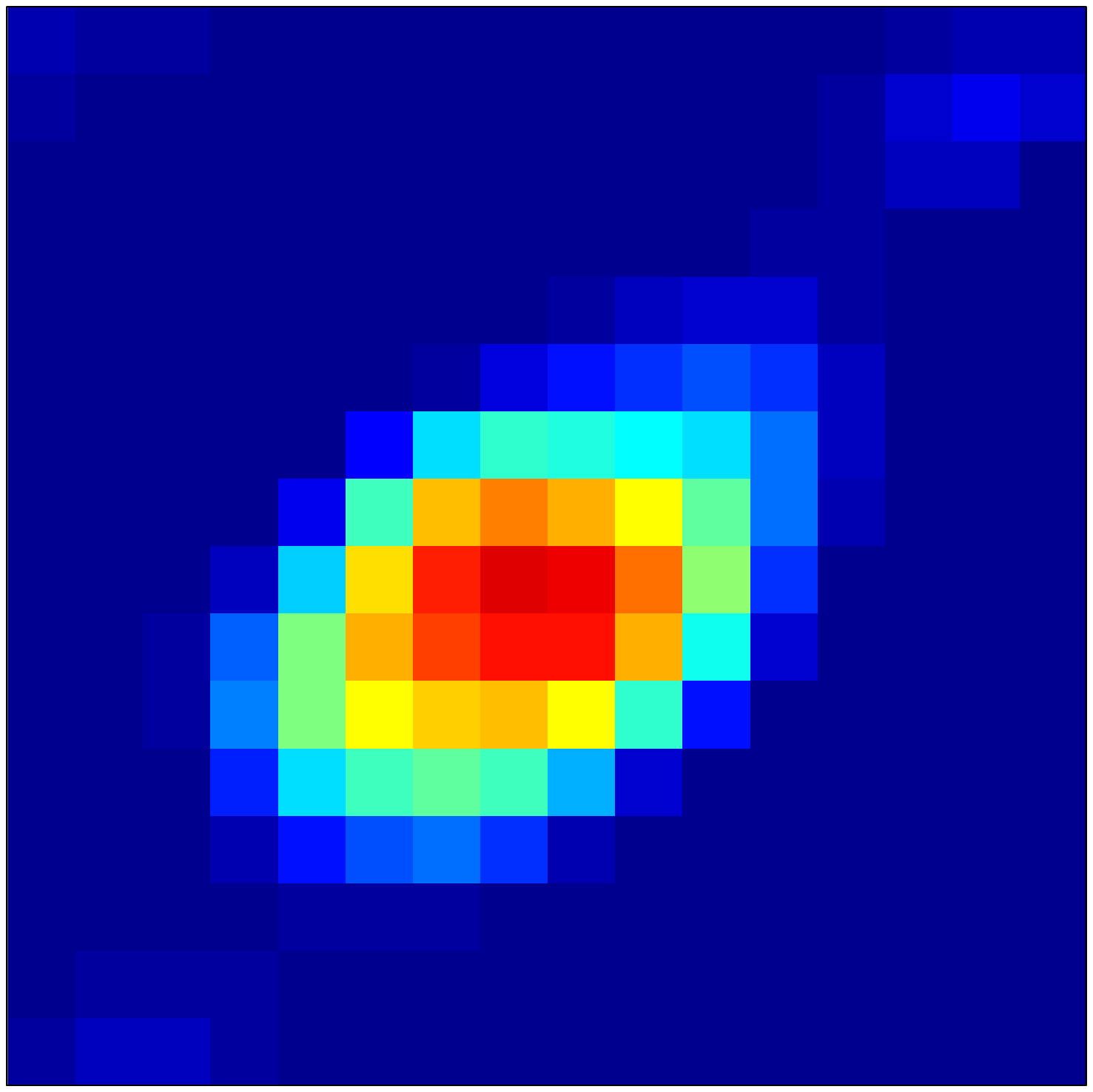}\label{fig:4e} \\
   \vspace{2mm}  
 \includegraphics[
    clip, keepaspectratio, width = 0.15\textwidth]{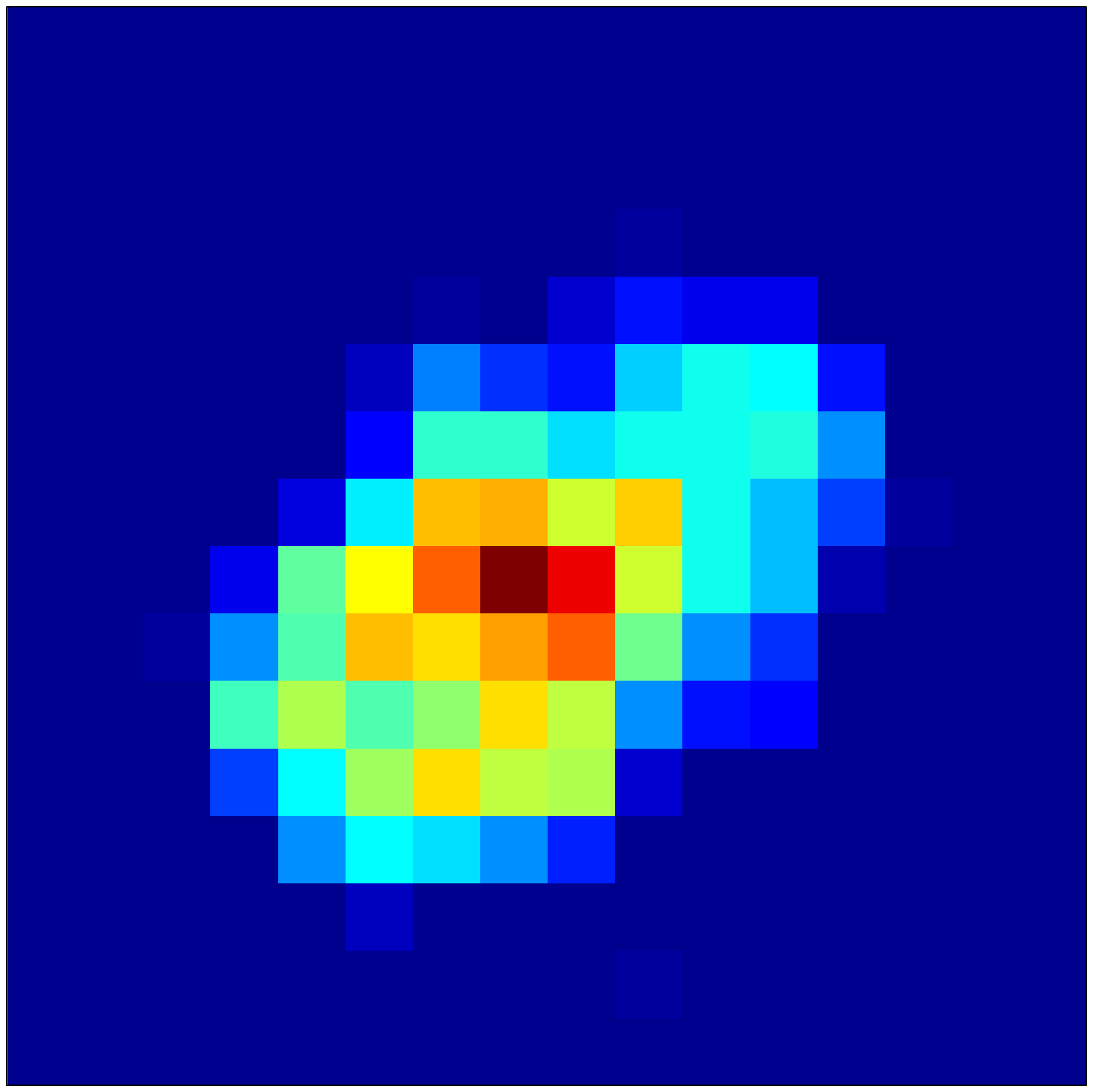}\label{fig:4f} 
\hfill \includegraphics[
     clip, keepaspectratio, width = 0.15\textwidth]{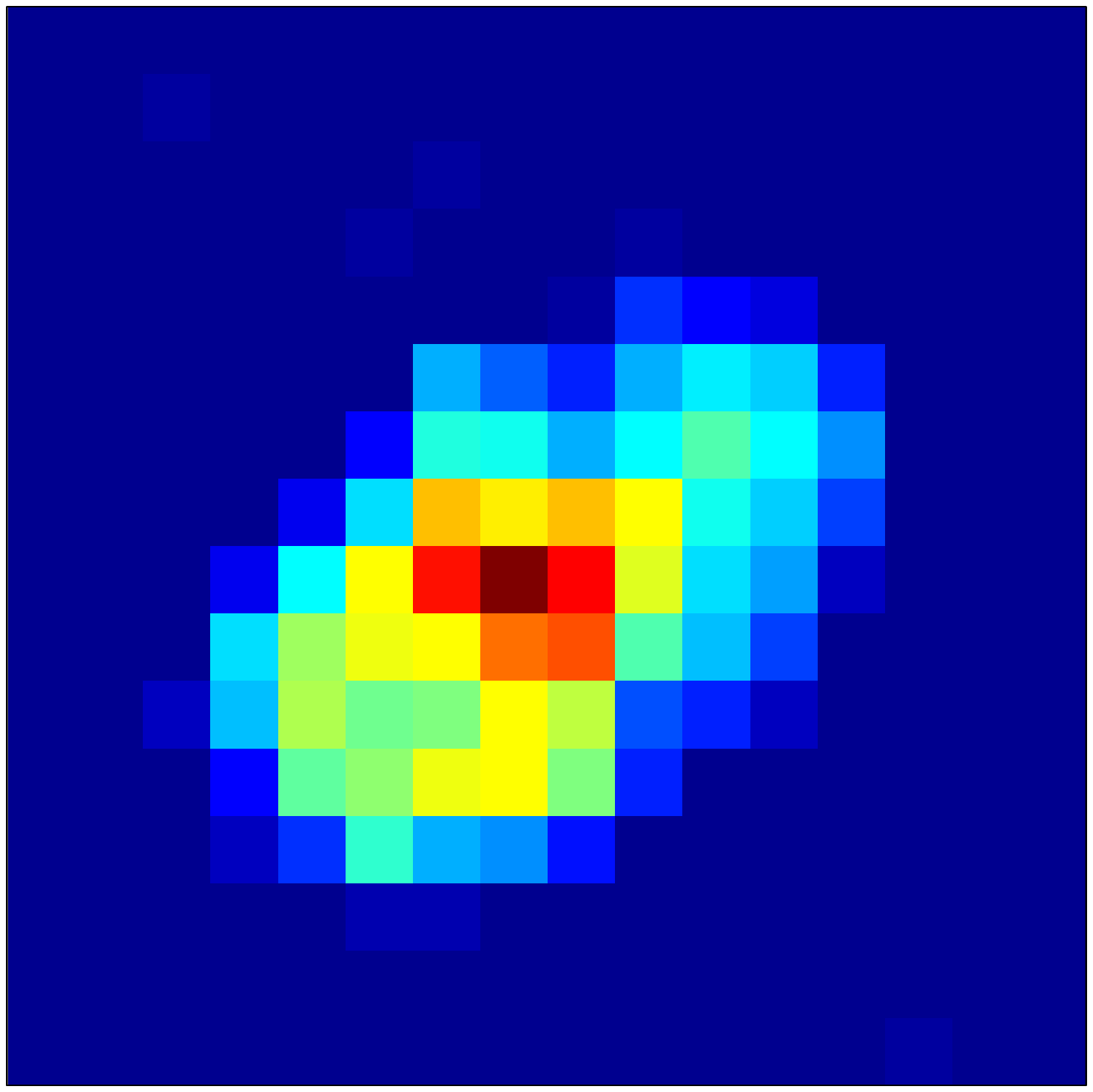}\label{fig:4g}
\hfill \includegraphics[
     clip, keepaspectratio, width = 0.15\textwidth]{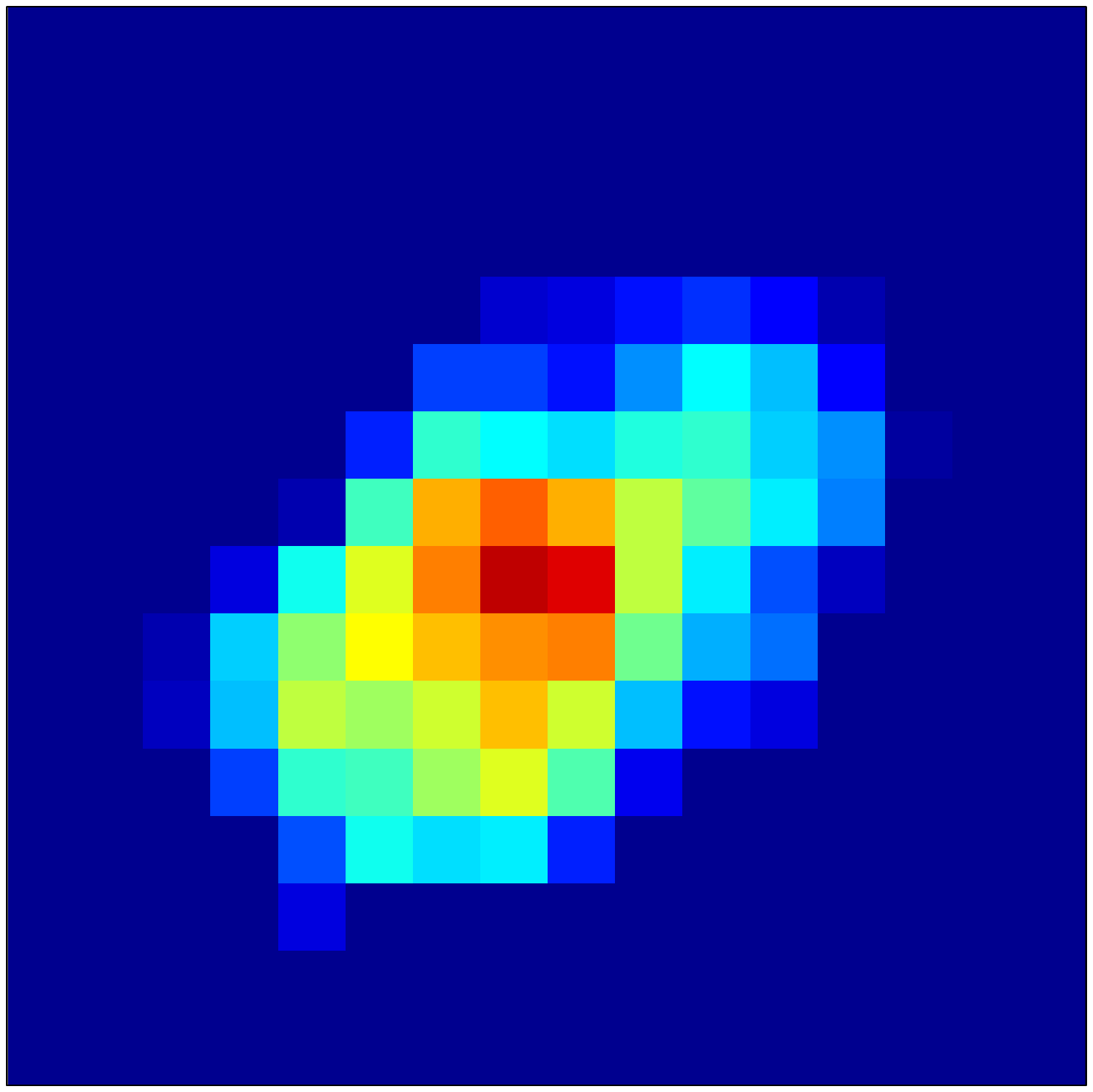}\label{fig:4h} \\
      \vspace{2mm}
     \includegraphics[
    clip, keepaspectratio, width = 0.15\textwidth]{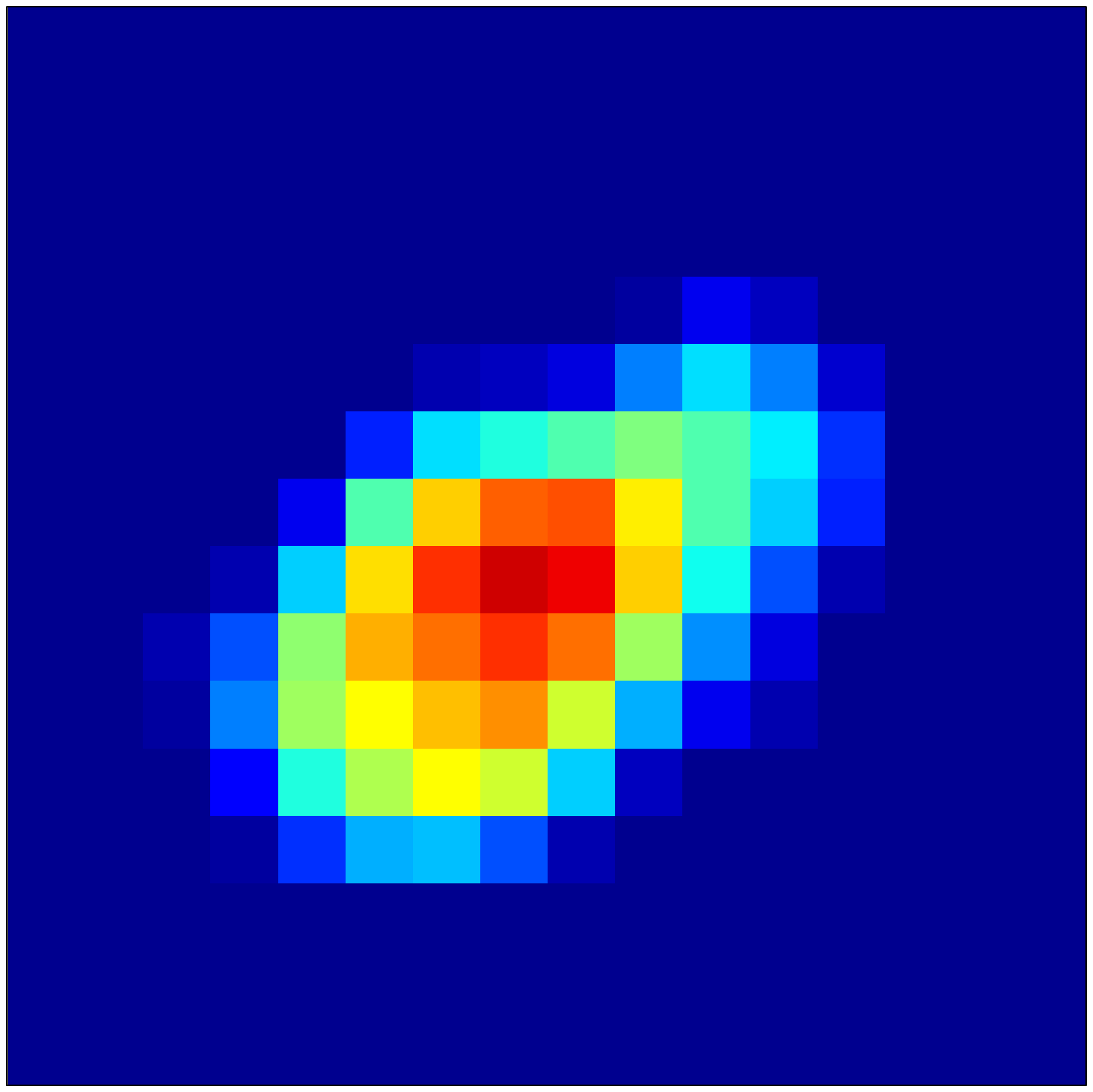}\label{fig:4i} 
\hfill \includegraphics[
     clip, keepaspectratio, width = 0.15\textwidth]{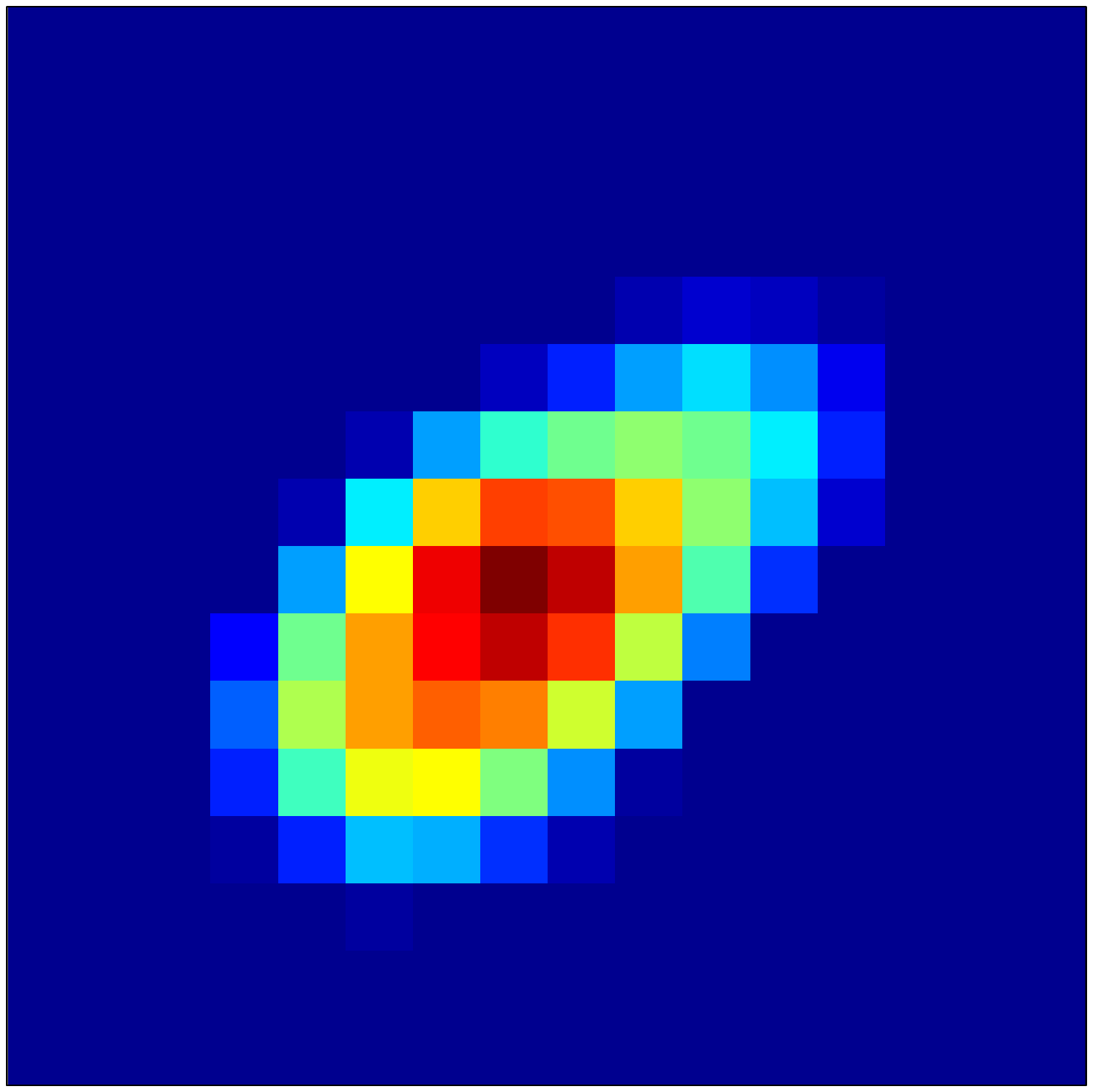}\label{fig:4j}
\hfill \includegraphics[
     clip, keepaspectratio, width = 0.15\textwidth]{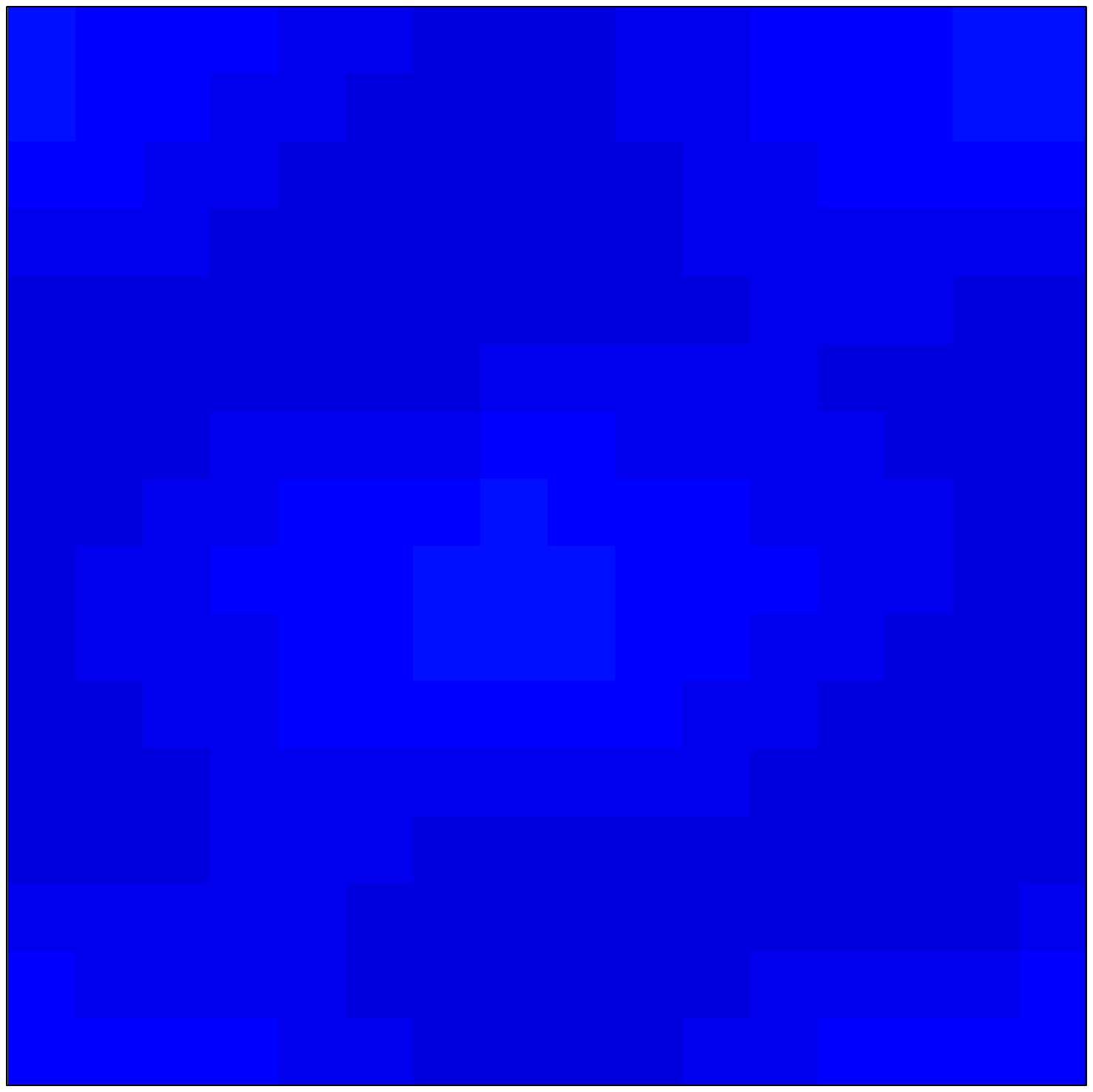}\label{fig:4k} \\
   \vspace{2mm} 
   \includegraphics[
    clip, keepaspectratio, width = 0.15\textwidth]{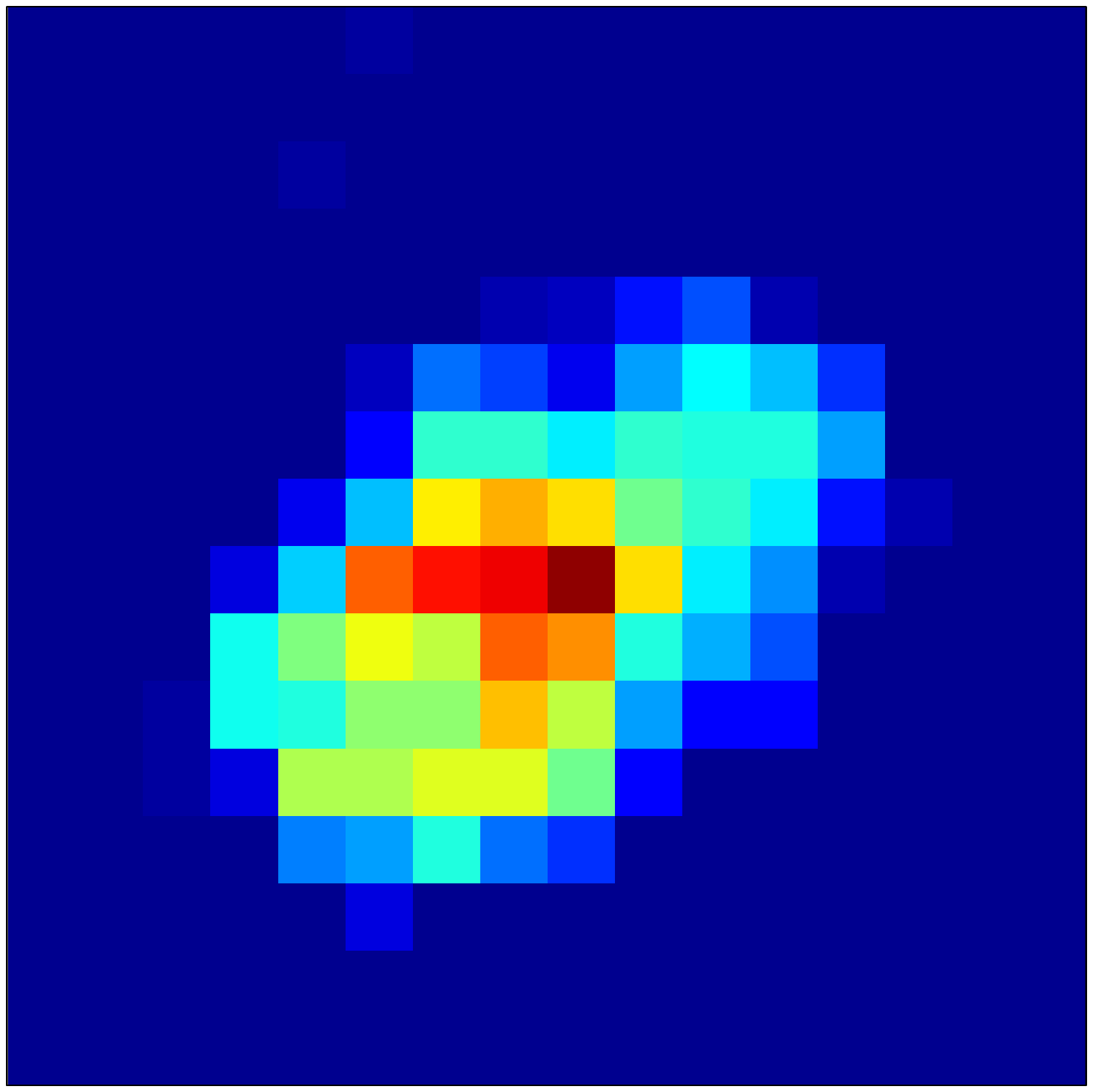}\label{fig:4l} 
\hfill \includegraphics[
     clip, keepaspectratio, width = 0.15\textwidth]{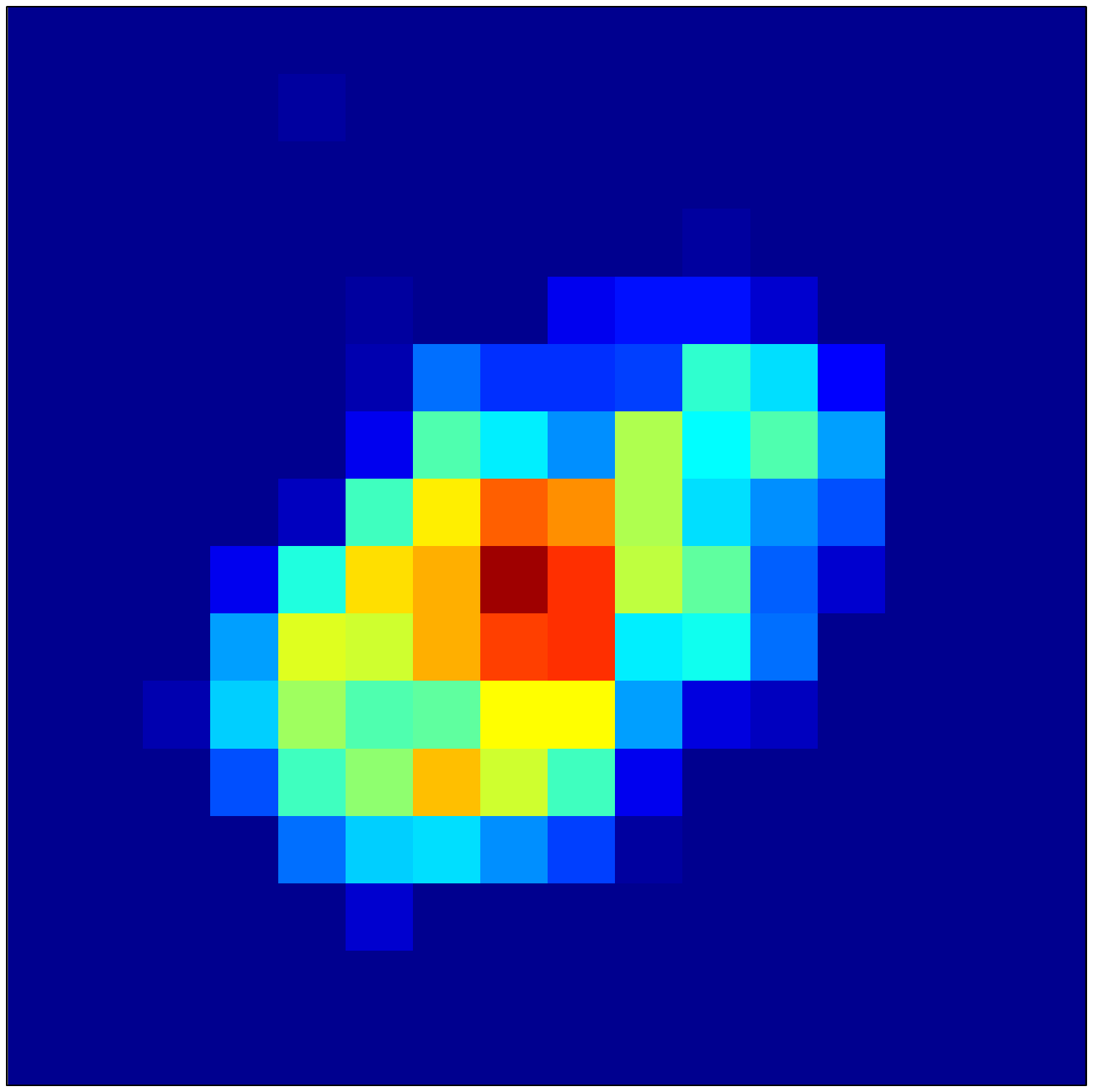}\label{fig:4m}
\hfill \includegraphics[
     clip, keepaspectratio, width = 0.15\textwidth]{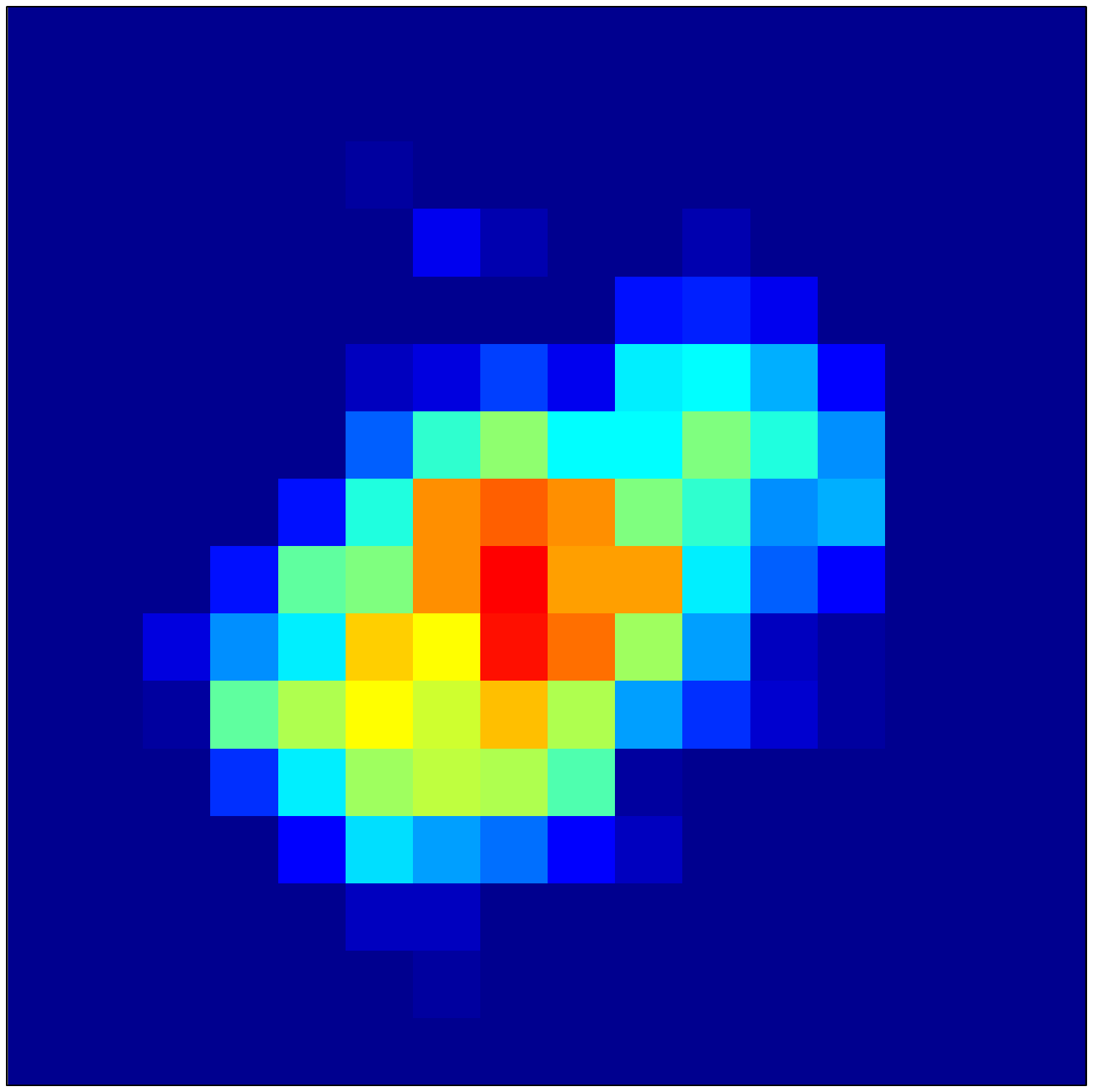}\label{fig:4n} \\

\caption{Eta Carinae star system illustration ($N=16^2$, ISNR $=$ 30dB). Top row: original image and SNR graph. The curves represent the average SNR values over multiple simulations ($50$ for AM and $10$ for NM) and corresponding 1-standard-deviation error bars. Second and third rows: NM (second) and AM for $n_{\rm ri}=5$ (third) reconstructions with best SNR for $M=N$ (left), $M=0.75N$ (centre) and $M=0.25N$ (right). Fourth and bottom rows: NM (fourth) and AM for $n_{\rm ri}=5$ (bottom) reconstructions  with median SNR for $M=N$ (left), $M=0.75N$ (centre) and $M=0.25N$ (right).}
\label{fig:eta}
\end{figure}

\begin{figure}

\centering

\raisebox{-.5\height}{\includegraphics[
     clip, keepaspectratio, width = 0.15\textwidth]{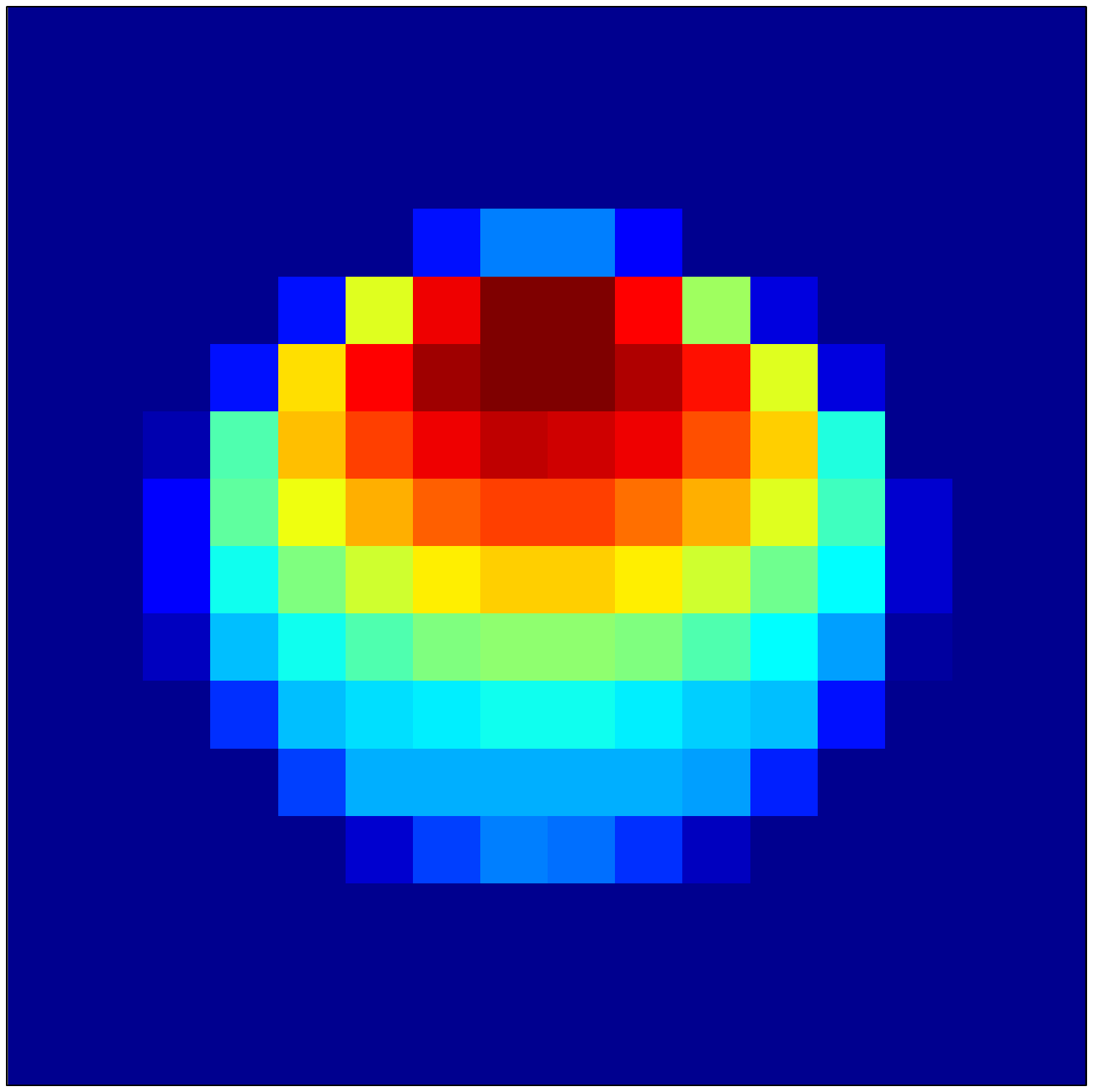}}\label{fig:5a}     
\hfill \raisebox{-.5\height}{\includegraphics[
     clip,keepaspectratio, height = 0.16\textwidth]{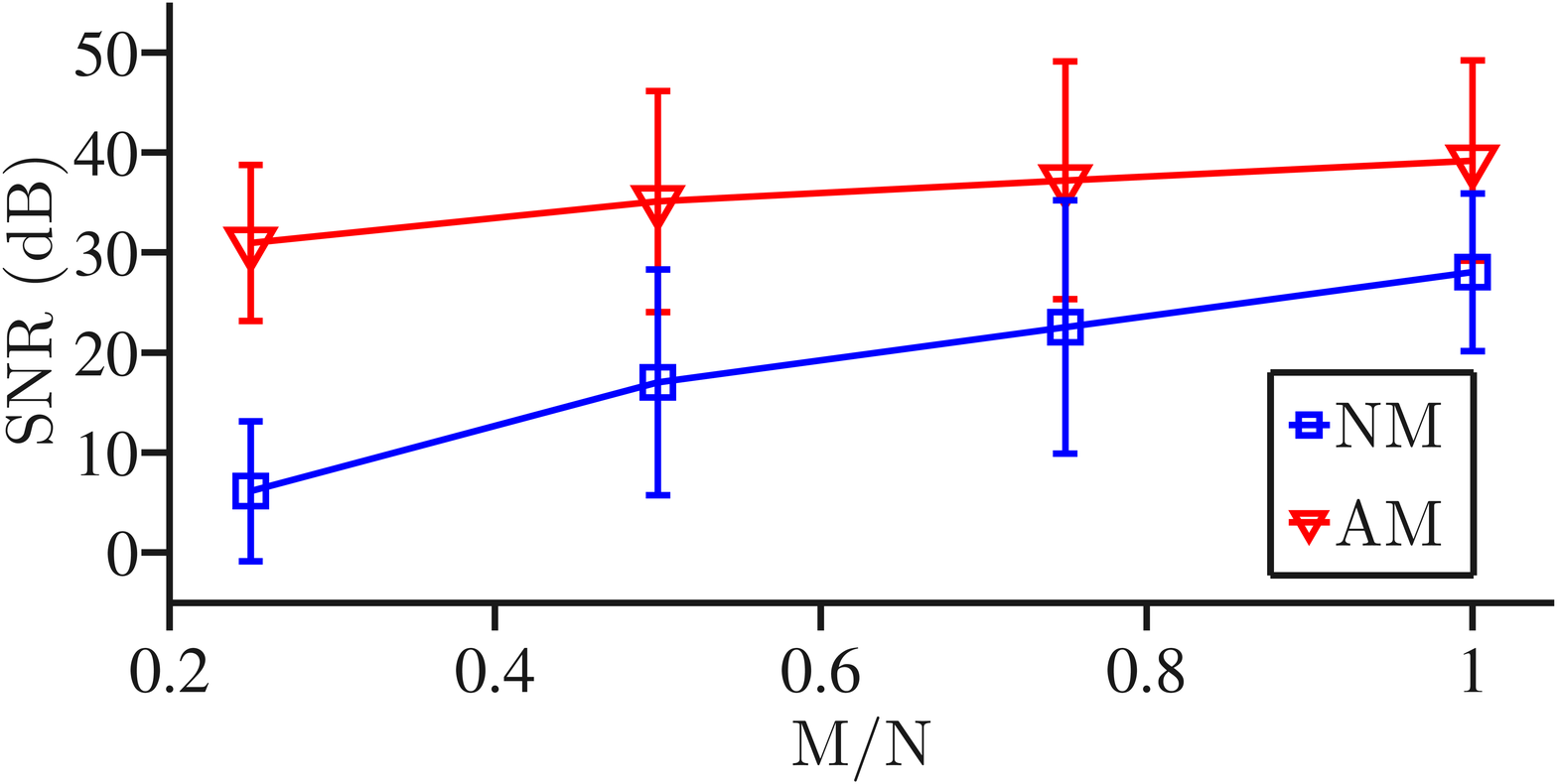}}\label{fig:3b} \\ 
\vspace{2mm}
\includegraphics[
    clip, keepaspectratio, width = 0.15\textwidth]{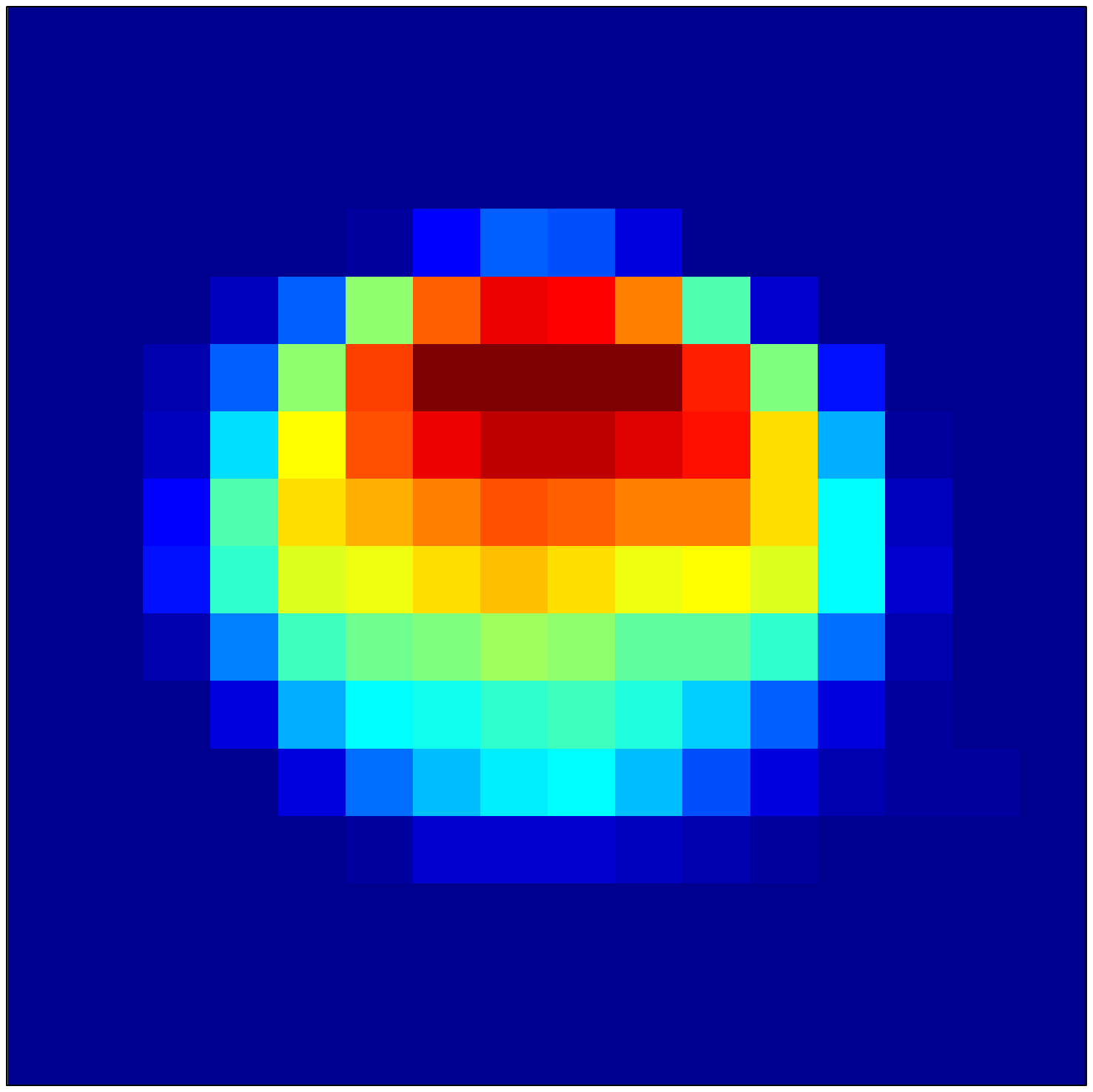}\label{fig:3b} 
\hfill \includegraphics[
     clip, keepaspectratio, width = 0.15\textwidth]{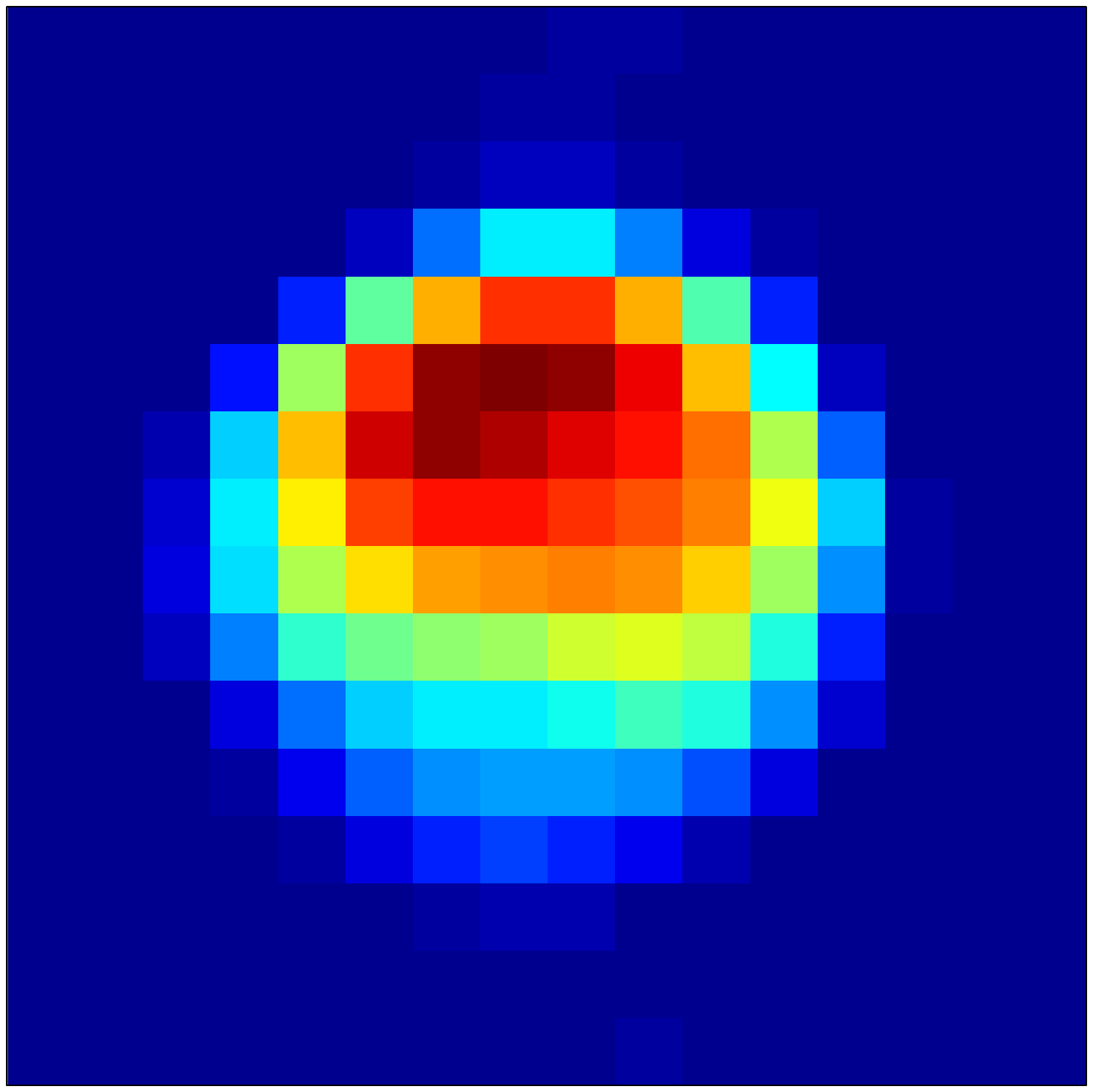}\label{fig:3d}
\hfill \includegraphics[
     clip, keepaspectratio, width = 0.15\textwidth]{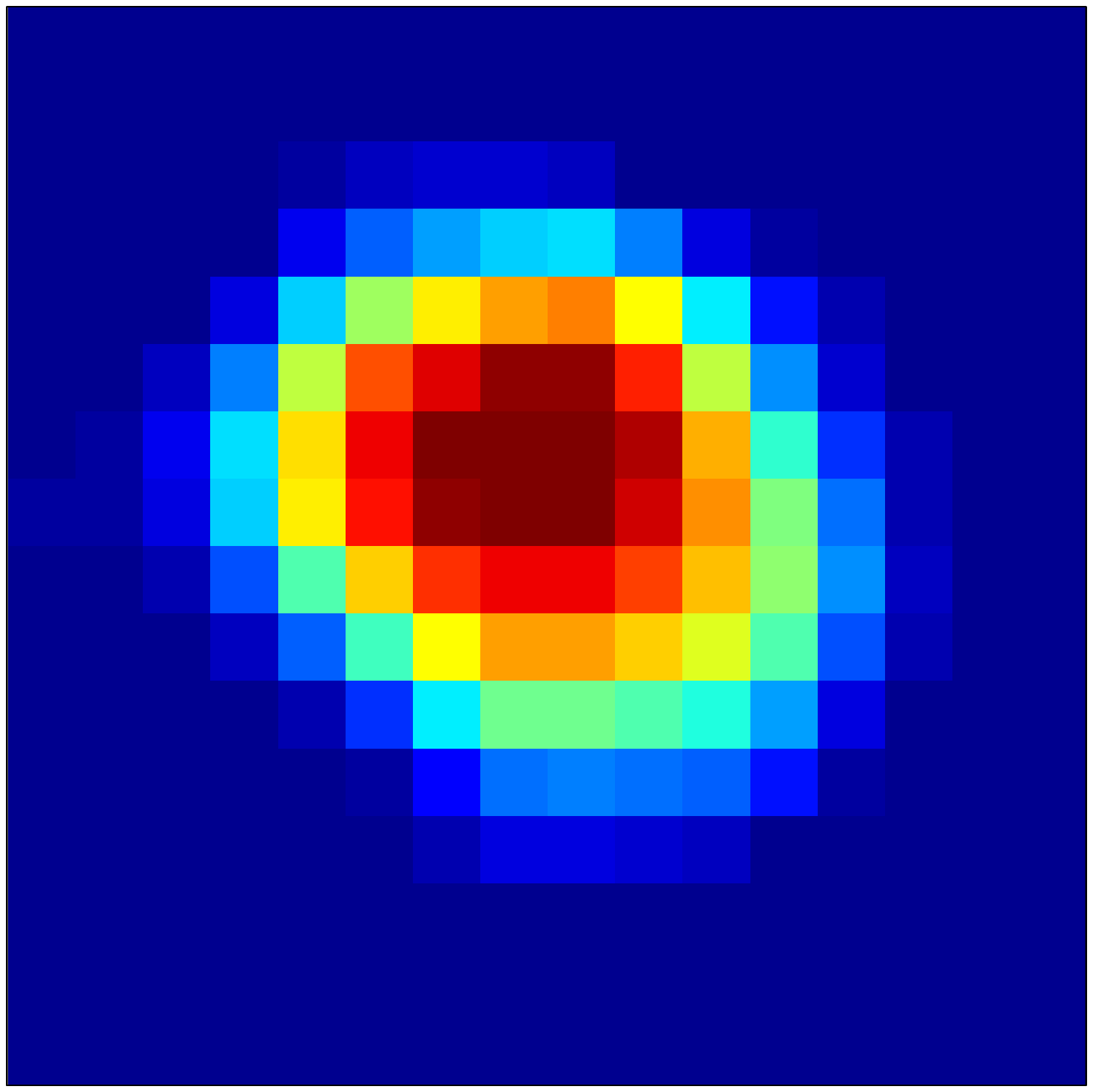}\label{fig:3e} \\
   \vspace{2mm}  
 \includegraphics[
    clip, keepaspectratio, width = 0.15\textwidth]{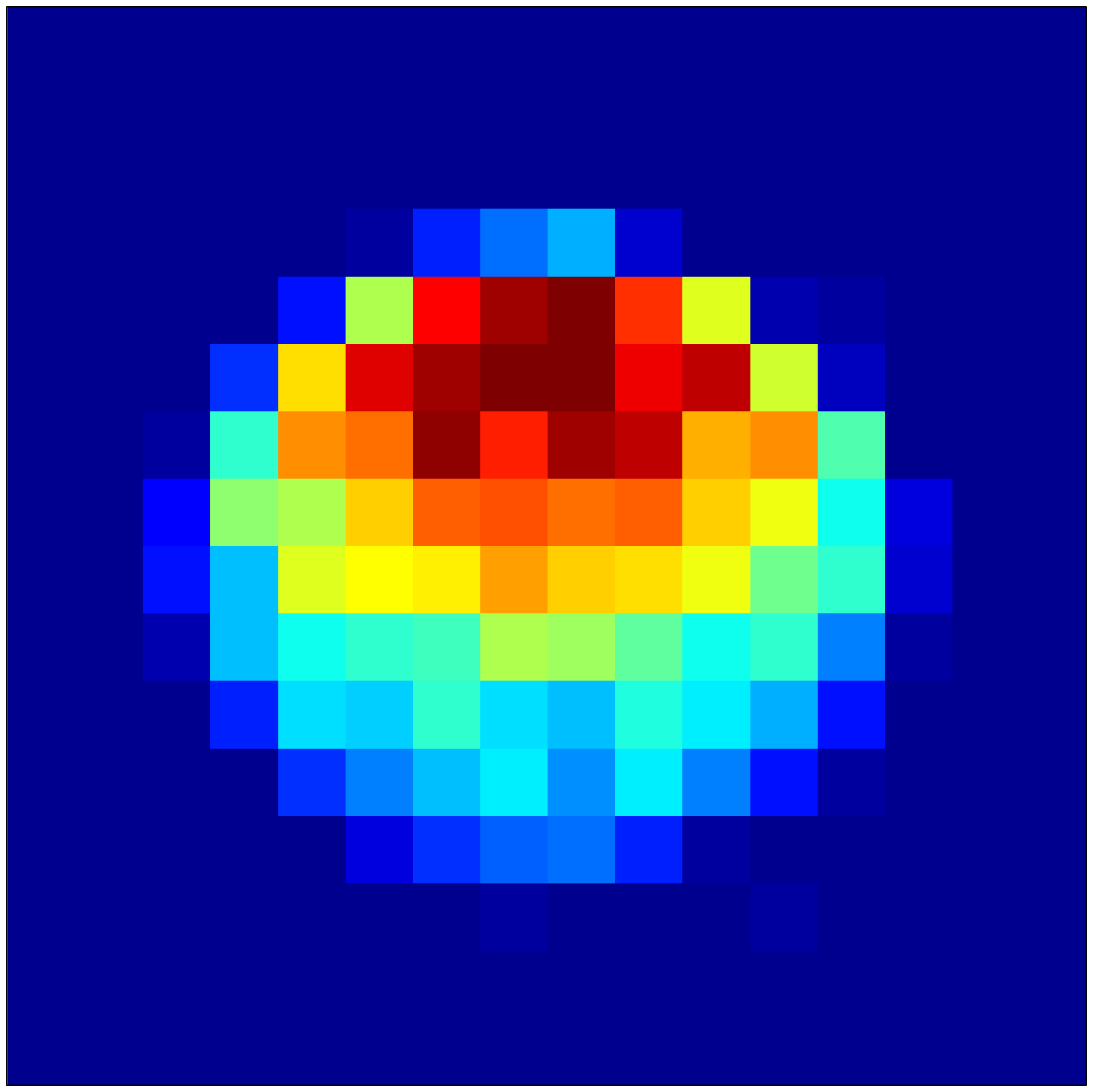}\label{fig:3b} 
\hfill \includegraphics[
     clip, keepaspectratio, width = 0.15\textwidth]{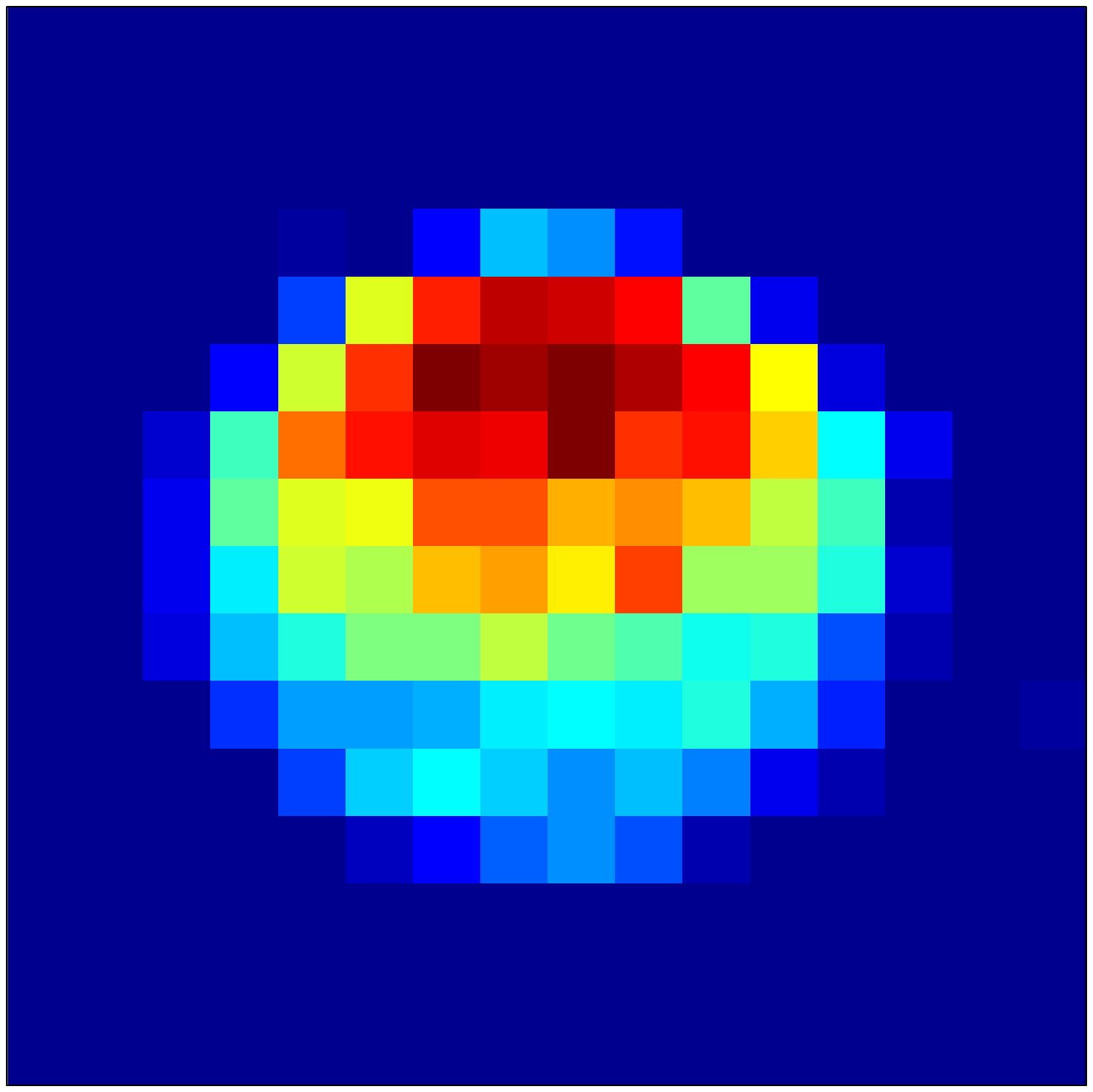}\label{fig:3d}
\hfill \includegraphics[
     clip, keepaspectratio, width = 0.15\textwidth]{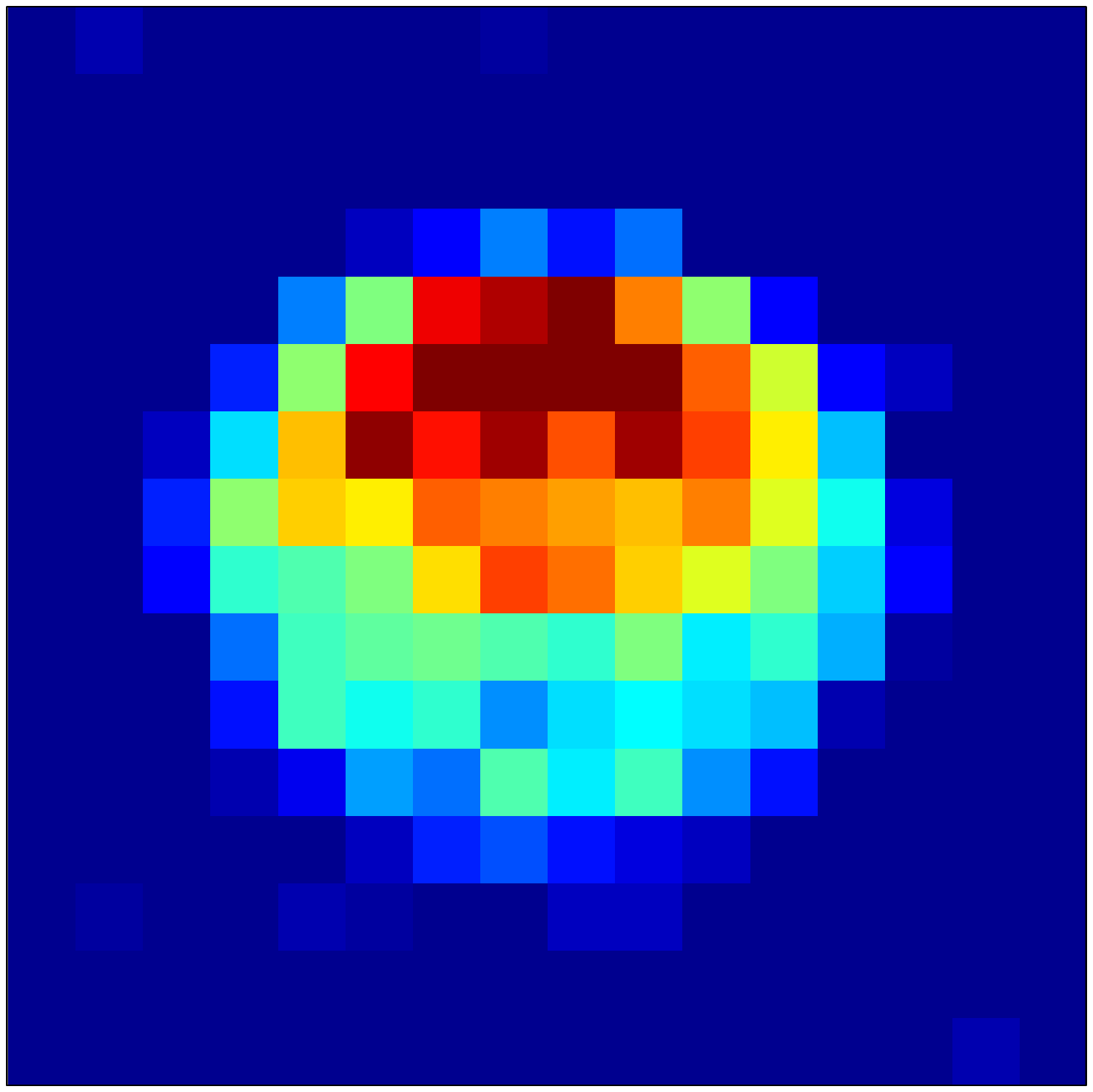}\label{fig:3e} \\
      \vspace{2mm}
     \includegraphics[
    clip, keepaspectratio, width = 0.15\textwidth]{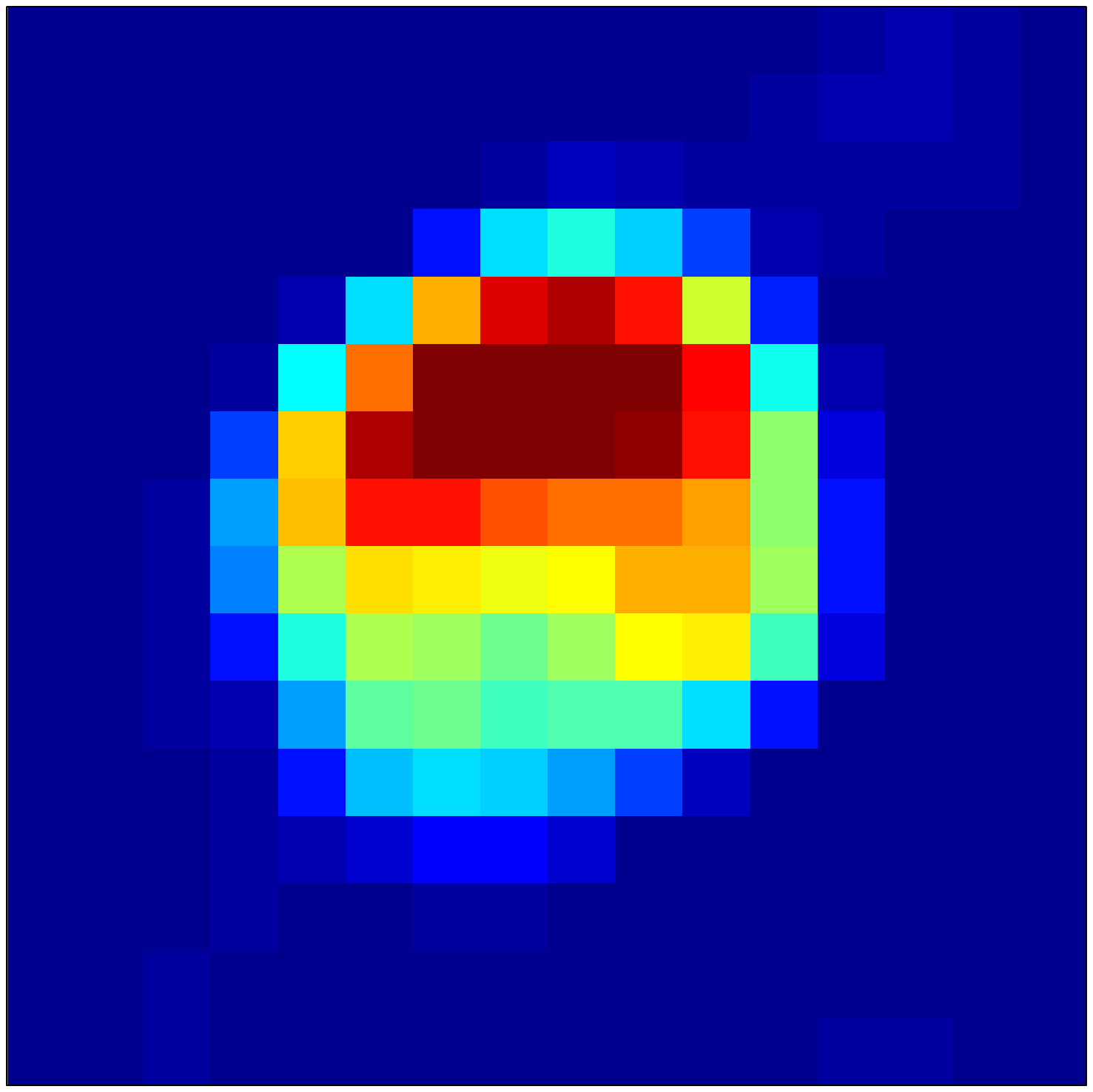}\label{fig:3b} 
\hfill \includegraphics[
     clip, keepaspectratio, width = 0.15\textwidth]{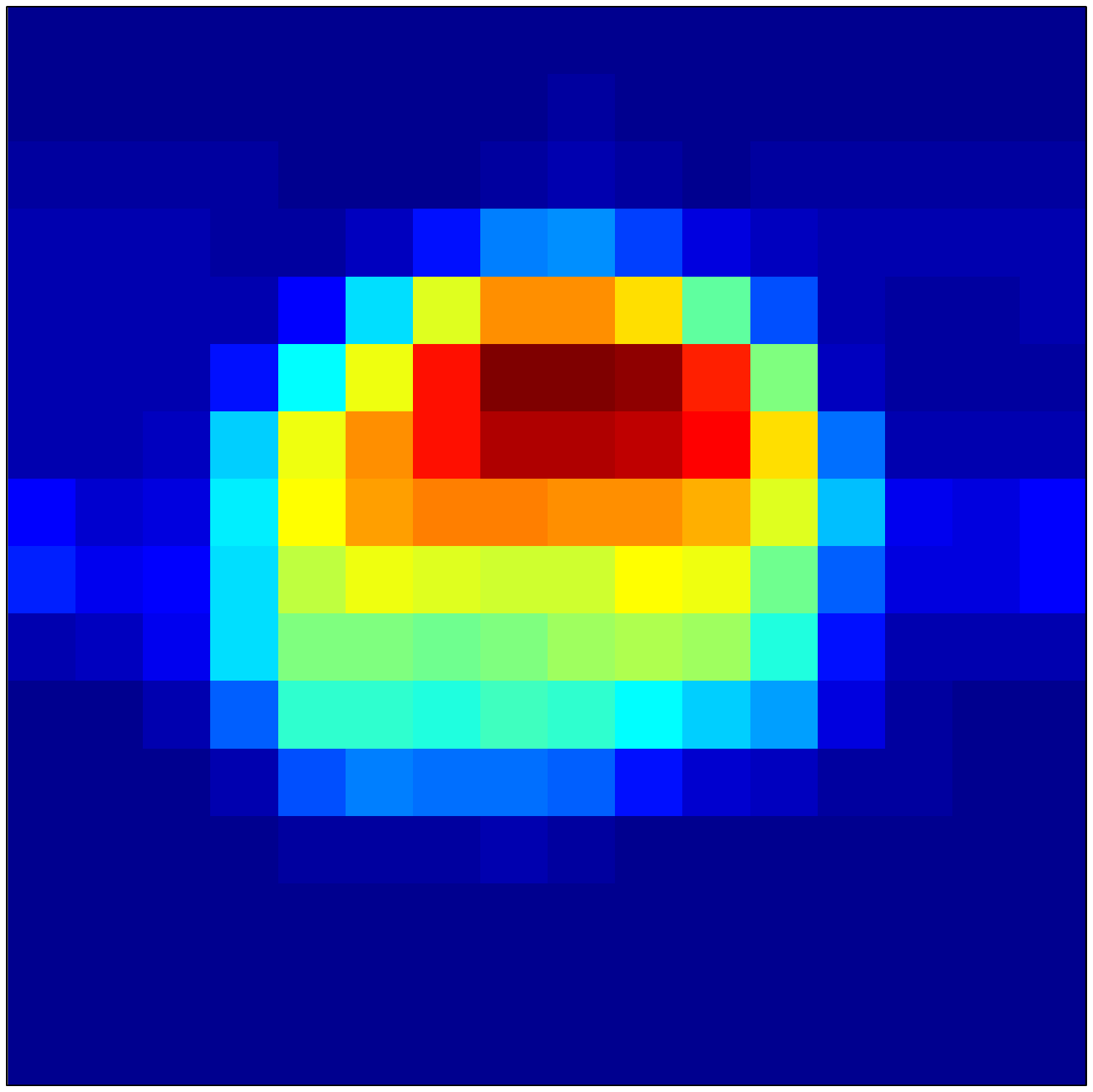}\label{fig:3d}
\hfill \includegraphics[
     clip, keepaspectratio, width = 0.15\textwidth]{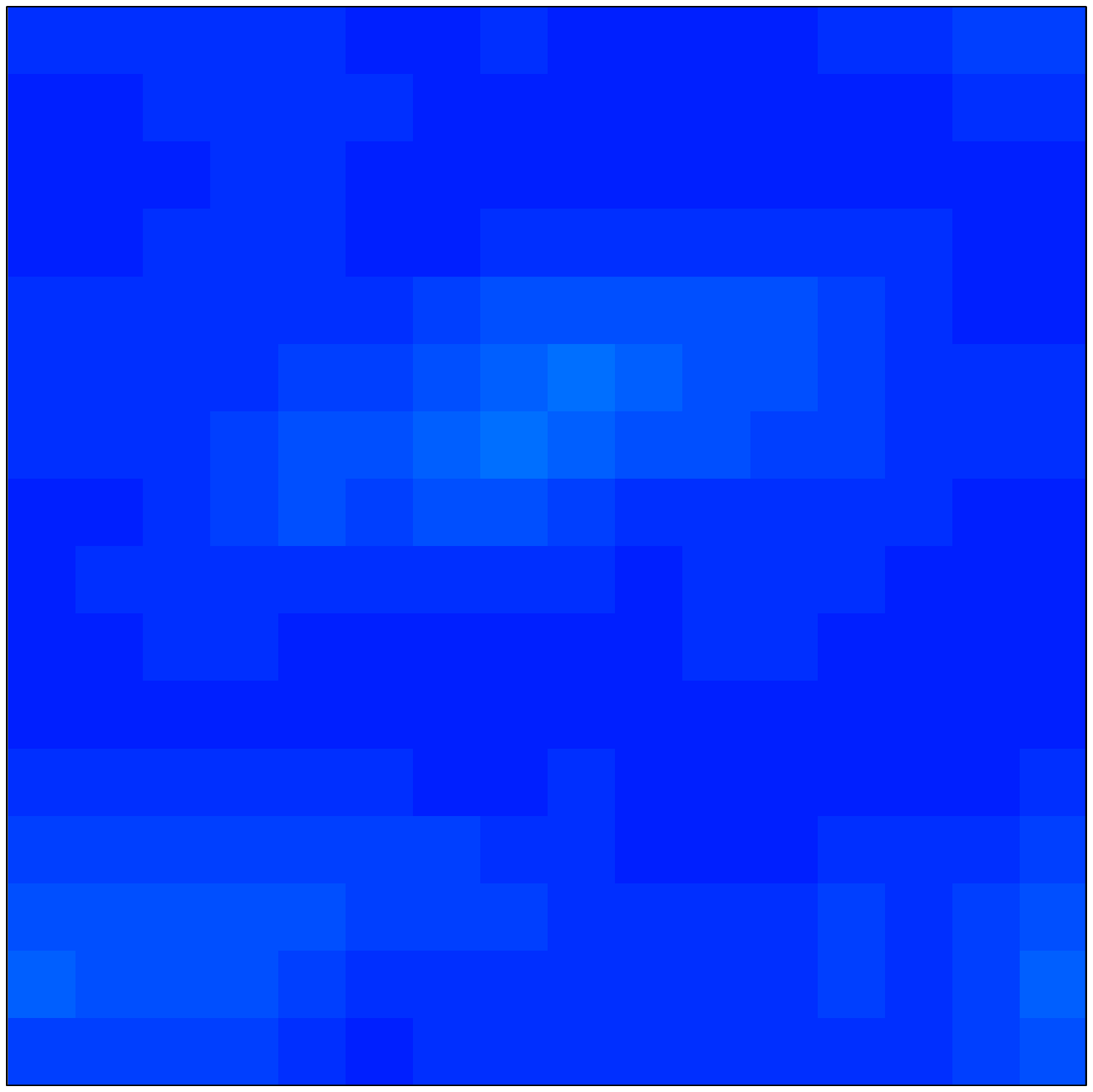}\label{fig:3e} \\
   \vspace{2mm} 
   \includegraphics[
    clip, keepaspectratio, width = 0.15\textwidth]{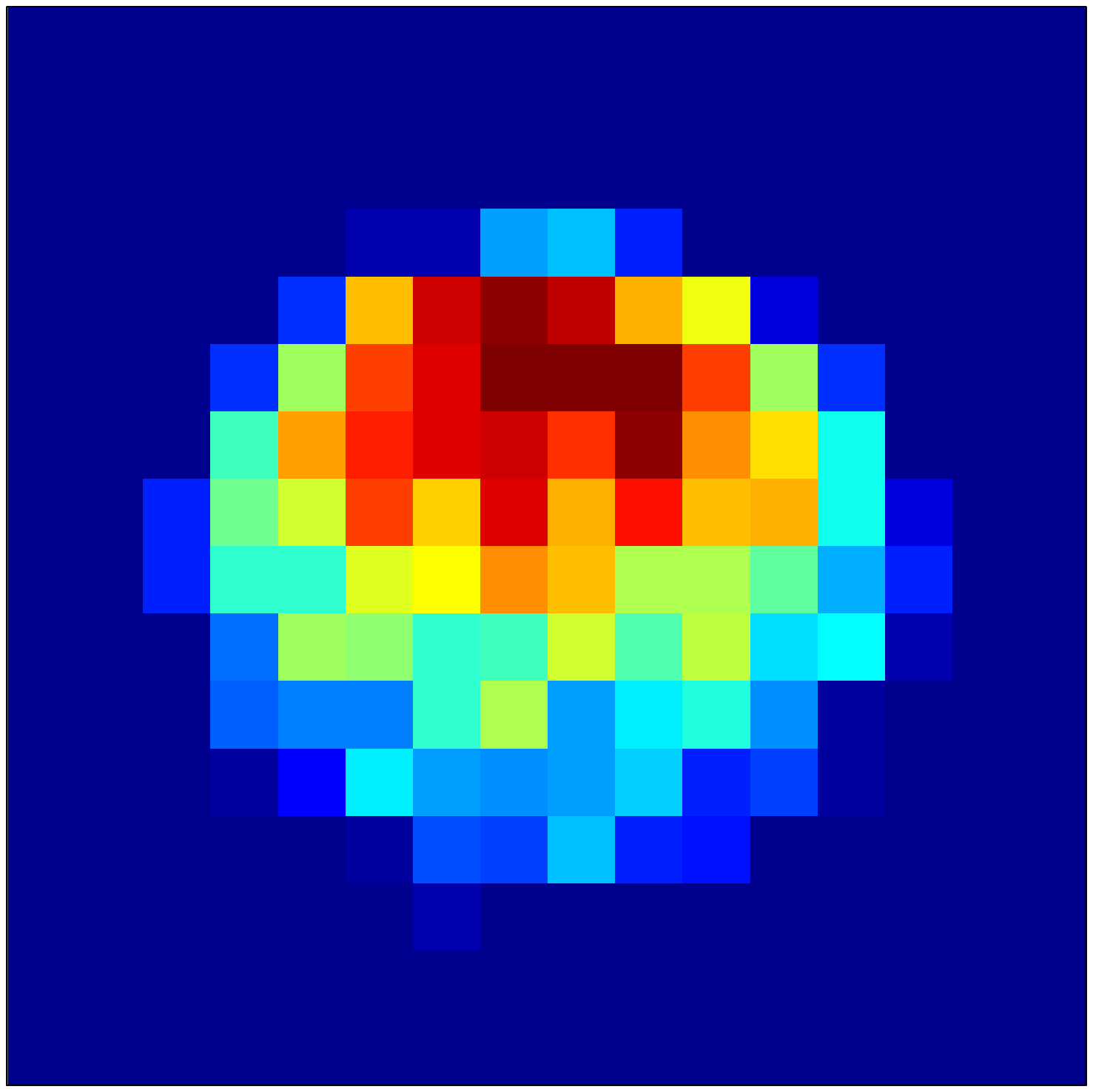}\label{fig:3b} 
\hfill \includegraphics[
     clip, keepaspectratio, width = 0.15\textwidth]{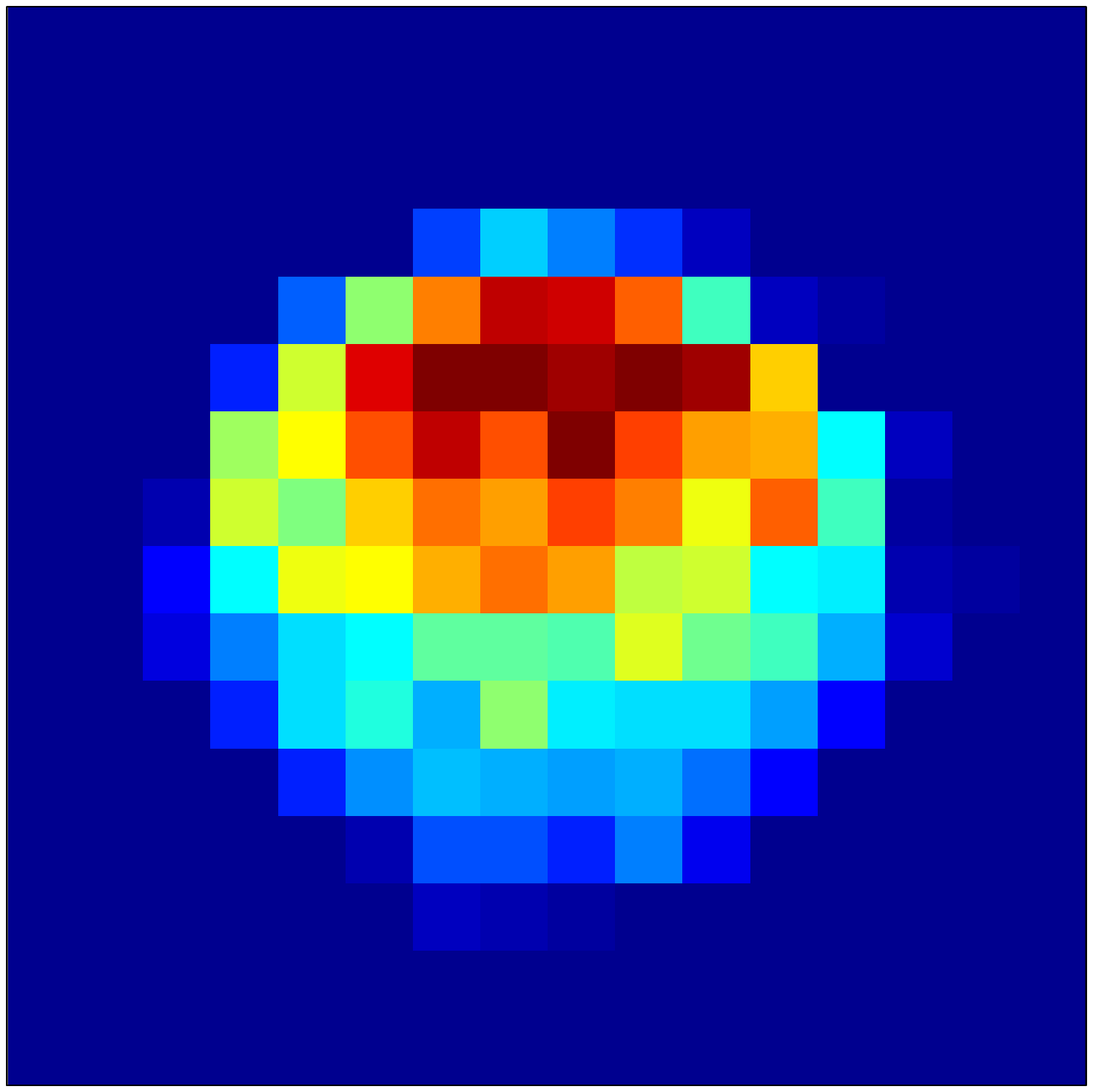}\label{fig:3d}
\hfill \includegraphics[
     clip, keepaspectratio, width = 0.15\textwidth]{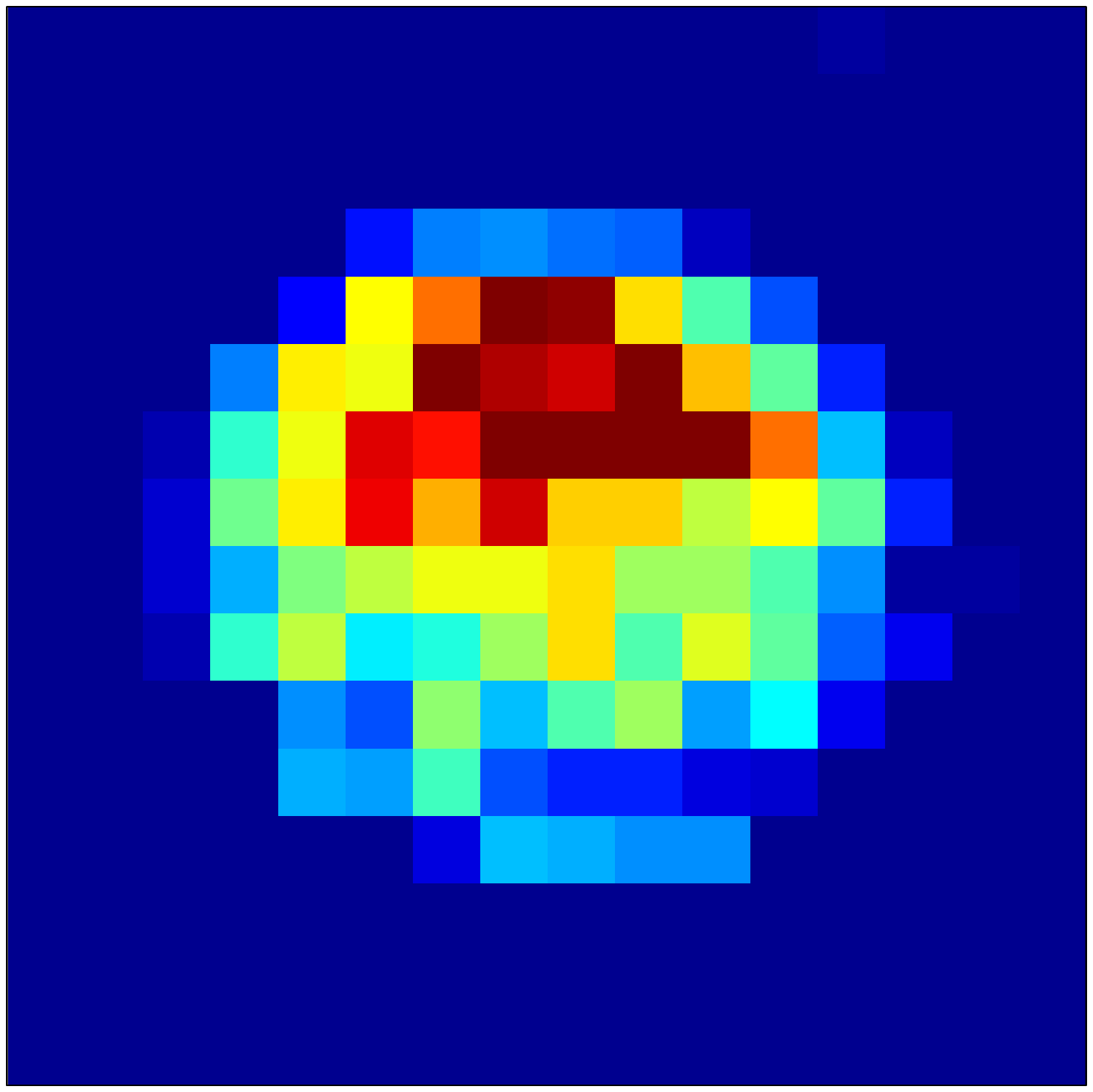}\label{fig:3e} \\

\caption{Rapidly rotating star illustration ($N=16^2$, ISNR $=$ 30dB). Top row: original image and SNR graph. The curves represent the average SNR values over multiple simulations ($50$ for AM and $10$ for NM) and corresponding 1-standard-deviation error bars. Second and third rows: NM (second) and AM for $n_{\rm ri}=5$ (third) reconstructions with best SNR for $M=N$ (left), $M=0.75N$ (centre) and $M=0.25N$ (right). Fourth and bottom rows: NM (fourth) and AM for $n_{\rm ri}=5$ (bottom) reconstructions  with median SNR for $M=N$ (left), $M=0.75N$ (centre) and $M=0.25N$ (right).}
\label{fig:rapid}
\end{figure}

\begin{figure}
\centering

\raisebox{-.5\height}{\includegraphics[
     clip, keepaspectratio, width = 0.15\textwidth]{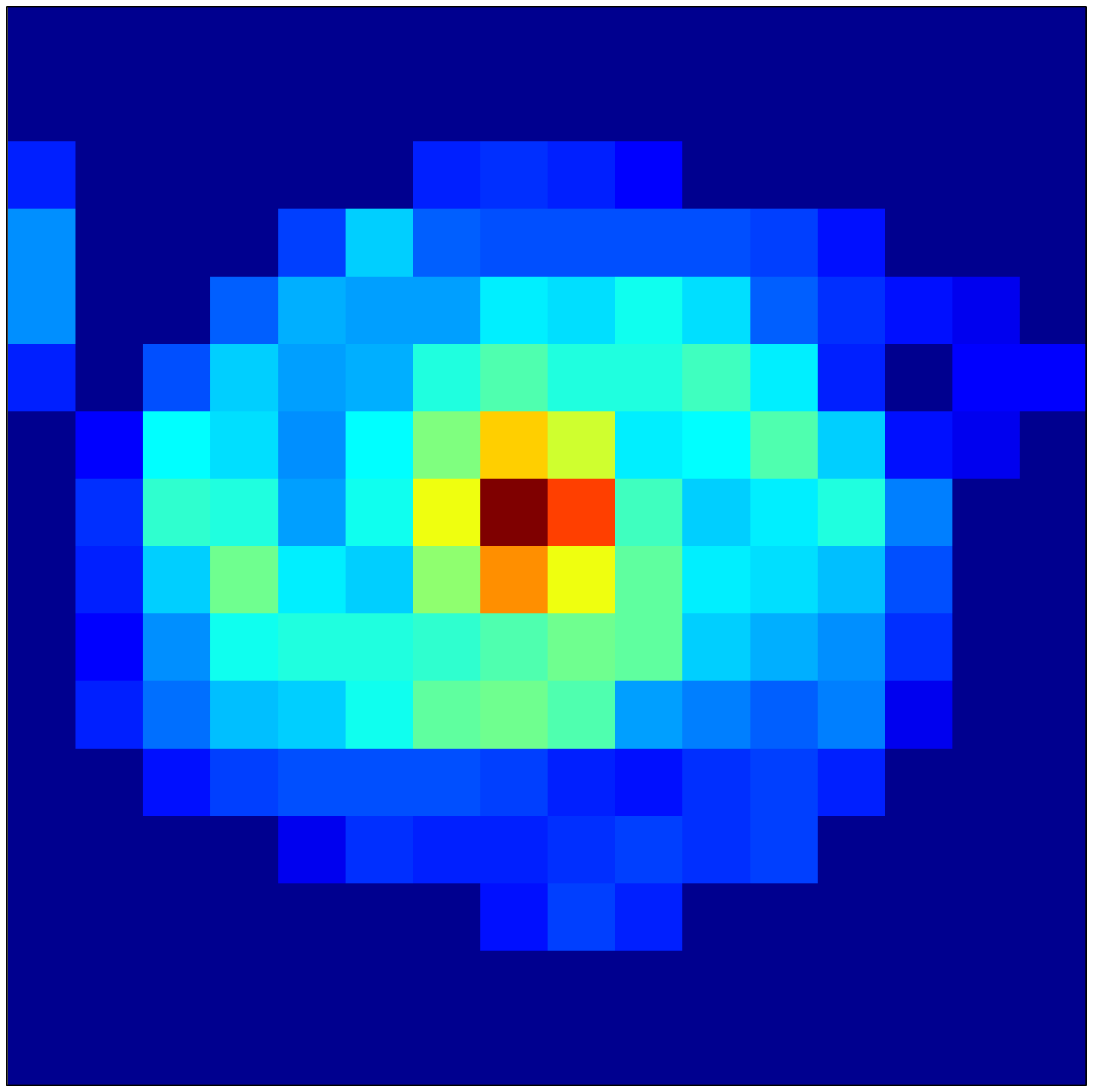}}\label{fig:6a}
\hfill \raisebox{-.5\height}{\includegraphics[
     clip,keepaspectratio, height = 0.16\textwidth]{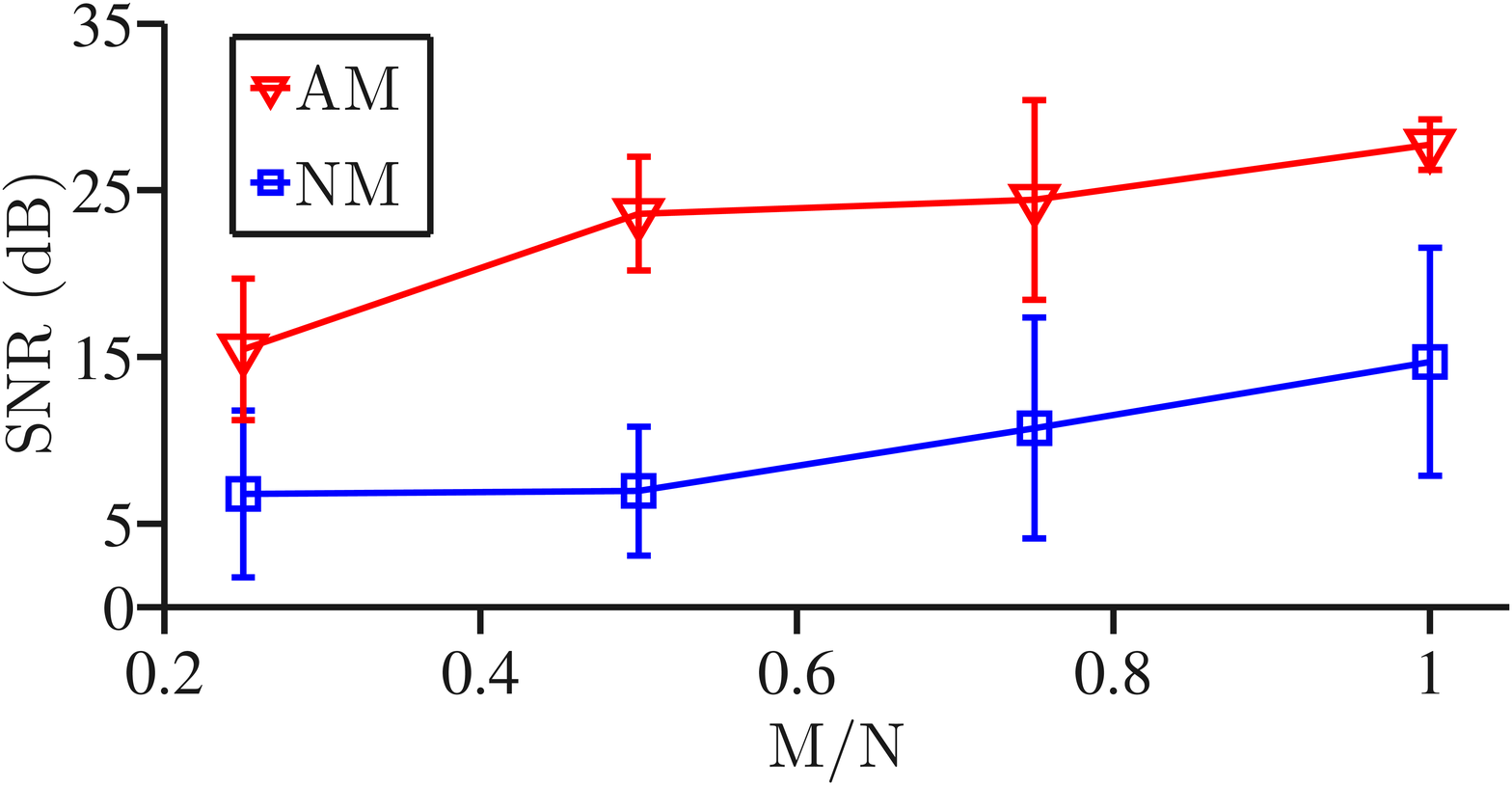}}\label{fig:3b} \\ 
\vspace{2mm}
\includegraphics[
    clip, keepaspectratio, width = 0.15\textwidth]{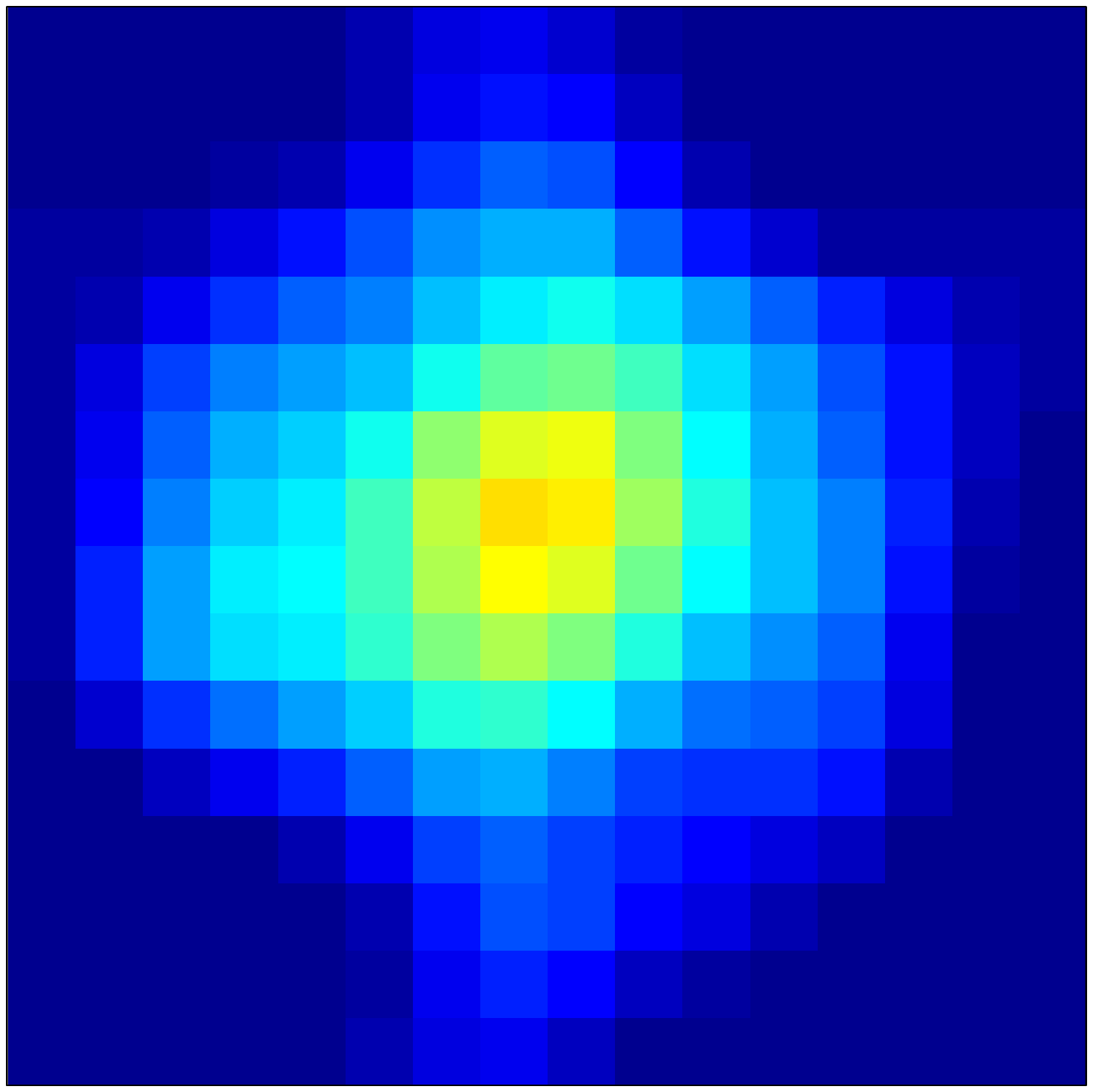}\label{fig:3b} 
\hfill \includegraphics[
     clip, keepaspectratio, width = 0.15\textwidth]{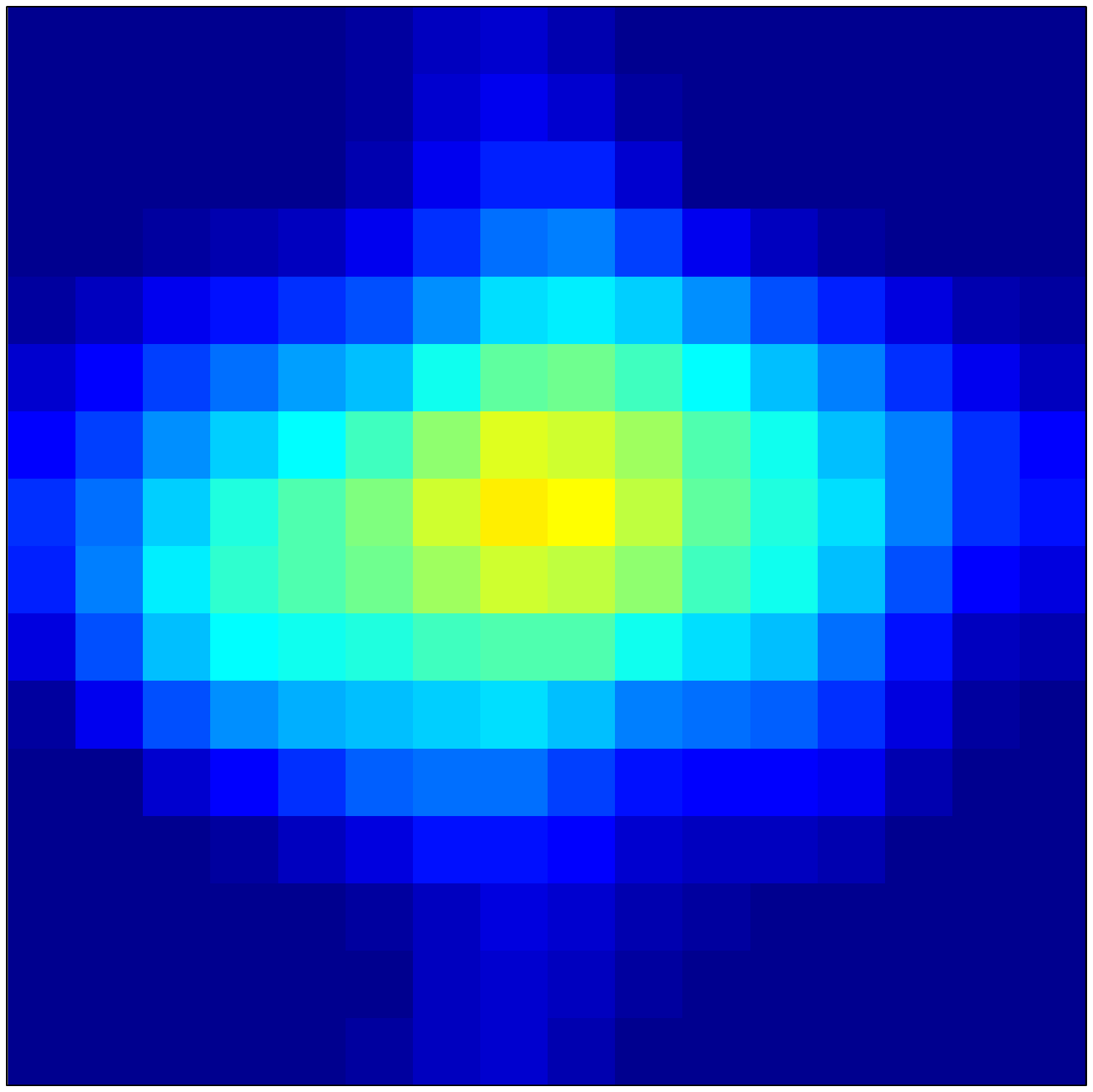}\label{fig:3d}
\hfill \includegraphics[
     clip, keepaspectratio, width = 0.15\textwidth]{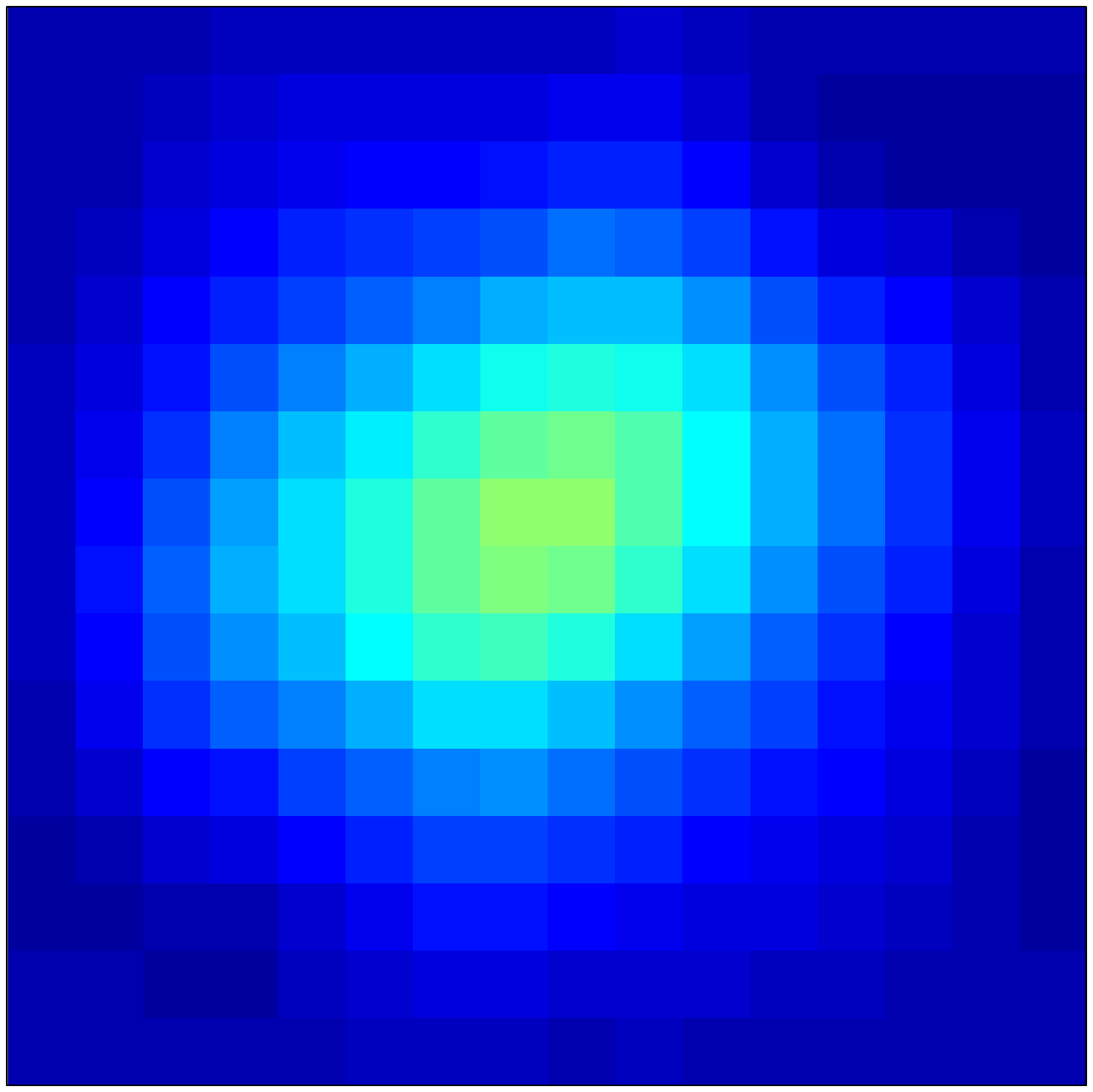}\label{fig:3e} \\
   \vspace{2mm}  
 \includegraphics[
    clip, keepaspectratio, width = 0.15\textwidth]{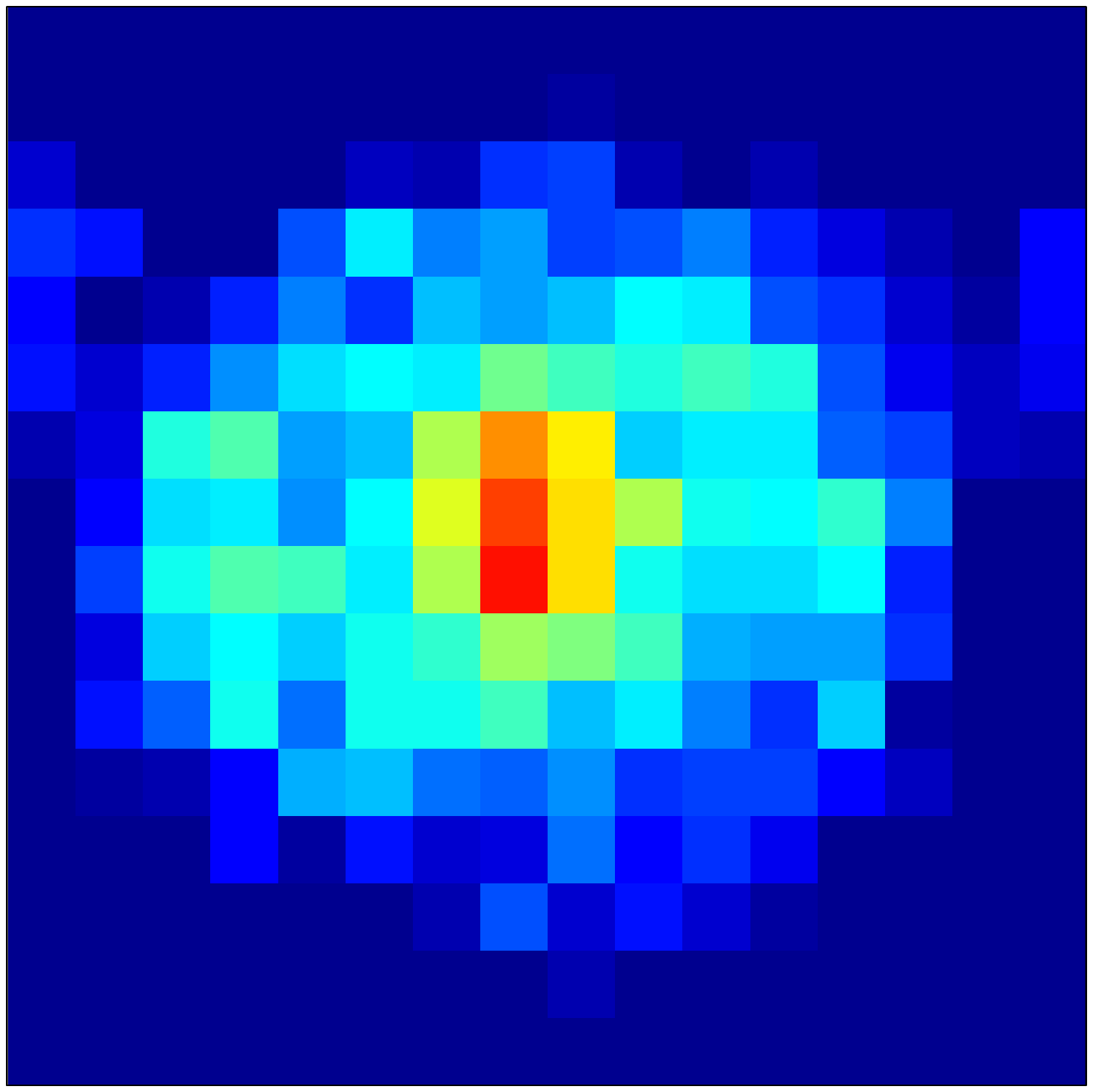}\label{fig:3b} 
\hfill \includegraphics[
     clip, keepaspectratio, width = 0.15\textwidth]{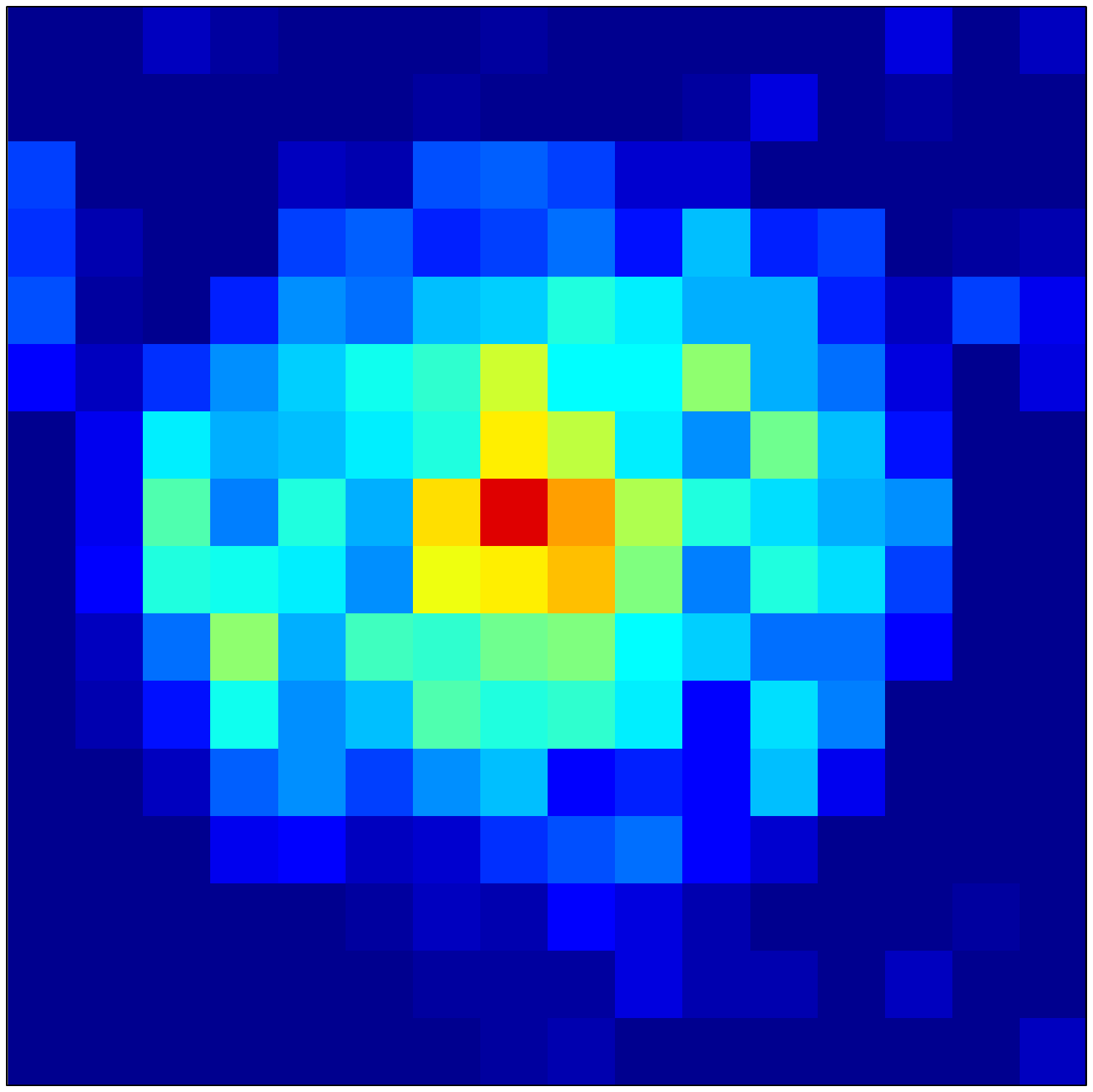}\label{fig:3d}
\hfill \includegraphics[
     clip, keepaspectratio, width = 0.15\textwidth]{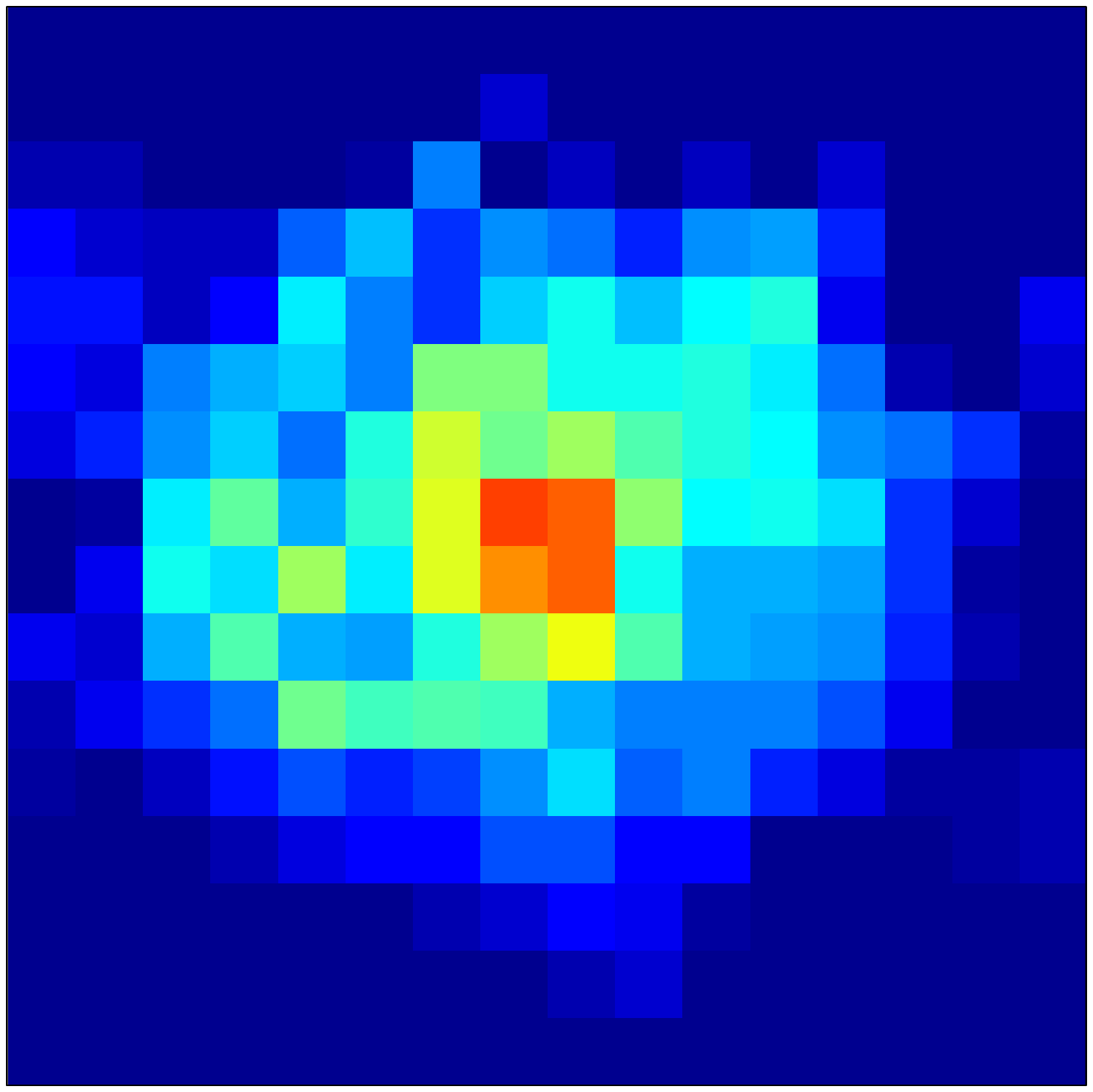}\label{fig:3e} \\
      \vspace{2mm}
     \includegraphics[
    clip, keepaspectratio, width = 0.15\textwidth]{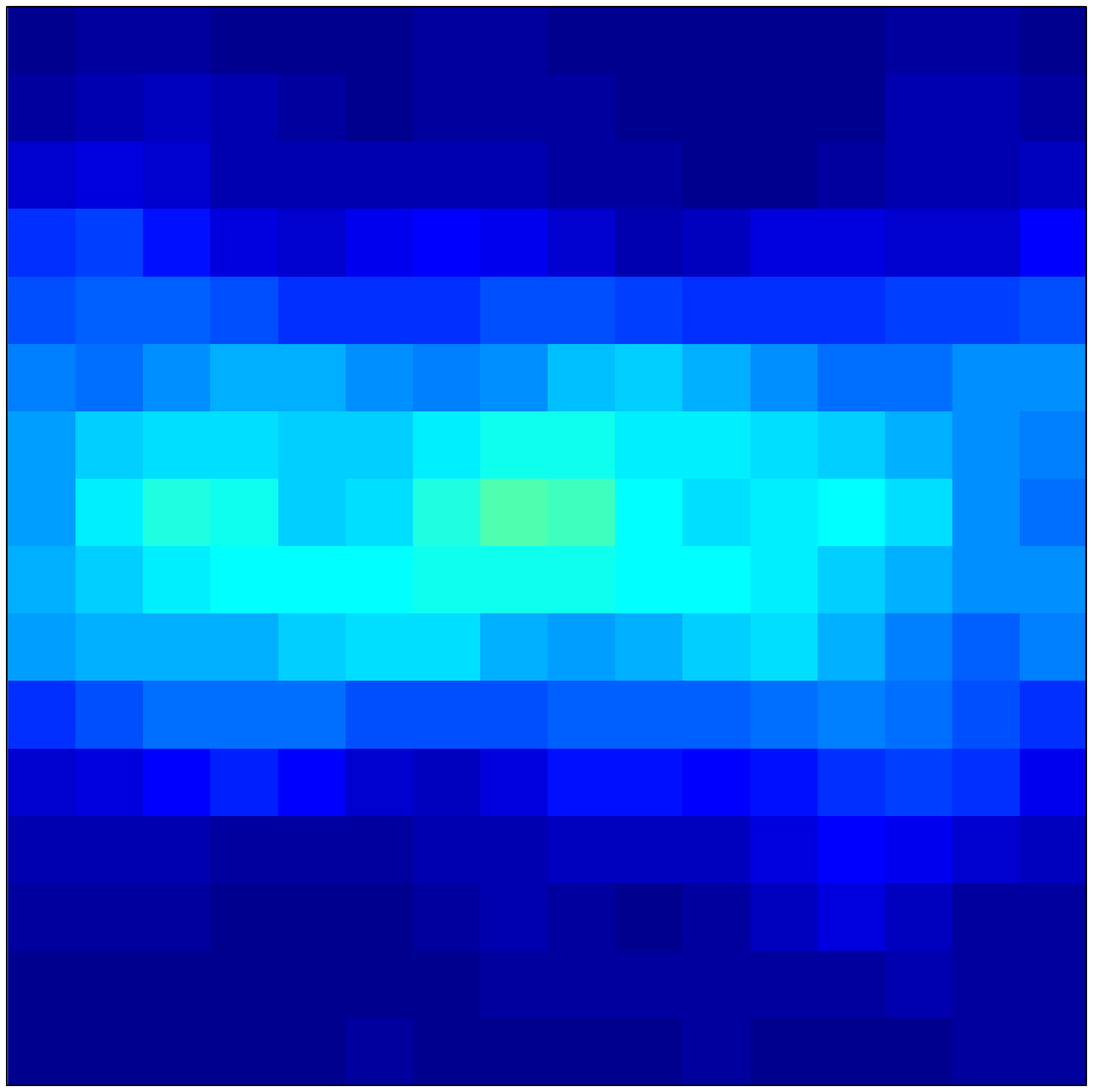}\label{fig:3b} 
\hfill \includegraphics[
     clip, keepaspectratio, width = 0.15\textwidth]{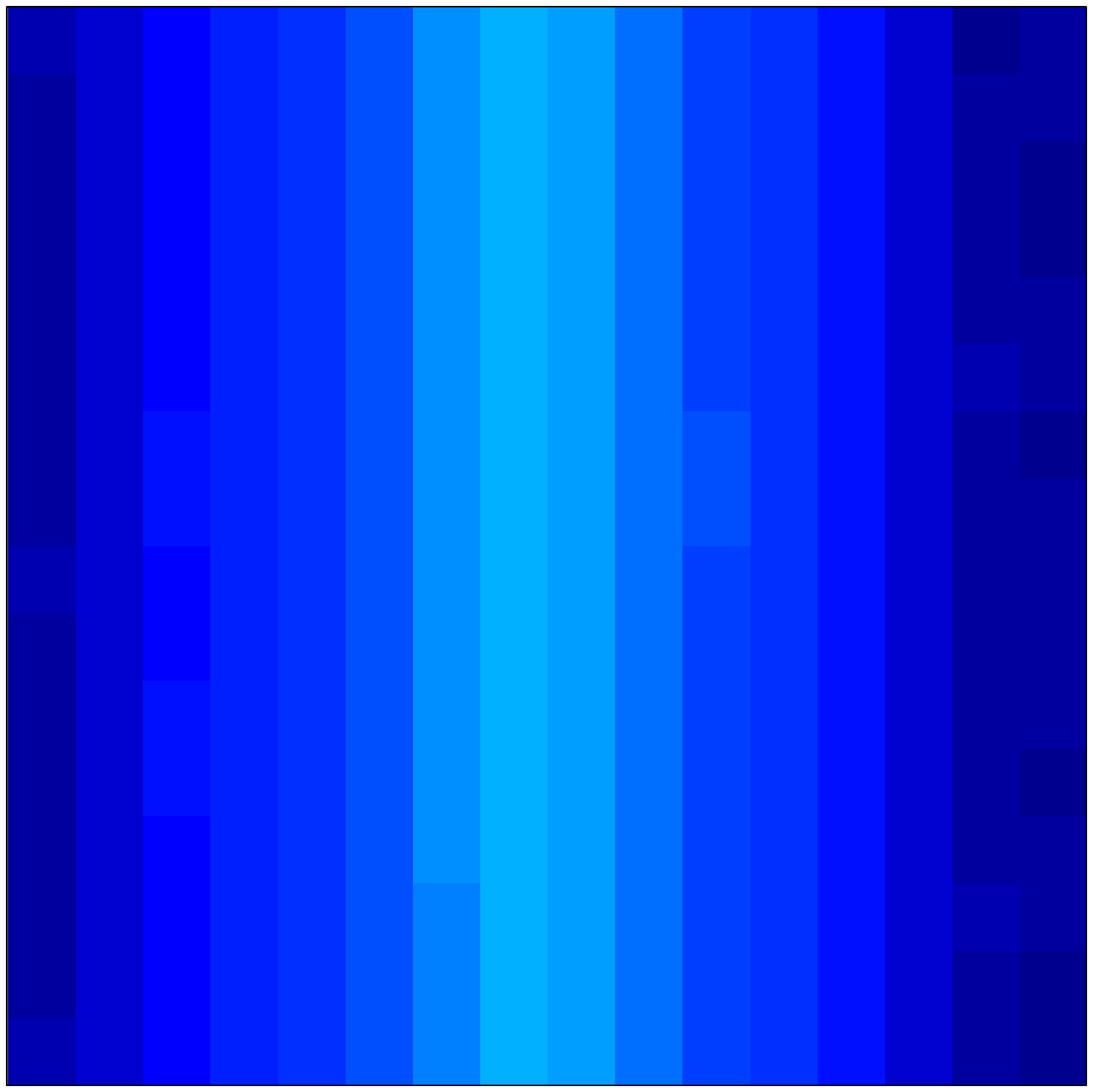}\label{fig:3d}
\hfill \includegraphics[
     clip, keepaspectratio, width = 0.15\textwidth]{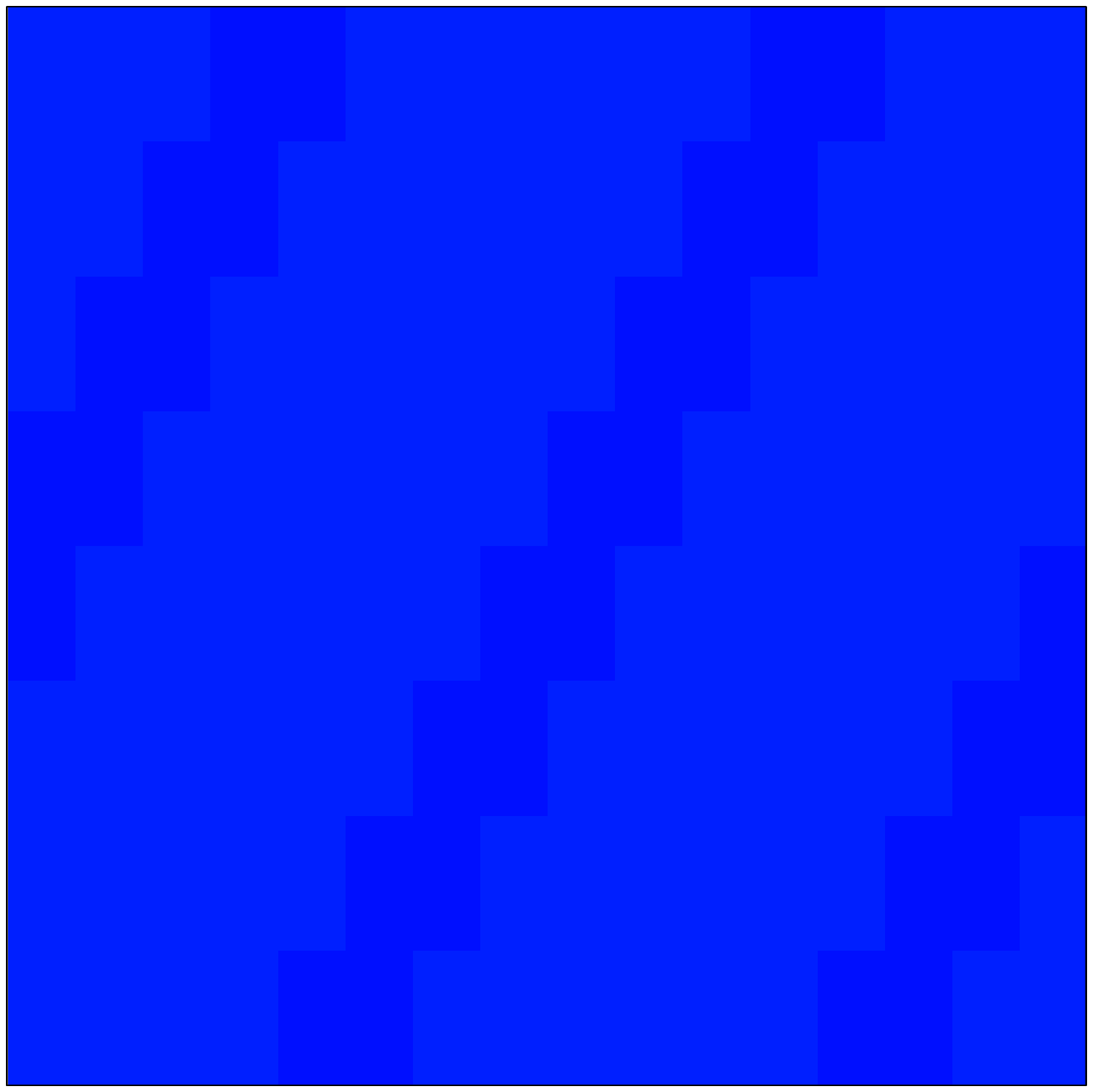}\label{fig:3e} \\
   \vspace{2mm} 
   \includegraphics[
    clip, keepaspectratio, width = 0.15\textwidth]{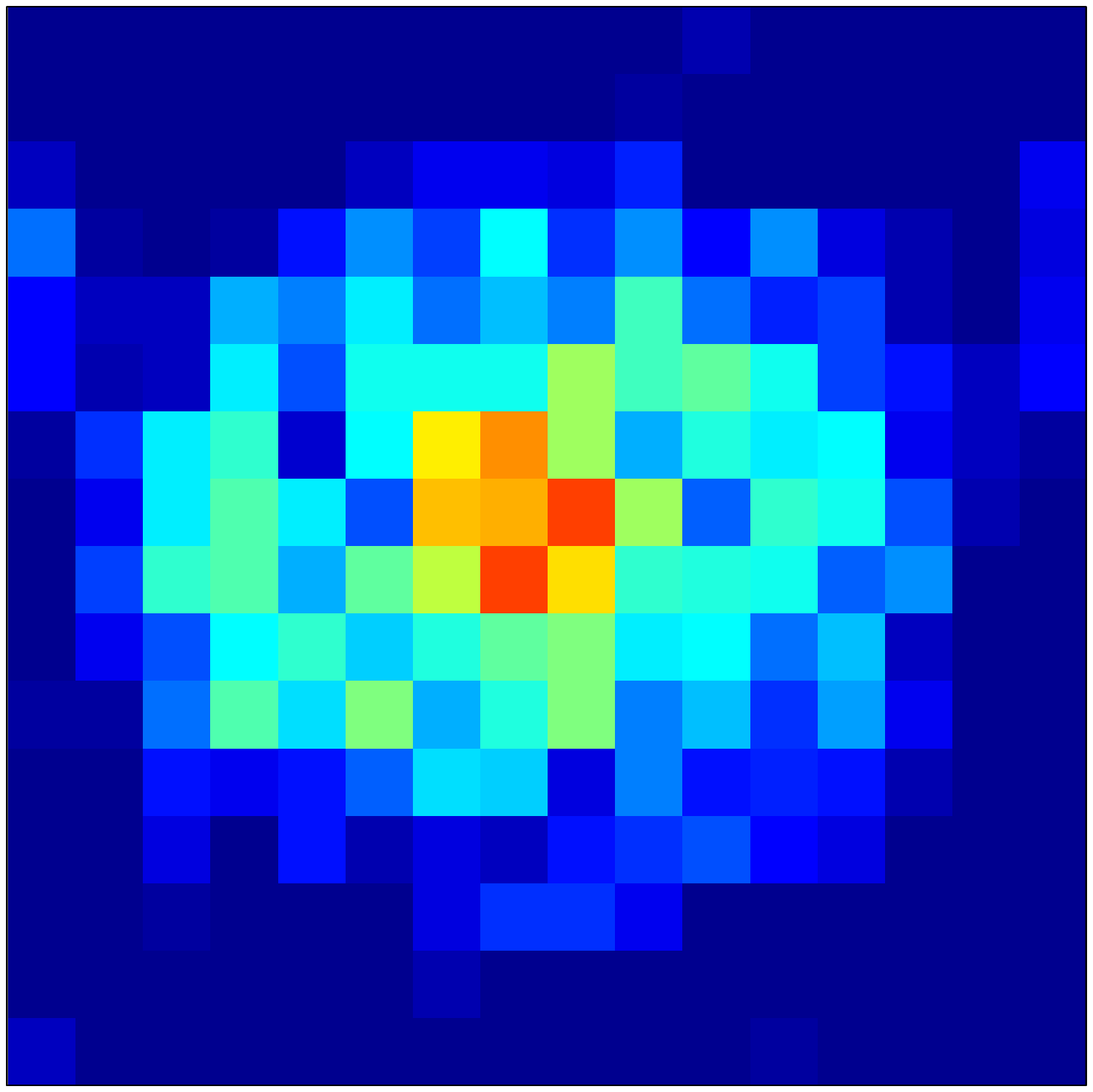}\label{fig:3b} 
\hfill \includegraphics[
     clip, keepaspectratio, width = 0.15\textwidth]{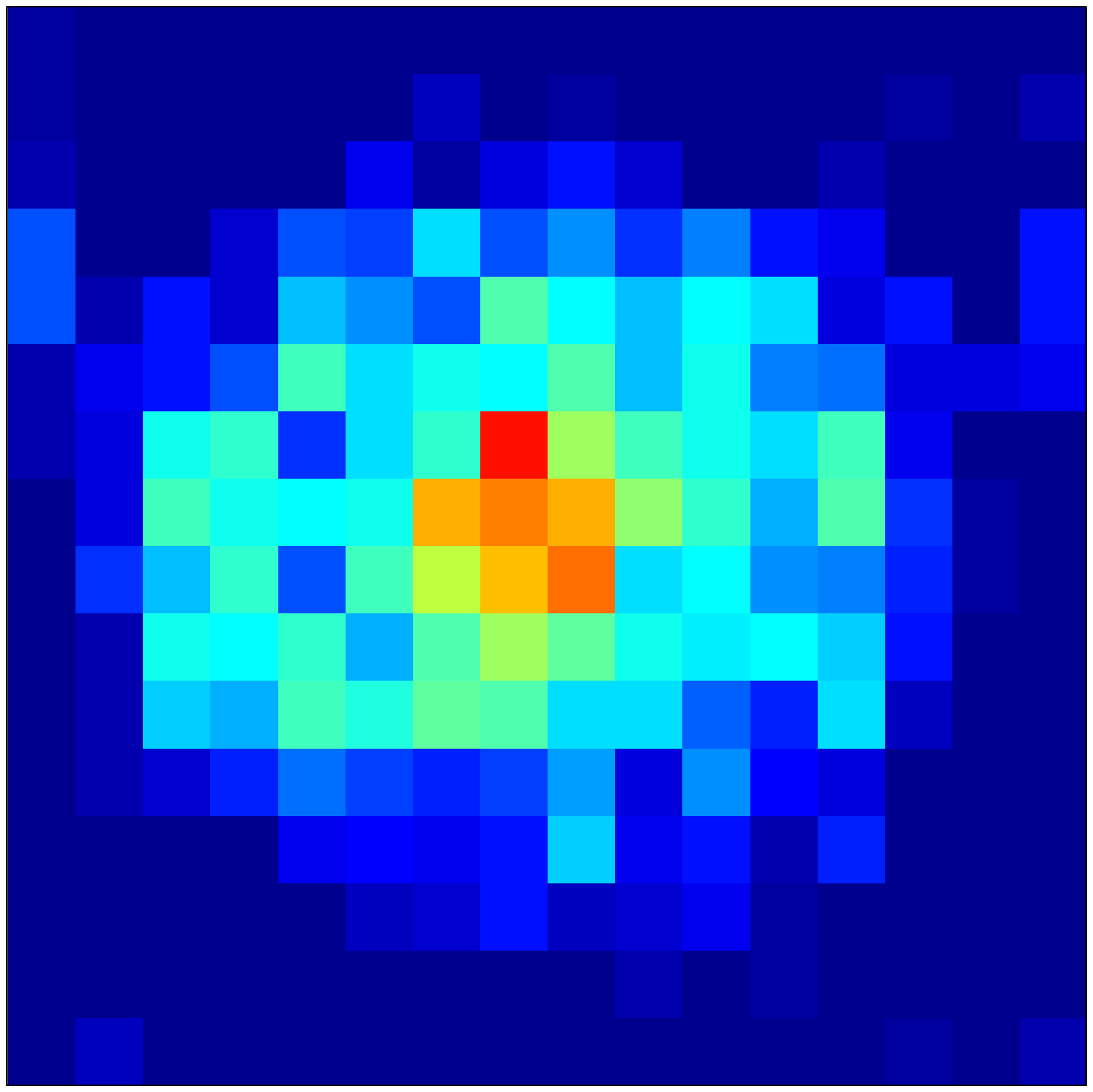}\label{fig:3d}
\hfill \includegraphics[
     clip, keepaspectratio, width = 0.15\textwidth]{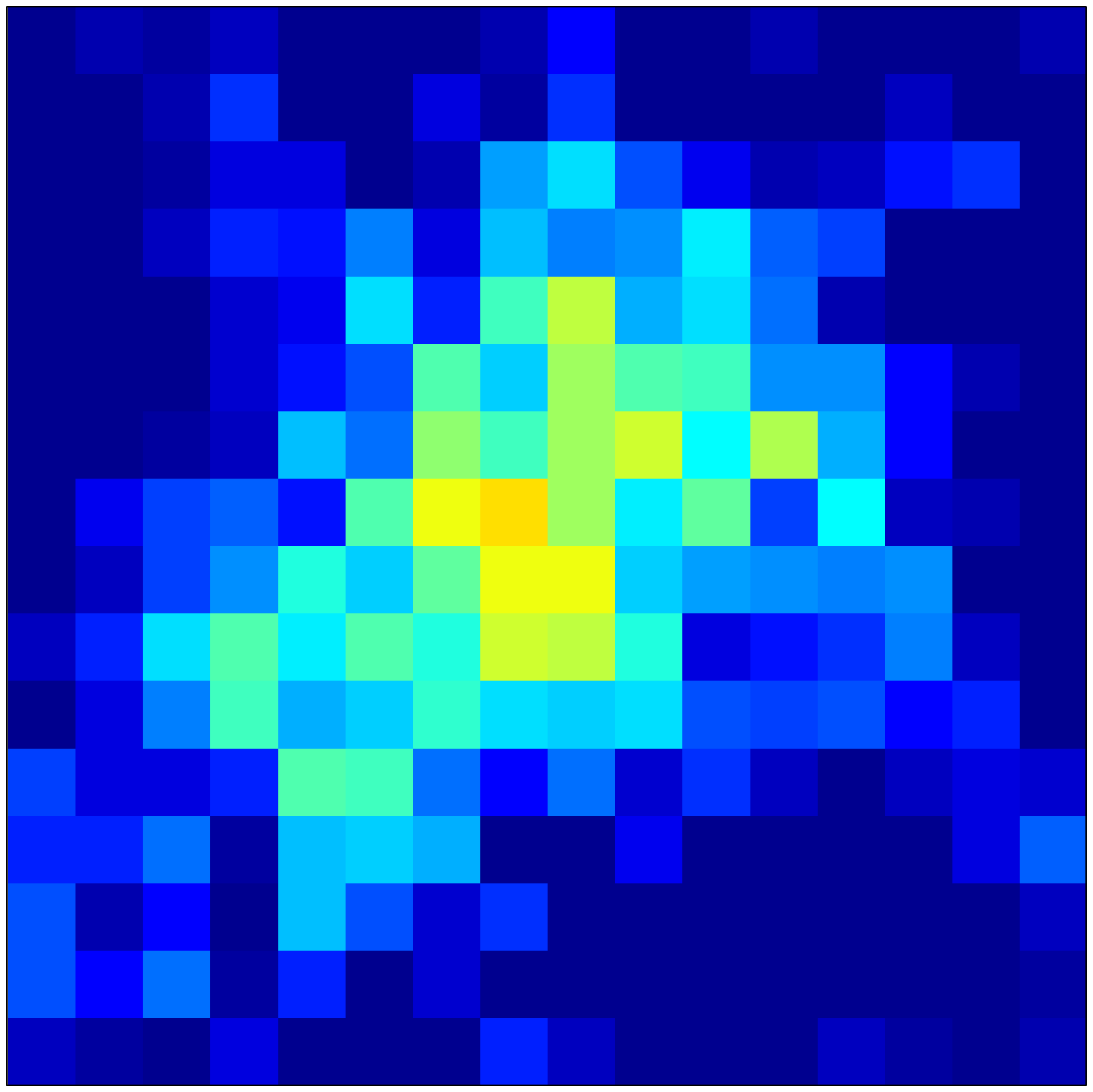}\label{fig:3e} \\
  
\caption{M51 Galaxy illustration ($N=16^2$, ISNR $=$ 30dB). Top row: original image and SNR graph. The curves represent the average SNR values over multiple simulations ($50$ for AM and $10$ for NM) and corresponding 1-standard-deviation error bars. Second and third rows: NM (second) and AM for $n_{\rm ri}=5$ (third) reconstructions with best SNR for $M=N$ (left), $M=0.75N$ (centre) and $M=0.25N$ (right). Fourth and bottom rows: NM (fourth) and AM for $n_{\rm ri}=5$ (bottom) reconstructions  with median SNR for $M=N$ (left), $M=0.75N$ (centre) and $M=0.25N$ (right).}
\label{fig:cm51}
\end{figure}

Secondly, simulations are performed in an identical setting on realistic images representing low-resolution versions of the Eta Carinae star system, of a simulated rapidly rotating star, and of the M51 Galaxy\footnote{Images from \cite{Renard:arXiv1106.4508} downloaded from the JMMC service at apps.jmmc.fr/oidata/shared/srenard/.}. The multiple simulations per point are performed by varying the sampling and noise realisations. The graphs in Figures \ref{fig:eta}, \ref{fig:rapid}, and \ref{fig:cm51} present the SNR curves as a function of undersampling  in the range $[0.25, 1]$ (AM only reported for $n_{\rm ri}=5$), confirming the previous results on random images. Reconstructed images are also reported, providing visual confirmation of the superiority of AM relative to NM over the full undersampling range. In both approaches, the visual quality difference between the reconstructions with, respectively, best and median SNR values illustrates the variability of the reconstruction quality. The NM approach suffers from a significantly larger visual degradation of median SNR value at $M=0.25N$ than AM. This degradation appears at larger sampling ratios for M51.

Let us highlight that, while only $5$ reinitialisations are necessary in the AM approach in low dimension to reach saturation, additional experimental tests on random signals of size $N=64^2$ show that $n_{\rm ri}=20$ or larger is necessary for a meaningful reconstruction, thereby emphasising the convergence problem due to nonconvexity in higher dimension. Also, computation time scales linearly with $n_{\rm ri}$ and can rapidly blow up in this context.

\section{Conclusion}
\label{sec:Conclusion}

We have proposed a novel linear formulation of the optical-interferometric imaging problem in terms of the supersymmetric rank-1 order-3 tensor formed by the tensor product of the vector representing the image sought with itself. In this context, we proposed a linear convex approach for tensor recovery with built-in supersymmetry, and regularising the inverse problem through nuclear norm minimisation. We have also studied a nonlinear nonconvex alternate minimisation approach where supersymmetry is relaxed while the rank-1 constraint is built-in. While the former approach is associated with drastically increased dimensionality of the unknown, the underlying convexity ensures essential properties of convergence to a global minimum of the objective function and independence to initialisation, justifying its analysis. Simulation results in low dimension show that the AM scheme provides significantly superior imaging quality than the NM approach, in addition to be much lighter in its memory requirements and computation complexity. Another set of results in higher dimension however suggests that the number of necessary reinitialisations for the nonconvex AM scheme rapidly increases with $N$. This state of things clearly calls for further considerations of a purely convex approach.

Future work should address sparsity constraints along the lines of the recent evolutions brought in radio interferometry \citep{wia09a, wia09b, MNR:MNR21605} and in optical interferometry \citep{thie2010, Renard:arXiv1106.4508}. Our approaches should also be studied in a more realistic setting with exact power spectrum and bispectrum measurements in the continuous domain and for different noise statistics, and explicitly compared to existing MiRA and WISARD implementations. Software and hardware optimisation will also be key to handle large-size images, e.g. using graphics processing units \citep{bar10}. 

\section*{Acknowledgments}

The authors thank G. Puy for insightful discussions on optimisation algorithms. A.~Aur{\'i}a and R. E.~Carrillo are supported by the Swiss National Science Foundation (SNSF) under grants 205321\_138311 and 200021\_140861. Y.~Wiaux is supported by the Center for Biomedical Imaging (CIBM) of the Geneva and Lausanne Universities, EPFL and the Leenaards and Louis-Jeantet foundations.

\bibliographystyle{mymnras_eprint}
\bibliography{bib}

\label{lastpage}

\end{document}